\documentclass[12pt]{iopart}
\usepackage[dvips]{graphicx}
\usepackage[dvips]{color}
\newcommand {\nc} {\newcommand}
\newcommand {\beq} {\begin{eqnarray}}
\nc {\eol} {\nonumber \\}
\nc {\eoln}[1] {\label {#1} \\}
\nc {\eeq} {\nonumber \end{eqnarray}}
\nc {\eeqn}[1] {\label {#1} \end{eqnarray}}
\newcommand{\cpg}{\mbox{$^{12}{\rm C}$(p$,\gamma)^{13}$N}}
\nc {\ve} [1] {\mbox{\boldmath $#1$}}
\nc {\rref} [1] {(\ref{#1})}
\nc {\uint} {\mbox{$u_{l}^{\rm int}$}}
\nc {\uext} {\mbox{$u_{l}^{\rm ext}$}}
\nc {\uintm} [1] {\mbox{$u_{l,#1}$}}
\nc {\uintmp} [1] {\mbox{$u'_{l,#1}$}}
\nc {\la} {\mbox{$\langle$}}
\nc {\ra} {\mbox{$\rangle$}}
\nc {\dem} {\mbox{$\frac{1}{2}$}}
\nc {\cL} {\mbox{${\cal L}$}}
\nc {\cM} {\mbox{${\cal M}$}}
\nc {\arrow} [2] {\mbox{$\ \mathop{\longrightarrow}
\limits_{#1 \rightarrow #2}$}\ }
\nc {\wiggle} [2] {\mbox{$\mathop{\sim}
\limits_{#1 \rightarrow #2}$}}
\begin{document}

\review[The $R$-matrix Theory]{The $R$-matrix Theory}
\author{P Descouvemont}

\address{Physique Nucl\'eaire Th\'eorique et Physique Math\'ematique, C.P. 229,\\
Universit\'e Libre de Bruxelles (ULB), B 1050 Brussels, Belgium}
\ead{pdesc@ulb.ac.be}
\author{D Baye}
\address{Physique Quantique, C.P. 165/82,  \\
Physique Nucl\'eaire Th\'eorique et Physique Math\'ematique, C.P. 229,\\
Universit\'e Libre de Bruxelles (ULB), B 1050 Brussels, Belgium}
\ead{dbaye@ulb.ac.be}

\begin{abstract}
The different facets of the $R$-matrix method are presented pedagogically in a general framework. 
Two variants have been developed over the years: 
$(i)$ The "calculable" $R$-matrix method is a calculational tool to derive scattering properties 
from the Schr\"odinger equation in a large variety of physical problems. 
It was developed rather independently in atomic and nuclear physics with too little mutual influence. 
$(ii)$ The "phenomenological" $R$-matrix method is a technique to parametrize various types 
of cross sections. 
It was mainly (or uniquely) used in nuclear physics. 
Both directions are explained by starting from the simple problem of scattering by a potential. 
They are illustrated by simple examples in nuclear and atomic physics. 
In addition to elastic scattering, the $R$-matrix formalism is applied to transfer and radiative-capture reactions. 
We also present more recent and more ambitious applications of the theory in nuclear physics. 
\end{abstract}

\submitto{\RPP}

\setcounter{equation}{0}
\renewcommand{\theequation}{1.\arabic{equation}}
\section{Introduction}
\label{sec:intro}

\subsection{Principle}

The $R$-matrix theory is a powerful tool of quantum physics, introduced by 
Wigner and Eisenbud \cite{Wi46,Wi46b,WE47} 
where they simplified an original idea of Kapur and Peierls \cite{KP38}. 
The advantage of their simplification is that the $R$ matrix only 
involves real energy-independent parameters. 
Initially the theory was aimed at describing resonances 
in nuclear reactions. 
However even the very first developments also contained 
the principle of a technique for solving coupled-channel 
Schr\"odinger equations in the continuum. 

At present, the main aim of the $R$-matrix theory is to describe scattering 
states resulting from the interaction of particles or systems of particles, 
which can be nucleons, nuclei, electrons, atoms, molecules. 
Its principle relies on a division 
of the configuration space into two regions: 
the internal and external regions. 
The boundary between these regions is defined by a parameter 
known as the channel radius. 
This radius is chosen large enough so that, in the external region, 
the different parts of the studied system interact only through 
known long-range forces and antisymmetrization effects can be neglected. 
The scattering wave function is approximated there by its 
asymptotic expression which is known except for some coefficients 
related to the scattering matrix. 
In the internal region, the system is considered as confined. 
Its eigenstates thus form a discrete basis which can be calculated. 
A scattering wave function at an arbitrary energy is expanded in the internal region 
over these square-integrable eigenstates. 
Then, the $R$ matrix, which is the inverse of the logarithmic derivative 
of the wave function at the boundary, can be calculated. 
A matching with the solution in the external region provides the scattering matrix. 
This method can also provide the bound states of the system. 
In this case, the external solution behaves as a decreasing exponential. 
Since the exponential decrease depends on the unknown binding energy, 
an iteration is then necessary. 

The $R$-matrix theory was developed into two different directions 
with little exchange between these variants. 
Many of its practitioners often ignore the progresses about the other aspect of this double-faced method. 

As already mentioned, the original goal was to provide an efficient 
theory for the treatment of nuclear resonances \cite{WE47,LT58}. 
From information on bound states and low-energy resonances, it soon became 
clear that the $R$-matrix theory offers an efficient way for accurately
parametrizing not only resonances but also the non-resonant part of low-energy cross sections
 with a small number of parameters \cite{LT58}. 
An important advantage is that most of these parameters have a physical meaning. 
This first variant of the method is still very important and much employed, 
in particular to parametrize the low-energy cross sections relevant in nuclear astrophysics. 
This version of the $R$-matrix theory will be called hereafter the phenomenological $R$ matrix. 
Its properties are reviewed in \cite{LT58,Br59}. 

The other aspect of the $R$-matrix theory is that it can provide a simple and elegant way 
for solving the Schr\"odinger equation. 
It is especially competitive in coupled-channel problems with large numbers 
of open channels where direct numerical integrations may become unstable. 
An additional advantage is that narrow resonances which can escape a purely 
numerical treatment are easily studied. 
This other facet of the $R$-matrix theory has been mostly developed 
in atomic physics although we shall see that it can also be very useful 
for nuclear-physics applications. 
This variant will be called hereafter the calculable or computational $R$ matrix. 
Its properties are reviewed in \cite{BBD73,BR75b,BRT83,AGL96}. 

A very comprehensive review of the phenomenological $R$-matrix method has been given 
in 1958 by Lane and Thomas \cite{LT58}. 
Their article contains most of the important aspects 
of the phenomenological applications of the $R$ matrix to nuclear physics. 
Many of their results can also be useful for the calculable $R$ matrix. 
However, in 1957, just before that review appeared in print, an important improvement of the method 
was published which is therefore not used in their review. 
Bloch introduced a singular operator defined on the boundary between the 
two regions, now known as the Bloch operator, which allows a more elegant and compact 
presentation of the method \cite{Bl57}. 
The main interest of the Bloch operator is that its use led to extensions of the method 
to more general treatments of the resolution of the Schr\"odinger equation in the internal region 
and opened the way to accurate methods of resolution in atomic and nuclear physics. 
Several reviews on the computational $R$ matrix have been published 
in the context of nuclear physics \cite{BBD73} and of atomic physics \cite{BR75b,BB93,AGL96}. 
Reference \cite{BRT83} deals with both aspects. 
An update is nevertheless timely. 

\subsection{The phenomenological $R$ matrix}

The phenomenological $R$ matrix and most of its applications were already 
exhaustively described fifty years ago in \cite{LT58}. 
Among these applications, let us mention a detailed study of resonances 
and an extension of the method to the description of electromagnetic 
processes. 
As far as we know, all these applications have been made in nuclear physics, 
i.e.\ for the scattering of neutrons on nuclei or of nuclei on nuclei 
with the presence of a repulsive Coulomb barrier. 
Nevertheless, using the method still revealed a number of difficulties. 
In a series of papers, Barker and collaborators provided 
practical solutions to the determination of the $R$-matrix parameters 
from experimental data \cite{BHT68,Ba72,Ba88} 
and applied this framework to the spectroscopy 
and reactions of light nuclei \cite{Ba87c,BF80}. 
They also explained non-intuitive effects such as the Thomas-Ehrman shift 
\cite{BF80}, ghosts of resonances \cite{BCM76} 
and extended the method to further processes 
such as radiative-capture reactions \cite{Ba80,Ba87b,BK91} and 
delayed $\beta$ decay \cite{Ba89}. 
The approach developed by Barker and collaborators has become a standard 
tool for the analysis of low-energy radiative-capture reactions useful 
in astrophysics. 
Recent progresses in the adjustment of $R$-matrix parameters have been performed 
in \cite{AD00,Br02b}. 

The $R$ matrix allows parametrizing various physical processes 
and its determination provides collision matrices and cross sections. 
For each set of good quantum numbers, i.e.\ total angular momentum and parity, 
the dimension of the phenomenological $R$ matrix is equal to the number 
of channels relevant to the physical properties. 
When a single channel is considered, the $R$ matrix for a partial wave 
with orbital momentum $l$ and total angular momentum $J$ 
is a function of the energy $E$ parametrized by the formula 
\beq
R_{lJ}(E) = \sum_{n=1}^N \frac{\gamma_{nlJ}^2}
{E_{nlJ} - E}.
\eeqn{1.1}
In principle, this function possesses an infinity of poles at the real energies $E_{nlJ}$ 
but only a limited number $N$ of such poles affect the low-energy cross sections. 
The lowest poles are closely related to bound states at negative energies 
or to narrow resonances at positive energies. 
Nevertheless, the poles and the energies of physical states are slightly different. 
Because of this shift, the determination of these parameters from data 
requires some skill. 
The real parameters $\gamma_{nlJ}$ are known as the reduced width amplitudes because 
their square is a crucial factor of the width of non-overlapping resonances. 
More precisely, we shall see in section \ref{sec:resbs} 
that the width is given as $\Gamma = 2 \gamma_{nlJ}^2 P_l$ 
where $P_l$ is the penetration factor which includes most of the effects 
of transmission through the Coulomb barrier. 
This factor depends on energy and the width thus also depends on energy. 

A serious drawback of the phenomenological $R$ matrix is that the 
poles and widths depend on the choice of channel radius, 
i.e. on a rather arbitrary value. 
This aspect of the $R$ matrix has been criticized by a number of authors 
and has led to further developments of the competing phenomenological $K$ matrix 
\cite{Hu90}. 
The $K$ matrix, which provides an alternative formulation of the collision matrix, is also
expanded in a series involving an infinity of poles.
This approach is based on a delicate treatment of Coulomb functions. 
In spite of the fact that the $K$ matrix does not contain an arbitrary parameter 
such as the channel radius, 
its parametrization is more difficult because its parameters 
may have a less direct physical interpretation. 

\subsection{The calculable $R$ matrix}

The aim of the calculable $R$ matrix is to provide an efficient way 
of solving the Schr\"odinger equation both at positive and negative energies. 
It was proposed in 1965 by Haglund and Robson and applied to a two-channel 
problem involving square-well potentials \cite{HR65}. 
An expansion over a finite basis was introduced by Buttle \cite{Bu67}. 
He performed the first realistic application on $^{12}$C + n scattering \cite{Bu67}. 
He also proposed a correction to the truncation of the $R$ matrix to a finite number 
of poles, that is now named after him. 
A more serious problem is a discontinuity of the derivative 
of the wave function at the boundary between the regions 
that occurs with the traditional choice of basis states inspired by the original ideas in \cite{WE47}. 
Various solutions to the lack of matching at the boundary have been suggested 
(see \cite{BRT83} for a review). 
This apparent problem has attracted a lot of attention even long after an efficient 
technique where it does not occur was introduced \cite{LR66,LR69}. 
By dropping an unnecessary condition as we will describe, the $R$-matrix method can be very accurate 
without matching problems and without need for a Buttle correction. 

Some users of the phenomenological $R$ matrix consider the channel radius 
as a parameter which must be optimized when fitting the data. 
Even if this dependence on a parameter without strong physical meaning is weak, 
this is a drawback that would not be acceptable 
when aiming at accurately solve the Schr\"odinger equation. 
Hence a crucial test of the results of the calculable $R$ matrix 
is an almost perfect independence with respect to the choice of channel radius. 
This test provides a measure of the accuracy of the calculations. 

In spite of its introduction for nuclear-physics problems \cite{HR65,Bu67}, 
this approach was first extensively developed to study electron (or positron) collisions 
on atoms and molecules \cite{BR75b,BB93,AGL96}. 
It allows describing the excitation and ionization of these systems. 
Photoionization, i.e. collisions with a photon leading to the single 
or double ionization of the atom, is also a well-studied application \cite{BJ03,St82}.

Important and difficult aspects of these atomic-physics problems are 
the non-locality of the interaction due to electron exchanges and  
the long-range nature of the interactions, due to the polarization interactions. 
The non-locality is well treated in the $R$-matrix approach. 
The long range of the force implies that the asymptotic behaviour of the solution 
is only reached for very large values of the interparticle distance. 
To avoid using a very large channel radius, propagation methods have been introduced 
\cite{LW76}. 
They involve an intermediate region where the interaction can be simplified, 
for example with an asymptotic expansion. 

Electron scattering on heavy atoms has required the introduction of relativistic 
corrections. 
The extension of the $R$ matrix to the Dirac equation has been introduced as 
early as in 1948 \cite{Go48} but its validity remained controversial during a 
long time. 
Relativistic corrections were thus first derived from the Breit-Pauli equation. 
The validity of the Dirac extension is now well established and elaborate 
relativistic codes have been developed. 
This aspect will not be covered here (see \cite{Gr08} for a recent review). 

In nuclear physics, the computational $R$ matrix is much less used although 
it should be very useful in large coupled-channel calculations. 
It has been much applied in microscopic cluster calculations in which the 
difficult antisymmetrization is taken into account in the internal region only 
\cite{BH74,BHL77}. 
The versatility of the $R$ matrix found interesting applications in processes 
where bound and scattering states are mixed, such as radiative capture or delayed 
$\beta$ decay \cite{BD83,BD88b}. 
The application of the $R$-matrix method to coupled-channel calculations 
has been simplified by its combination with the Lagrange-mesh method 
which avoids calculating matrix elements of the potentials \cite{Ma94,BHS98,HSV98}. 
Recently this approach has been extended to non-local interactions \cite{HRB02}. 

Other approaches to the same problem present a number of similarities. 
The variational $K$-matrix method \cite{Ka77} has a very similar spirit. 
The Kohn variational principle for the logarithmic derivative is equivalent 
to the calculable $R$ matrix \cite{Me94}. 
Practical implementations of the Gamow-state method \cite{TON97} 
are also exactly equivalent to $R$-matrix calculations \cite{BGS02}.

\subsection{Outline}

In this review, we present both calculable and phenomenological versions of the $R$-matrix theory. 
Very few papers deal with both aspects simultaneously \cite{HCL98,BDL05}. 
Since many excellent reviews already exist, we try to make an introductory 
presentation, illustrated with simple numerical examples. 
We also show the parallel evolutions of the method in atomic and nuclear physics 
and try to shed light on some common misunderstandings or controversies about the $R$-matrix methods. 
The numerical examples that we display are tailored to allow a motivated reader 
to test his/her understanding by reproducing them with limited effort. 
Because of our background as nuclear physicists, most examples (but not all) 
correspond to nuclear applications. 
Finally, we also review state-of-the-art calculations in nuclear physics 
where the $R$ matrix proves useful. 

Contrary to tradition, we start with the calculable $R$ matrix 
on a finite basis, which provides a convenient numerical approach. 
Taking the limit for an infinite complete basis will introduce the theoretical $R$ matrix 
which leads after truncation to the phenomenological approximation \rref{1.1}. 
For the sake of simplicity, we detail potential scattering in the single-channel 
case so avoiding the unpedagogical definitions of channel wave functions. 
We only provide the main steps for the multichannel case. 

The bibliography about the $R$ matrix is enormous and can not be fully covered here. 
We have tried to quote papers that we think significant or useful for further bibliographic research. 

In section \ref{sec:coll}, we introduce the necessary basics of scattering theory 
with radiative capture as a more elaborate application. 
The calculable $R$ matrix is presented in section \ref{sec:comp} 
and relatively simple numerical applications in section \ref{sec:apcomp}. 
The phenomenological $R$ matrix and and its applications are presented in section \ref{sec:phen}. 
Recent elaborate calculations in nuclear physics are reviewed in section \ref{sec:recent}.  
Concluding remarks are made in section \ref{sec:conc}. 

\setcounter{equation}{0}
\renewcommand{\theequation}{2.\arabic{equation}}
\section{Summary of scattering theory}
\label{sec:coll}

\subsection{Coulomb scattering}

Consider the collision of two particles with respective masses $m_1$ and $m_2$ 
and charges $Z_1e$ and $Z_2e$ at a positive energy $E$ in the centre-of-mass frame. 
The wavenumber is defined as 
\beq
k = \sqrt{2\mu E}/\hbar,
\eeqn{2.1}
where $\mu = m_1 m_2/(m_1+m_2)$ is the reduced mass. 

Let us start with some definitions about pure Coulomb scattering. 
In this case, a Bohr radius can be defined as 
\beq
a_B = \frac{\hbar^2}{\mu |Z_1 Z_2| e^2}.
\eeqn{2.2}
A useful parameter is the dimensionless Sommerfeld parameter 
\beq
\eta = \frac{Z_1 Z_2 e^2}{\hbar v} = \frac{{\rm sgn}(Z_1 Z_2)}{a_B k}
\eeqn{2.3}
where $v = \hbar k/\mu$ is the relative velocity. 
Parameter $\eta$ measures the importance of Coulomb effects at a given energy. 
The neutral case is recovered with $\eta = 0$.

For a central potential, a wave function can be factorized 
in spherical coordinates $\ve{r} = (r,\Omega)$ as 
$\psi(\ve{r}) = r^{-1} u_l (r) Y_l^m (\Omega)$. 
The spherical harmonics $Y_l^m (\Omega)$ depend on the orbital and magnetic quantum 
numbers $l$ and $m$, and on the angles $\Omega = (\theta,\varphi)$. 
They are defined according to the convention of Condon and Shortley. 
The radial Schr\"odinger equation for the Coulomb problem in partial wave $l$ then reads 
\beq
\left( \frac{d^2}{dr^2} - \frac{l(l+1)}{r^2} 
- \frac{2k \eta}{r} + k^2 \right) u_{l}(r) = 0.
\eeqn{2.4}
Its solutions are combinations of the regular and irregular Coulomb functions 
$F_l (\eta,kr)$ and $G_l (\eta,kr)$ \cite{AS72}. 
The regular function vanishes at the origin and is normalized at infinity 
according to 
\beq
F_l (\eta,x) \arrow{x}{\infty} \sin (x - \dem l \pi - \eta \ln 2x + \sigma_l),
\eeqn{2.5}
where appears the Coulomb phase shift 
\beq
\sigma_l = \arg \Gamma (l + 1 + i\eta)
\eeqn{2.6}
involving the Euler function $\Gamma$. 
The irregular function $G_l (\eta,x)$ is unbound at the origin (except for $\eta = l = 0$) 
and is fixed by its asymptotic behaviour 
\beq
G_l (\eta,x) \arrow{x}{\infty} \cos (x - \dem l \pi - \eta \ln 2x + \sigma_l).
\eeqn{2.7}
Also very useful are the conjugate functions 
\beq
I_l (\eta,x) = G_l - i F_l, \qquad O_l (\eta, x) = G_l + i F_l,
\eeqn{2.8}
which behave asymptotically like incoming and outgoing waves, respectively, 
\beq
I_l (\eta,x) \arrow{x}{\infty} e^{-i(x - \dem l \pi - \eta \ln 2x + \sigma_l)},
\qquad
O_l (\eta,x) \arrow{x}{\infty} e^{i(x - \dem l \pi - \eta \ln 2x + \sigma_l)}.
\eeqn{2.8a}
Notice that we do not follow here the same phase convention as Lane and Thomas \cite{LT58}. 
In the neutral case, one has $F_l (0,x) = x j_l (x)$ and $G_l (0, x) = x n_l (x)$ 
where $j_l$ and $n_l = -y_l$ are spherical Bessel functions \cite{AS72}. 

In some applications, solutions of \rref{2.4} are also needed at negative energies. 
The real solution decreasing at infinity is the Whittaker function $W_{-\eta_B,l+\frac{1}{2}}(2\kappa r)$ 
\cite{AS72}. 
It depends on the wave number $\kappa = \sqrt{-2\mu E}/\hbar$ 
and on the Sommerfeld parameter $\eta_B = {\rm sgn}(Z_1 Z_2)/a_B \kappa$ of the bound state. 
Whittaker functions behave asymptotically as 
\beq
W_{-\eta_B,l+\frac{1}{2}}(x) \arrow{x}{\infty} x^{-\eta_B} e^{-x/2}.
\eeqn{2.8b}
They are singular at the origin. 

A bounded solution of the three-dimensional Schr\"odinger equation at a positive energy 
with the same Coulomb potential is given by \cite{Me62} 
\beq
\psi^+_C (\ve{r}) = (2\pi)^{-3/2} e^{-\pi \eta/2} \Gamma(1+i\eta) 
e^{ikz}\, {_1F_1} (-i\eta, 1, ik(r-z))
\eeqn{2.9a}
where ${_1F_1}$ is the confluent hypergeometric function \cite{AS72}. 
This wave function has the asymptotic behaviour of an outgoing scattering state 
\beq
\psi^+_C (\ve{r}) \arrow{|r-z|}{\infty} (2\pi)^{-3/2} 
\left( e^{i[kz + \eta \ln k(r-z)]} 
+ f_C (\Omega)\, \frac{e^{i(kr - \eta \ln 2kr)}}{r} \right).
\eeqn{2.10}
The coefficient of the second term is the Coulomb scattering amplitude, 
\beq
f_C (\Omega) = -\frac{\eta}{2k\sin^2 \dem\theta}\, 
e^{2i(\sigma_0 - \eta \ln\sin\dem\theta)}.
\eeqn{2.11}
The square of the modulus of $f_C$ provides the Rutherford cross section. 
Function $\psi^+_C$ can be expanded in partial waves as 
\beq
\psi^+_C (\ve{r}) = (2\pi)^{-3/2} (kr)^{-1} 
\sum_{l=0}^{\infty} (2l+1) i^l e^{i\sigma_l} P_l (\cos \theta) F_l(\eta,kr),
\eeqn{2.9}
where $P_l$ is a Legendre polynomial \cite{AS72}. 

\subsection{Scattering by a potential}

Consider a potential $V$ tending to the Coulomb potential faster than $r^{-2}$, 
\beq
V(r) \arrow{r}{\infty} \frac{Z_1 Z_2 e^2}{r} + o\left(\frac{1}{r^2}\right).
\eeqn{2.19}
The radial Schr\"odinger equation in partial wave $l$ reads 
\beq
\left( \frac{d^2}{dr^2} - \frac{l(l+1)}{r^2} - \frac{2\mu V(r)}{\hbar^2} 
+ k^2 \right) u_{l}(r) = 0
\eeqn{2.20}
with the condition at the origin 
\beq
u_l (0) = 0. 
\eeqn{2.20a}
A real solution at positive energy $E$ behaves asymptotically as 
\beq
u_{l} (r) \arrow{r}{\infty} \cos \delta_l F_l (\eta,kr) + \sin \delta_l G_l (\eta,kr),
\eeqn{2.21b}
up to a normalization factor. 
The important physical quantity is the phase shift $\delta_l$. 
For later use, it is however more convenient to write the solution as 
\beq
u_{l} (r) \arrow{r}{\infty} C_l [I_l (\eta,kr) - U_l O_l (\eta,kr)],
\eeqn{2.21a}
where $C_l$ can be chosen in various ways. 
For example, \rref{2.21b} is recovered with $C_l = i \exp(-i\delta_l)/2$ 
and the normalization of the $u_l$ with respect to $\delta(k-k')$ is obtained 
with $C_l = i \exp(-i\delta_l)/\sqrt{2\pi}$.
The collision or scattering `matrix' $U_l$ is given by 
\beq
U_l = e^{2i\delta_l}.
\eeqn{2.24}

With the different partial solutions, one can construct an outgoing stationary solution 
\beq
\Psi^+ (\ve{r}) = (2\pi)^{-3/2} (2kr)^{-1} 
\sum_{l=0}^{\infty} (2l+1) i^{l+1} e^{i\sigma_l} P_l (\cos \theta) C_l^{-1} u_{l} (r)
\eeqn{2.22}
behaving asymptotically as a Coulomb wave \rref{2.9a} propagating in the $z$ direction 
plus an outgoing spherical wave
\beq
\Psi^+ (\ve{r}) \arrow{r}{\infty} \psi^+_C (\ve{r}) 
+ (2\pi)^{-3/2} f(\Omega)\, \frac{e^{i(kr-\eta \ln 2kr)}}{r}.
\eeqn{2.22a}
The coefficient of the second term in this asymptotic expression determines the additional scattering amplitude 
\beq
f (\Omega) = \frac{1}{2ik} \sum_{l=0}^{\infty} 
(2l+1) e^{2i\sigma_l} (U_l-1) P_l (\cos \theta).
\eeqn{2.23}
From \rref{2.23}, one obtains the elastic cross section 
\beq
\frac{d\sigma}{d\Omega} = |f_C(\Omega) + f (\Omega)|^2
\eeqn{2.25}
also involving the Coulomb amplitude \rref{2.11}. 
The scattering wave function \rref{2.22} is useful in various types 
of reactions. 
We illustrate it below with radiative capture. 

\subsection{Collisions in a many-body system}

An $N$-body system is described with the microscopic Hamiltonian 
\beq
H = T + V = \sum_{i=1}^N T_i + \sum_{i>j=1}^N V_{ij}
\eeqn{2.38}
where for simplicity we only display two-body forces. 
This Hamiltonian is invariant under rotations, translations and reflections. 
We do not display nor discuss the removal of the centre of mass. 

At positive excitation energies, several channels may be open. 
In each channel, the particles are divided into various groups. 
Each such division is known as a partition. 
A given partition is denoted as $\alpha$. 
For simplicity, we only consider here channels 
where the particles form only two subsystems: $N = N_{\alpha}^{(1)} + N_{\alpha}^{(2)}$. 
Both subsystems of partition $\alpha$ are described with an internal Hamiltonian $H_\alpha^{(i)}$ 
($i = 1, 2$) which has the form \rref{2.38} with $N$ replaced by $N_\alpha^{(i)}$. 

A channel $c$ is defined by specifying in addition the energy of each subsystem, 
i.e. a certain eigenvalue for each internal Hamiltonian. 
The exact or approximate energies $E^{(i)}_{c}$ and wave functions $\phi^{(i)}_{c}$ 
are related by 
\beq
E^{(i)}_{c} = \la \phi^{(i)}_{c} | H^{(i)}_c | \phi^{(i)}_{c} \ra.
\eeqn{2.39}
Here and in what follows, subscript $c$ has a variable symbolic meaning 
depending on the considered quantity. 
In $H^{(i)}_c \equiv H_\alpha^{(i)}$, it represents the set of internal coordinates 
of subsystem $i$. 
In $E^{(i)}_{c}$, it represents a set of quantum numbers. 
In $\phi^{(i)}_{c}$, it means both. 
When $\phi^{(i)}_{c}$ is not an exact eigenfunction of $H^{(i)}_c$, 
\rref{2.39} remains valid from a variational perspective. 

To each partition $\alpha$ of the system into two subsystems 
may correspond several channels differing by their internal states. 
Each channel $c$ has a threshold energy 
\beq
E_c = E^{(1)}_{c} + E^{(2)}_{c}
\eeqn{2.40}
defined with respect to some common reference energy. 
A channel is open or closed according to whether 
$E_c$ is smaller or larger than the total energy $E$ of the system. 
In each open channel, one can define a wave number 
$k_c = \sqrt{2\mu_c (E - E_c)}/\hbar$ and 
a relative velocity $v_c = \hbar k_c/\mu_c$, 
$\mu_c$ being the reduced mass of partition $\alpha$. 
In each closed channel, one can define a wave number 
$\kappa_c = \sqrt{2\mu_c (E_c - E)}/\hbar$. 

For each partition $\alpha$, the relative coordinate $\ve{r}_c \equiv \ve{r}_\alpha$ 
is the difference between the centre-of-mass coordinates of the subsystems. 
The eigenstates of each subsystem $i=1,2$ are characterized by their energy $E^{(i)}_{c}$ 
and by their good quantum numbers, i.e.\ the total angular momentum $I_i$ 
(usually called spin) and its projection $M_i$. 
Under the time reversal operator $K$ \cite{Me62}, they are assumed to transform according to 
the convention 
\beq
K |JM\rangle = (-1)^{J-M} |J-M\rangle. 
\eeqn{2.40a}
A channel state is an eigenstate of the total angular momentum of the full system 
resulting from the coupling of both internal states 
$\phi^{(1)}_{c I_1 M_1}$ and $\phi^{(2)}_{c I_2 M_2}$ 
with a spherical harmonics depending on the angles $\Omega_c$ 
defining the orientation of the relative coordinate $\ve{r}_c$. 
More precisely, a channel state is represented as 
\beq
| c \ra = i^{l_c} \left[[\phi^{(1)}_{c I_1} \otimes \phi^{(2)}_{c I_2}]_{I_c} 
\otimes Y_{l_c}(\Omega_c)\right]^{JM\pi},
\eeqn{2.41}
where $I_c$ is the channel spin resulting from the coupling of $I_1$ and $I_2$, 
$l_c$ is the orbital momentum of the relative motion in channel $c$, 
$J$ is the total angular momentum quantum number of the many-body system, 
$M$ is its projection, and $\pi$ is the total parity. 
Thanks to the phase factor $i^{l_c}$, the channel states transform under time reversal 
according to \rref{2.40a}. 
For simplicity, we do not display the parity quantum numbers 
$\pi_{c1}$ and $\pi_{c2}$ of the subsystems. 
They are related to the total parity by 
\beq
\pi = \pi_{c1} \pi_{c2} (-1)^{l_c}.
\eeqn{2.41b}
Channel states are assumed orthogonal to each other and normed,  
\beq
\la c | c' \ra = \delta_{cc'}.
\eeqn{2.41a}
The orthogonality is not obvious for different partitions $\alpha$ and $\alpha'$. 
In that case, it is only true asymptotically. 

Since Hamiltonian $H$ is invariant under rotation and reflection, 
$J$, $M$ and $\pi$ are good quantum numbers. 
A partial wave of the total wave function of the system at energy $E$ can be written as  
\beq
\Psi_{(c_0)}^{JM\pi} = \sum_{c} {\cal A} |c \rangle r_{c}^{-1} u_{c(c_0)} (r_{c}),
\eeqn{2.42}
where indices $c_0$ and $c$ may represent either all quantum numbers appearing (or understood) 
in the right-hand side of \rref{2.41} or a relevant subset of them. 
Projector ${\cal A}$ performs any antisymmetrization due to the indistinguishability 
of some identical particles. 
These can be electrons in atomic physics or nucleons in nuclear physics 
within the isospin formalism. 
The subscript $(c_0)$ recalls the entrance channel as explained below. 
This wave function is an approximate eigenstate of the full many-body Hamiltonian 
\beq
H \Psi_{(c_0)}^{JM\pi} = E \Psi_{(c_0)}^{JM\pi}.
\eeqn{2.43}
To complete the definition of $\Psi_{(c_0)}^{JM\pi}$, one must specify its asymptotic behaviour. 

We shall describe the asymptotic behaviour in terms of the collision or scattering matrix 
which is usually denoted as $\ve{U}$ in the $R$-matrix context. 
This matrix is interesting physically because it is directly related to cross sections 
but it leads to complex radial wave functions. 
Calculations only involving real radial wave functions are also possible (for real potentials). 
They are dominantly used in atomic physics \cite{BNS87} but are also encountered in nuclear physics 
\cite{BRT83}. 
The relation between both approaches is summarized in appendix \ref{sec:A}. 

The asymptotic behaviour in open channels generalizing \rref{2.21a} is given by  
\beq
u_{c(c_0)} (r_{c}) \arrow{r_{c}}{\infty} C_{c_0} v_{c}^{-1/2} 
[\delta_{cc_0} I_{c} (k_{c} r_{c}) - U_{cc_0} O_{c} (k_{c} r_{c})].
\eeqn{2.44}
In \rref{2.44}, $I_c$ and $O_c$ are defined as in \rref{2.8} and $C_{c_0}$ is arbitrary 
(see appendix \ref{sec:A}). 
The asymptotic form is chosen in such a way that incoming flux only 
occurs in the entrance channel $c_0$. 
Taking all possible entrance channels $c_0$ into account, 
the coefficients $U_{cc_0}$ of the outgoing waves form the collision matrix $\ve{U}$. 
It is defined for each angular momentum $J$ and parity $\pi$. 
Its dimension is given by the number of open channels at energy $E$. 
Closed channels may also appear in expansion \rref{2.42} 
but the asymptotic behaviour of the corresponding radial functions 
is exponentially decreasing according to \rref{2.8b}. 

For real potentials, thanks to the introduction of coefficients $v_{c}^{-1/2}$, 
current conservation imposes that the collision matrix is unitary, 
\beq
\ve{U} \ve{U}^\dagger = \ve{U}^\dagger \ve{U} = \ve{1}.
\eeqn{2.44a}
Because of time-reversal invariance, it is also symmetric,
\beq
\ve{U} = \ve{U}^{\rm T}
\eeqn{2.44b}
where T means transposition. 
This property imposes the phase $i^{l_c}$ in the definition \rref{2.40} 
of the channel states. 
As shown by Huby \cite{Hu54}, this factor is missing in some important references \cite{WE47,BB52} 
and some phases must be corrected accordingly. 
This correction is not necessary if the symmetry property \rref{2.44b} is never used. 
A matrix with properties \rref{2.44a} and \rref{2.44b} can be diagonalized with an orthogonal (real) matrix $S$, 
\beq
S \ve{U} S^{\rm T} = e^{2 i \ve{\delta}},
\eeqn{2.44c}
where $\ve{\delta}$ is a diagonal matrix whose elements are the eigenphases $\delta_n$. 

Introducing \rref{2.42} in the Schr\"odinger equation \rref{2.43} and projecting 
over a channel wave function $|c\ra$ leads to the coupled equations 
\beq
\left[T_c + V_c(r) + E_{c} - E \right] u_{c(c_0)}(r) 
+ \sum_{c'} \int_0^\infty W_{cc'}(r,r') u_{c'(c_0)}(r') dr' = 0.
\eeqn{2.45}
They involve the kinetic-energy operators 
\beq
T_c = -\frac{\hbar^2}{2\mu_c} \left( \frac{d^2}{dr^2} - \frac{l_c(l_c+1)}{r^2} \right).
\eeqn{2.46}
The local or direct potentials are defined by 
\beq
V_{c} (r_c) = \la c | V | c \ra
\eeqn{2.46a}
where $V$ is the total potential appearing in the many-body Hamiltonian \rref{2.38} 
and the integration is performed over the internal coordinates of the subsystems 
appearing in channel $c$. 
The non-local potentials 
\beq
W_{cc'} (r,r') = \la c | \delta(r_c-r) V  {\cal A} \delta(r_{c'}-r') | c' \ra 
- V_{c} (r) \delta_{cc'}\delta(r-r')
\eeqn{2.46b}
occur because of antisymmetrization 
and/or because several partitions are taken into account. 

Mathematically, system \rref{2.45} can be written as a function of a single coordinate $r$ 
(which becomes $r'$ in the integrals of the non-local terms). 
Wave functions however depend on several relative coordinates $r_c$ 
when several partitions are taken into account. 
Channels $c$ differ either by the nature of the subsystems 
or by their level of excitation. 
To simplify the presentation, we now consider that a single partition is taken 
into account or that all coordinates $r_c$ are approximated by a single one. 
Interactions may still be non local but this does not raise major problems 
as long as the non-local terms are short-ranged. 

The colliding systems have initial orientations specified by the spin projections $M_1$ and $M_2$ 
in the entrance channel now denoted as $c$. 
One is looking for a solution of the Schr\"odinger equation with the asymptotic behaviour 
\beq
\Psi^+_{(c M_1M_2)} (\ve{r}) \arrow{r}{\infty}  
\psi^+_C (\ve{r}) \phi^{(1)}_{c I_1 M_1} \phi^{(2)}_{c I_2 M_2} 
\eol
+ (2\pi)^{-3/2} \sum_{c' M'_1 M'_2} \frac{e^{i(k_{c'} r-\eta_{c'} \ln 2k_{c'} r)}}{r} 
f_{c' M'_1M'_2}^{(c M_1M_2)} (\Omega)\, \phi^{(1)}_{c' I'_1 M'_1} \phi^{(2)}_{c' I'_2 M'_2}.
\eeqn{2.51}
Several scattering amplitudes $f_{c'M'_1M'_2}^{(c M_1M_2)}$ appear. 
The partial wave functions \rref{2.42} read with $|c\ra$ representing $|(\alpha I_1I_2)IlJM\ra$, 
\beq
\Psi_{(c)}^{JM\pi} (\ve{r}) = \sum_{c'} | c' \ra r^{-1} u_{c'(c)} (r).
\eeqn{2.50}
A stationary scattering wave function is constructed as 
\beq
\Psi_{(cM_1M_2)}^+ (\ve{r}) & = & i (2\pi)^{-3/2} \pi^{1/2} k^{-1} 
\sum_{J\pi} \sum_{I l} C_{c}^{-1} (2l+1)^{1/2} 
e^{i\sigma_{l}} 
\eol && \times
(I_1I_2M_1M_2|I M) (I l M0|JM) \Psi_{(c)}^{JM\pi} (\ve{r})
\eeqn{2.53}
with $M=M_1+M_2$. 
From the outgoing waves of the asymptotic form of \rref{2.53}, 
one deduces the scattering amplitudes 
\beq
f_{c'M'_1M'_2}^{(cM_1M_2)} (\Omega) & = & i \frac{\sqrt{\pi}}{k} 
\sum_{J\pi} \sum_{I l} \sum_{I'l'} (2l+1)^{1/2} 
e^{i(\sigma_{l}+\sigma_{l'})} (I_1I_2M_1M_2|I M) 
\eol && \times
(I l M0|JM) (I'_1I'_2M'_1M'_2|I'M') (I'l'M'M-M'|JM)
\eol && \times
(\delta_{c'c} \delta_{I'I} \delta_{l'l} -U^{J\pi}_{c'I'l',c I l}) Y_{l'}^{M-M'} (\Omega) 
\eeqn{2.54}
where $M'=M'_1+M'_2$. 
The elastic cross section averaged over initial orientations 
and summed over final orientations reads 
\beq
\frac{d\sigma_{\rm el.}}{d\Omega} & = & \frac{1}{(2I_1+1)(2I_2+1)} 
\eol
&& \times \sum_{M_1M_2} \sum_{M'_1M'_2} 
|f_C (\Omega) \delta_{M'_1M_1} \delta_{M'_2M_2} 
+ f_{cM'_1M'_2}^{(cM_1M_2)} (\Omega)|^2
\eeqn{2.55}
where $f_C$ is defined in \rref{2.11}. 
Inelastic or reaction cross sections are given by 
\beq
\frac{d\sigma_{c \rightarrow c'}}{d\Omega} & = & \frac{1}{(2I_1+1)(2I_2+1)} 
\sum_{M_1M_2} \sum_{M'_1M'_2} |f_{c'M'_1M'_2}^{(cM_1M_2)} (\Omega)|^2
\eeqn{2.55a}
for $c' \ne c$. 
The summations can be performed analytically.
A long but simple calculation provides \cite{BB52} 
\beq
\frac{d\sigma_{c \rightarrow c'}}{d\Omega}= 
\frac{\pi}{k^2}\frac{1}{(2I_1+1)(2I_2+1)}\sum_\lambda B_\lambda(E) \, P_\lambda(\cos \theta),
\eeqn{2.55b}
where the anisotropy coefficients $B_\lambda(E)$ are given by 
\beq
B_\lambda(E) & = & \frac{1}{4\pi}\sum_{J\pi} \sum_{I l L} \sum_{J'\pi'} \sum_{I' l' L'} (-1)^{I-I'} 
e^{i(\sigma_l + \sigma_{l'} - \sigma_L - \sigma_{L'})}\, Z(l J L J',I \lambda) 
\eol && \times
Z(l' J L' J',I' \lambda) U^{J\pi}_{c' I'l',c I l}(E) \, U^{J'\pi'*}_{c' I'L',c I L}(E).
\eeqn{2.55c}
The real coefficients $Z$ defined in \cite{BB52} are modified here for consistency 
with the symmetry property \rref{2.44b} of the collision matrix as \cite{Hu54} 
\beq
Z(l J L J',I \lambda) & = & (-1)^{J+J'} [(2\lambda+1)(2J+1)(2J'+1)(2l+1)(2L+1)]^{1/2} 
\eol && \times 
\left( \begin{array}{ccc} l & L & \lambda \\ 0 & 0 & 0 \end{array} \right)
\left \{ \begin{array}{ccc} l & L & \lambda \\ J' & J & I \end{array} \right \}.
\eeqn{2.55d}
They verify the symmetry relation
\beq
Z(l J L J',I \lambda) = Z(L J' l J,I \lambda).
\eeqn{2.55f}
From $B_0$, one deduces the integrated inelastic or reaction cross sections 
\beq
\sigma_{c \rightarrow c'} = \frac{\pi}{k^2} \frac{1}{(2I_1+1)(2I_2+1)} 
\sum_{J\pi} (2J+1) \sum_{I l } \sum_{I' l'}  |U^{J\pi}_{c' I' l',c I l}(E)|^2.
\eeqn{2.55e}
For the elastic cross section, the summation over the orientations can also be performed analytically \cite{BB52} 
and provides three contributions: nuclear and Coulomb cross sections, as well as an interference term. 
However, in practical applications, it turns out that definition (\ref{2.55}) is more direct to use. 

\subsection{Radiative capture}
\label{subsec:rc}

In nuclear physics, radiative capture is an important process 
because of its astrophysical applications \cite{RR88}. 
In this reaction, the two colliding nuclei fuse 
into a nucleus with mass $m$ with the emission of a photon. 
In stars, this process often occurs at very low scattering energies 
and requires that the reaction has a positive threshold energy 
$Q = (m_1 + m_2 - m) c^2$. 
The analog reaction in atomic physics is electron capture. 
However, much more attention is paid in that field to the reversed process, 
namely photoionization. 
Here we shall proceed with radiative capture in a nuclear physics context
but the formulas can be easily adapted to photodissociation or to photoionization 
by using the detailed balance. 

Radiative capture is an electromagnetic transition from a scattering state 
to a bound state. 
The electromagnetic aspects of this process can be studied 
at first order of perturbation theory \cite{LT58}, 
with an outgoing scattering state $\Psi^+_{(cM_1 M_2)}$ as initial state 
at positive energy $E$ and a bound state in partial wave $J_f\pi_f$ as final state 
at negative energy $E_f$. 
The final energy $E_f$ is equal to $-Q+E_x$ where $E_x$ 
is the excitation energy of the final level with respect to the ground state. 
The radiative capture cross section to this state is then given by 
\beq
\sigma_{J_f \pi_f} (E) & = & \frac{64\pi^4}{\hbar v} \frac{1}{(2I_1+1)(2I_2+1)} 
\sum_{\sigma\lambda} \frac{k_{\gamma}^{2\lambda+1}}{[(2\lambda + 1)!!]^2} 
\frac{\lambda + 1}{\lambda} 
\eol && \times \sum_{M_1M_2M_f\mu} 
|\la \Psi^{J_f M_f \pi_f} | \cM^{\sigma \lambda}_\mu | \Psi^+_{(cM_1 M_2)} (E) \ra|^2 
\eeqn{2.30}
where the symbols $\sigma \lambda$ label the electric (E$\lambda$) and magnetic 
(M$\lambda$) multipoles and the corresponding multipole operators 
are denoted as $\cM^{\sigma \lambda}_\mu$ ($\mu = -\lambda, +\lambda$) \cite{RB67}. 
The photon wave number is given by 
\beq
k_{\gamma} = (E - E_f)/\hbar c.
\eeqn{2.31}
In practice, the sum over $\sigma \lambda$ can usually be restricted to the 
dominant electric multipole (E1, or E2 when E1 is forbidden) because 
$k_{\gamma}$ is small with respect to the inverse of the nucleus dimension. 
Below the Coulomb barrier, the cross section strongly depends on energy. To reduce the
energy dependence, it is often converted into the astrophysical $S$-factor, defined as
\beq
S_{J_f \pi_f}(E)=E \exp(2\pi\eta) \sigma_{J_f \pi_f}(E),
\eeqn{2.59}
where $\eta$ is the Sommerfeld parameter. 

Let us restrict the scattering wave function to a single channel. 
The spins $I_1$ and $I_2$ are then fixed and only $I$ and $l$ are needed 
to specify the entrance channel. 
By using expansion (\ref{2.53}) in this particular case, 
the cross section (\ref{2.30}) can be written as
\beq
\sigma_{J_f \pi_f}(E) =\sum_{\sigma \lambda}
\sigma_{J_f\pi_f}^{\sigma \lambda}(E),
\eeqn{2.60}
where the partial cross section of multipolarity $\sigma\lambda$ reads, in analogy with (\ref{2.55e}),
\beq \fl
\sigma_{J_f\pi_f}^{\sigma \lambda}(E)=\frac{\pi}{k^2}
\frac{1}{(2I_1+1)(2I_2+1)} \sum_{J \pi} (2J+1) 
\sum_{Il} \left| \tilde{U}^{\sigma \lambda}_{Il}(E,J \pi \rightarrow J_f\pi_f) \right| ^2.
\eeqn{2.61}
In (\ref{2.61}), $\tilde{U}^{\sigma \lambda}_{Il}$ is dimensionless 
and proportional to a matrix element of the electromagnetic operator 
between the final state and the initial partial wave. 
From (\ref{2.53}) and (\ref{2.30}), it is given by
\beq
\tilde{U}^{\sigma \lambda}_{Il}(E,J \pi \rightarrow J_f\pi_f) &=&
\left( \frac{2J_f+1}{2J+1} \right)^{1/2}
\left(\frac{8\pi(\lambda+1)k_{\gamma}^{2\lambda+1}}{\hbar v\lambda(2\lambda+1)!!^2} 
\right)^{1/2}\nonumber \\
&&\times \frac{1}{C_{Il}} \la \Psi^{J_f\pi_f}||{\cal M}^{\sigma \lambda}||\Psi^{J\pi}_{(Il)}(E) \ra,
\eeqn{2.62}
where the reduced matrix element is defined by
\beq
\la \Psi^{J_f M_f\pi_f}|{\cal M}^{\sigma \lambda}_{\mu}|\Psi^{JM\pi}_{(Il)} \ra 
= ( J \lambda M \mu|J_f M_f ) 
\la \Psi^{J_f\pi_f}||{\cal M}^{\sigma \lambda}||\Psi^{J\pi}_{(Il)} \ra.
\eeqn{2.63}

For a number of reactions involving light nuclei, capture 
mainly occurs at distances where the wave functions of the colliding nuclei overlap weakly. 
This situation can be described by a simple model where the internal 
structure of the colliding nuclei is neglected and the physics of the process 
is modeled by a local potential $V$ depending on the distance $r$ between 
the centres of mass of the nuclei. 
In this case, the asymptotic form of the initial state is described by \rref{2.51} where 
the internal states $\phi^{(1)}_{c I_1 M_1}$ and $\phi^{(2)}_{c I_2 M_2}$ 
reduce to spinors $|I_1 M_1 \ra$ and $|I_2 M_2 \ra$. 

The electric operators $\cM^{{\rm E}\lambda}_\mu$ are given in this 
simple model by 
\beq
\cM_{\mu}^{E\lambda} = e Z_{\rm eff}^{({\rm E}\lambda)} r^{\lambda} 
Y_{\lambda}^{\mu} (\Omega)
\eeqn{2.32}
where $Z_{\rm eff}^{({\rm E}\lambda)}$ is the effective charge 
\beq
Z_{\rm eff}^{({\rm E}\lambda)} = Z_1 \left(-\frac{m_2}{m} \right)^{\lambda} 
+ Z_2 \left(\frac{m_1}{m} \right)^{\lambda}.
\eeqn{2.33}
The radiative-capture cross section to a final bound state with angular momentum $J_f$ 
can be calculated in this model. 
The initial scattering state with quantum numbers $JM$ is defined by \rref{2.53} 
where \rref{2.50} is replaced by
\beq
\Psi^{J M} (\ve{r}) = \sum_{I l} i^l | (I_1I_2)IlJM \ra r^{-1} u_{Il}^{J} (r)
\eeqn{2.56}
and radial functions are normalized according to \rref{2.44}. 
The normed final bound state with quantum numbers $J_fM_f$ is assumed to be approximated 
by expression \rref{2.56} with the normalization 
\beq
\sum_{I_f l_f} \int_0^\infty  \left[ u^{J_f}_{I_fl_f} (r) \right]^2 dr = 1.
\eeqn{2.57}
The reduced matrix element reads 
\beq
&& \la \Psi^{J_f\pi_f}||{\cal M}^{E \lambda}||\Psi^{J\pi}_{(Il)}(E) \ra 
= eZ_{\rm eff}^{({\rm E}\lambda)} [4\pi (2J_f+1)]^{-1/2} 
\eol && \times 
\sum_{I_f l_f l_i} (-1)^{I_f - J} Z(l_f J_f l_i J,I_f \lambda) 
\int_0^\infty u^{J_f}_{I_fl_f} (r) r^{\lambda} u^{J}_{I_fl_i(Il)} (r) dr.
\eeqn{2.58}
In practice, \rref{2.58} must often be corrected empirically by multiplicative factors 
called spectroscopic factors to take an approximate account of the internal structure 
of the nuclei \cite{Ro73}. 

\setcounter{equation}{0}
\renewcommand{\theequation}{3.\arabic{equation}}
%
%
%
%
\section{The calculable $R$ matrix}
\label{sec:comp}
\subsection{Introduction}
The two variants of the $R$ matrix method mainly differ by their types of applications. 
In the calculable $R$ matrix, the aim is to accurately solve a given Schr\"odinger equation
mostly in the continuum, i.e. for positive energies. 
In the phenomenological $R$ matrix, the goal is to parametrize scattering data; 
it is thus essential to know the analytical form of the $R$ matrix. 
Of course, both variants have much in common and it is a matter of taste to start 
with one or the other. 
Historically, in nuclear physics, the emphasis has first been put on the phenomenological variant.
Conversely, most applications in atomic physics are related to the calculable $R$ matrix.
Here we will start with a general formalism leading to the calculable version and 
then deduce the properties allowing the phenomenological use. 

First, we restrict ourselves to a single channel for an arbitrary partial wave. 
This assumption is often valid, and allows simple notations. 
We thus attempt to find approximate solutions of the Schr\"odinger equation 
for the relative motion of two particles with reduced mass $\mu$ 
interacting via a central potential $V$. 
At large relative distances $r$, the interaction reduces to the Coulomb 
interaction $V_C$. 

After separation of the angular part, the radial Schr\"odinger equation \rref{2.20} 
for partial wave $l$ can be written as 
\beq
(H_l - E) u_l = 0.
\eeqn{31.2}
In this expression, the radial Hamiltonian $H_l$ is defined as 
\beq
H_l = T_l + V(r),
\eeqn{31.3}
where $T_l$ is given by \rref{2.46}. 
We are interested in bounded solutions $u_l(r)$ of \rref{31.2} verifying condition \rref{2.20a} 
at the origin.
Bound-state solutions at negative energies are square integrable over $(0,\infty)$ 
and can be normed. 
Scattering solutions at positive energies are assumed to be normalized according to 
\rref{2.21a} with the scattering matrix $U_l$ defined in \rref{2.24}.
We will essentially deal with real potentials for which the phase shifts are real 
and the scattering matrix is unitary. 
The generalization to complex potentials is straightforward. 
\subsection{Definition and calculation of $R$ matrix}
\label{sec:form}
In the $R$-matrix method, the configuration space is divided at the channel 
radius $a$ into an internal region and an external region. 
The channel radius is chosen large enough so that $V$ can be approximated by $V_C$ 
in the external region at the required accuracy. 
This means that the channel radius can in principle always be increased 
although often at a cost of computational time.  
At each energy $E$, the wave function is defined by different expressions in these regions. 
In the external region, the wave function $u_l(r)$ is approximated by the exact asymptotic 
expression \rref{2.21a}\footnote{From now on, the Sommerfeld parameter $\eta$ is implied.}, 
\beq
\uext (r) = C_l [I_l (kr) - U_l O_l (kr)].
\eeqn{32.1a}
In the internal region, the wave function $\uint(r)$ is expanded over some finite basis 
involving $N$ linearly independent functions $\varphi_j$ as 
\beq
\uint (r) = \sum_{j=1}^N c_j \varphi_j (r).
\eeqn{32.4}
The functions $\varphi_j$ must vanish at the origin but are not necessarily orthogonal. 
In contrast with some traditional presentations of the $R$-matrix theory \cite{BR75b}, 
we do not assume that they satisfy specific boundary conditions at $r = a$. 
Various choices are possible, as exemplified in section \ref{sec:apcomp}.
The internal and external pieces of the radial functions will be connected at the boundary $a$ 
by the continuity of the wave function $u_l$ and of its first derivative. 

The $R$ matrix at energy $E$ is defined through\footnote{
As defined here, the $R$ matrix is dimensionless. 
In some works, it has the dimension of a length and differs from the present definition by a factor $a$. 
The definition \rref{33.5} of the reduced width amplitudes must then be modified accordingly. 
}
\beq
u_l (a) = R_l(E) [a u_l' (a) - B u_l (a)].
\eeqn{32.8}
A dimensionless boundary parameter $B$ is included for later convenience. 
Its choice will be discussed later. 
The inverse of the $R$ matrix is thus the difference between the logarithmic derivative 
of the radial wave function at the boundary between both regions, and the boundary parameter $B$. 
This matrix has dimension 1 in a single-channel case and is just a function of energy. 
In multichannel problems, the dimension of the $R$ matrix is equal to the number of channels 
(see section \ref{sec:Rmulti}). 
The principle of the method relies on the facts that the $R$ matrix can be calculated 
from properties of the Hamiltonian in the internal region and that its knowledge allows determining 
the scattering matrix in the external region. 

The operator $H_l$ is not Hermitian over the internal region $(0,a)$. 
This property is not convenient for practical resolutions of the Schr\"odinger equation. 
This problem is elegantly solved with the help of the surface operator 
introduced by Bloch \cite{Bl57}
\beq
\cL(B) = \frac{\hbar^2}{2\mu} \, \delta (r-a) 
\left( \frac{d}{dr} - \frac{B}{r} \right).
\eeqn{32.1}
The operator $H_l + \cL(B)$ is Hermitian over $(0,a)$ when $B$ is real \cite{Bl57}. 
Moreover it has a fully discrete spectrum as the self-adjoint problem 
is defined over a finite interval. 

The Schr\"odinger equation in the internal region is approximated by the 
inhomogeneous Bloch-Schr\"odinger equation 
\beq
(H_l + \cL(B) - E) \uint = \cL(B) \uext,
\eeqn{32.2}
where the external solution is used in the right-hand member. 
The mathematical problem is complemented with the continuity condition 
\beq
\uint (a) = \uext (a).
\eeqn{32.3}
Until now, the approximation only consists in using in the right-hand side of {\rref{32.2} 
the asymptotic form \rref{2.21a} which is known except 
for the value of the scattering matrix $U_l$. 
The main advantage of the $R$-matrix method is that an expansion 
in square-integrable functions can now be used in the internal region. 

Because of the Dirac function in the Bloch operator, \rref{32.2} and  \rref{32.3} are equivalent 
to the Schr\"odinger equation \rref{31.2} restricted to the interval $(0,a)$ 
supplemented by the continuity condition at $r = a$ \cite{Bl57}, 
\beq
\uint'(a) = \uext'(a)
\eeqn{32.3a}
for any $B$. 
Hence, beyond making $H_l + \cL(B)$ Hermitian, the Bloch operator enforces 
the continuity of the derivative of the wave function. 
The importance of this aspect of the Bloch operator has often been underestimated in the literature. 
Condition \rref{32.3a} needs not be imposed to the basis functions $\varphi_j$ 
since the Bloch operator will impose it to the physical solution $u_l$. 
For historical reasons, a lot of confusion about the $R$ matrix arose from the 
misunderstanding of this property as we shall see in section \ref{sec:bc}. 

Formally, the inhomogeneous Bloch-Schr\"odinger equation \rref{32.2} 
can be solved with the Green function defined by 
\beq
(H_l + \cL(B) - E) G_l(r,r') = \delta(r-r')
\eeqn{33.7a}
and $G_l(0,r) = 0$. 
The solution reads  
\beq
\uint (r) = \int_0^a G_l(r,r') \cL(B) \uext(r') dr'.
\eeqn{33.7c}
With (\ref{32.1}) and (\ref{32.8}), the $R$ matrix is thus given by 
\beq
R_l(E) = \frac{\hbar^2}{2\mu a} G_l (a,a).
\eeqn{33.7b}
The calculable $R$-matrix method consists in solving the Bloch-Schr\"odinger equation 
with an approximate Green function expanded over a finite basis. 

To obtain a practical expression for \rref{33.7b}, expansion \rref{32.4} is introduced in \rref{32.2} 
and the resulting equation is projected on $\varphi_i (r)$, giving 
\beq
\sum_{j=1}^N C_{ij}(E,B) c_j 
= \frac{\hbar^2}{2\mu a} \varphi_i(a) \left(a \uext' (a) - B \uext(a)\right).
\eeqn{32.5}
The elements of the symmetric matrix $\ve{C}$ are defined as 
\beq
C_{ij}(E,B) = \la \varphi_i | T_l + \cL(B) + V - E | \varphi_j \ra.
\eeqn{32.6}
Dirac brackets correspond here to one-dimensional integrals over the variable $r$ from 0 to $a$. 
Because of the Bloch operator, the right-hand side of (\ref{32.5}) only involves values at $r=a$. 

Coefficients $c_j$ are obtained by solving system \rref{32.5}. 
Introducing them in (\ref{32.4}) at $r=a$ and comparing with (\ref{32.8}) provides 
the calculable $R$ matrix 
\beq
R_l (E,B) = \frac{\hbar^2}{2\mu a} 
\sum_{i,j=1}^N \varphi_i(a) (\ve{C}^{-1})_{ij} \varphi_j(a).
\eeqn{32.9}
This expression is nothing but a finite-basis approximation of \rref{33.7b}. 

The wave function in the internal region is given by 
\beq
\uint (r) = \frac{\hbar^2}{2\mu a R_l(E,B)} \uext(a) 
\sum_{j=1}^N \varphi_j (r) \sum_{i=1}^N (\ve{C}^{-1})_{ij} \varphi_i(a).
\eeqn{32.19}
We shall see in section \ref{sec:scat} that it does not depend on $B$. 
\subsection{Properties of the $R$ matrix}
\label{sec:prop}
Temporarily, in this section, 
the basis functions $\varphi_i (r)$ are assumed to be orthonormal. 
The matrix of overlaps $\la\varphi_i|\varphi_j\ra$ is thus the unit matrix. 
Let us consider the eigenvalues $E_{nl}$ and the corresponding normalized eigenvectors 
$\ve{v}_{nl}$ of matrix $\ve{C}(0,B)$, 
\beq
\ve{C}(0,B) \ve{v}_{nl} = E_{nl} \ve{v}_{nl}
\eeqn{33.1}
with the orthonormality property  
\beq
\ve{v}_{nl}^{\rm T} \ve{v}_{n'l} = \delta_{nn'}.
\eeqn{33.2}
With the spectral decomposition 
\beq
[\ve{C}(E,B)]^{-1} = \sum_{n=1}^{N} \frac{\ve{v}_{nl} \ve{v}_{nl}^{\rm T}}{E_{nl} - E},
\eeqn{33.3}
the $R$ function \rref{32.9} becomes 
\beq
R_l(E,B) = \sum_{n=1}^N \frac{\gamma_{nl}^2}{E_{nl} - E}
\eeqn{33.4}
with 
\beq
\gamma_{nl} = \left( \frac{\hbar^2}{2\mu a} \right)^{1/2} \phi_{nl} (a)
\eeqn{33.5}
and 
\beq
\phi_{nl} (r) = \sum_{i=1}^N v_{nl,i} \varphi_i(r), 
\eeqn{33.6}
where $v_{nl,i}$ is the $i$th component of $\ve{v}_{nl}$. 
The $\gamma_{nl}$ are known as the reduced width amplitudes 
and their squares $\gamma_{nl}^2$ as the reduced widths \cite{LT58}. 
Their interpretation is simple. 
They are proportional to the value at the channel radius of variational approximations $\phi_{nl}$ 
of the eigenfunctions of the Hermitian operator $H_l + \cL (B)$. 
Those corresponding to the lowest energies thus represent approximate eigenfunctions 
of the physical problem confined over the interval $(0,a)$ with logarithmic derivative $B$ 
at $r = a$. 

Expression \rref{33.4} looks familiar to practitioners of the $R$-matrix theory. 
It is however obtained here with a finite basis. 
The traditional expression for the $R$ matrix is obtained when $N$ tends towards 
infinity in a complete basis as 
\beq
R_l(E,B) = \sum_{n=1}^\infty \frac{\gamma_{nl}^2}{E_{nl} - E}.
\eeqn{33.7}
The energies $E_{nl}$ are now the exact eigenvalues of the operator $H_l + \cL (B)$ 
and the reduced width amplitudes $\gamma_{nl}$ are related to the values at $r=a$ 
of its exact eigenfunctions. 

The $R$ matrix is a real function when $V$ and $B$ are real. 
It has an infinity of real simple poles, bounded from below. 
Its derivative is always positive at regular points. 
It is a meromorphic function of the energy when the energy is considered 
as a complex variable \cite{LT58}. 
All residues are negative and given by minus the reduced widths $\gamma_{nl}^2$. 
\subsection{Scattering matrix and phase shifts}
\label{sec:scat}
Since the $R$ matrix is known, the external function (\ref{32.1a}) can be introduced 
in relation (\ref{32.3}) to determine the scattering matrix for the $l$th partial wave as 
\beq
U_l = e ^{2i\phi_l} \, 
\frac{1 - (L_l^* - B) R_l(E,B)}{1 - (L_l - B) R_l(E,B)}.
\eeqn{32.12}
In this expression, 
\beq
L_l = ka \frac{O'_l(ka)}{O_l(ka)}
\eeqn{32.10}
is the dimensionless logarithmic derivative of $O_l$ at the channel radius, 
$L_l^*$ is the conjugate of $L_l$, 
and 
\beq
\phi_l = \arg I_l(ka) = -\arctan [F_l(ka)/G_l (ka)]
\eeqn{32.11}
is the hard-sphere phase shift. 
Note that the same notation $\phi_l$ in \cite{LT58} represents the opposite 
of the hard-sphere phase shift. 

Expression (\ref{32.12}) has the striking property that it 
does not depend on the boundary parameter $B$, independently of the size of the basis. 
Indeed, with the matrix relation (\ref{A.4}) in Appendix \ref{sec:B}, one deduces 
from (\ref{32.9}) and (\ref{32.6}) 
\beq
\frac{1}{R_l(E,0)} = \frac{1}{R_l(E,B)} + B
\eeqn{32.13}
for any $B$, real or complex. 
Expression (\ref{32.13}) means that the logarithmic derivative of the internal 
wave function at the boundary is independent of $B$. 
Introducing relation (\ref{32.13}) into (\ref{32.12}) shows 
that any $B$ value leads to the same scattering matrix as for $B=0$. 
Equation (\ref{32.13}) is well known in $R$-matrix theory 
(see equation (IV.2.5a) of \cite{LT58}). 
It is also valid for the phenomenological $R$ matrix with a finite number of poles \cite{Ba72}. 
However its validity for the approximation (\ref{32.12}) 
for any basis size \cite{LR69,Mo72} is sometimes overlooked. 

Like the scattering matrix $U_l$ and the external wave function $\uext (r)$, 
the internal function \rref{32.19} does not depend on the choice for $B$. 
Indeed, with the help of relation (\ref{A.3}), 
one easily shows that, for any $B$, it is equal to the similar 
expression where $B$ is replaced by zero. 

For a better physical interpretation of the results which is important 
in applications, it is convenient to introduce some definitions. 
To this end, $L_l$ is separated into its real and imaginary parts as 
\beq
L_l = S_l +i P_l. 
\eeqn{34.3}
The real part $S_l$ and imaginary part $P_l$ of $L_l$ 
are called the shift and penetration factors, respectively. 
They depend on energy and on the channel radius. 
The penetration factor can be written with the Wronskian relation 
$I_l O'_l - I'_l O_l = 2i$ as 
\beq
P_l (E) = \frac{ka}{|O_l(ka)|^2} = \frac{ka}{F_l(ka)^2 + G_l (ka)^2}.
\eeqn{34.4}
It is always positive and increasing \cite{LT58}. 
The shift factor reads 
\beq
S_l (E) = P_l (E) [F_l(ka) F_l'(ka) + G_l (ka) G_l'(ka)].
\eeqn{34.5}
It is always negative for $\eta \ge 0$ \cite{LT58}. 
Although we do not know a proof that $S_l$ is always increasing in the same case, 
we could not find numerically a counterexample. 
As shown below, none of these properties is valid in the attractive case. 

\begin{figure}[ht]
\begin{center}
\includegraphics[width=0.5\textwidth,clip]{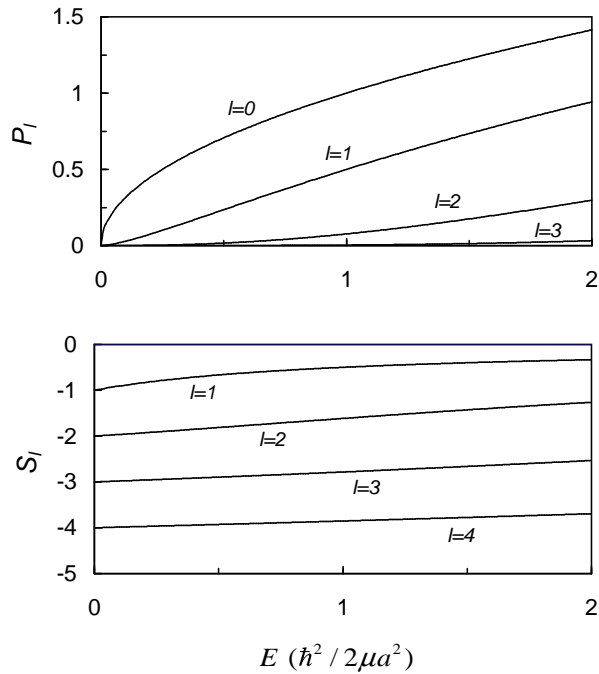} 
\caption{Penetration factors $P_l(E)$ (upper panel) and shift factors $S_l(E)$ (lower panel) 
in the neutral case ($\eta = 0$) as a function of $E$ in units of $\hbar^2/2\mu a^2$.}
\label{fig3.1}
\end{center}
\end{figure}
In the neutral case  ($\eta = 0$), the penetration factors have 
simple analytical expressions such as 
\beq
P_0 (E) = ka, \hspace*{1 cm} 
P_1 (E) = \frac{(ka)^3}{1 + (ka)^2}, \hspace*{1 cm} \dots
\eeqn{34.8}
They do not vary very fast with energy (see Fig.~\ref{fig3.1}). 
This figure is universal, i.e., independent of the collision. 
Notice that the derivative of $P_0$ with respect to energy 
is infinite at the origin. 
This property leads to the special behaviour of neutron scattering in the $s$ wave. 
Penetration factors decrease with the orbital momentum $l$ as expected 
from the occurrence of an increasing centrifugal barrier. 
The shift factors read 
\beq
S_0 (E) = 0, \hspace*{1 cm} 
S_1 (E) = - \frac{1}{1 + (ka)^2}, \hspace*{1 cm} \dots
\eeqn{34.9}
They vary smoothly with energy, starting from the integer values $-l$ 
(see Fig.~\ref{fig3.1}). 
This weak energy dependence is the origin of the Thomas approximation \cite{LT58}
where the shift factor is assumed to vary linearly in a limited energy range.
\begin{figure}[ht]
\begin{center}
\includegraphics[width=0.5\textwidth,clip]{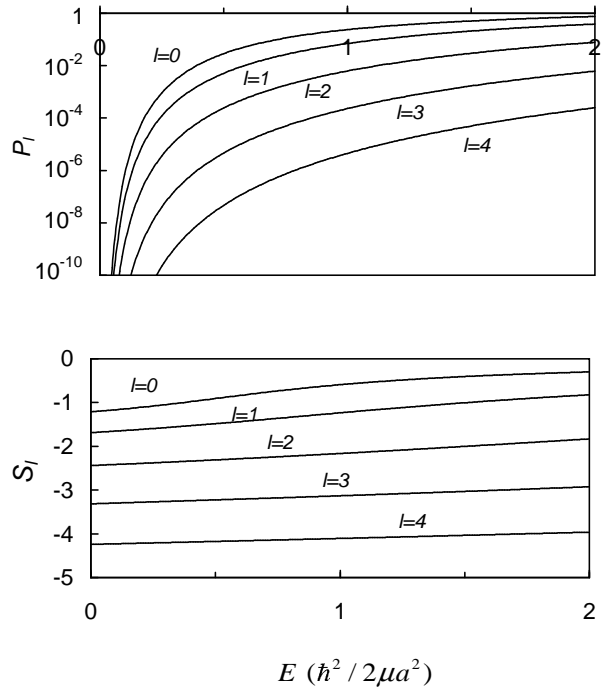} 
\caption{Penetration factors $P_l(E)$ (upper panel) and shift factors $S_l(E)$ (lower panel) 
in the repulsive charged case for $a = a_B$ as a function of $E$ in units of $\hbar^2/2\mu a^2$.}
\label{fig3.2}
\end{center}
\end{figure}

The penetration factors are very different according to whether 
both particles are charged or not. 
In the repulsive charged case, the energy dependence of the penetration factors is much stronger 
(notice the logarithmic scale in Fig.~\ref{fig3.2}). 
Here we have to choose the strength of the Coulomb interaction. 
Figure \ref{fig3.2} corresponds to a channel radius $a$ equal to the Bohr radius $a_B$. 
The strong dependence at low energies is due to the difficulty of penetrating a Coulomb barrier 
when the scattering energy becomes much smaller than the Coulomb barrier. 
Beyond $l = 1$, increasing from $l$ to $l+1$ decreases the penetration factors 
by more than an order of magnitude. 
In contrast, shift factors are much more similar in all cases. 
In the repulsive charged case, the shift factors are quite similar to those of the 
neutral case, except for $l = 0$ (see Fig.~\ref{fig3.2}). 

In the attractive charged case, the energy dependence of the penetration factors is 
displayed in Fig.~\ref{fig3.3} for $a = a_B$. 
It is rather similar to the neutral case except at low energies 
where it starts from a finite value at energy zero. 
Shift factors displayed in Fig.~\ref{fig3.3} resemble other cases. 
Notice however that $S_0$ is positive, and decreasing. 
\begin{figure}[ht]
\begin{center}
\includegraphics[width=0.5\textwidth,clip]{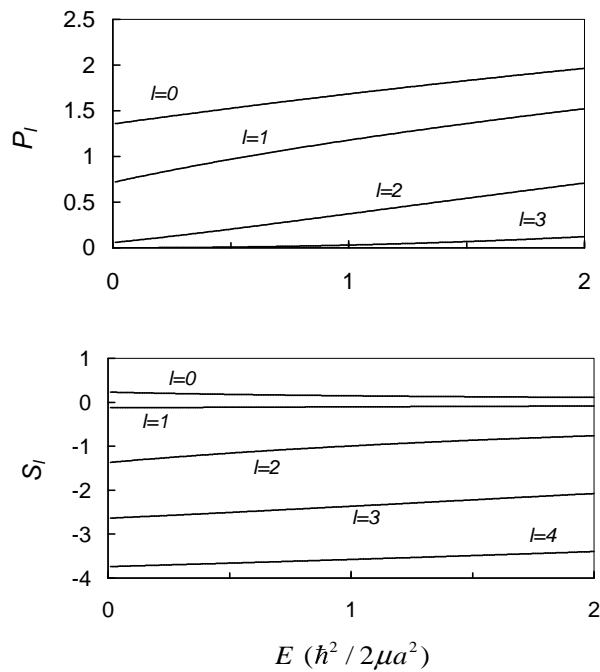} 
\caption{Penetration factors $P_l(E)$ (upper panel) and shift factors $S_l(E)$ (lower panel) 
in the attractive charged case for $a = a_B$ as a function of $E$ in units of $\hbar^2/2\mu a^2$.}
\label{fig3.3}
\end{center}
\end{figure}

With definition \rref{34.3}, the collision matrix \rref{32.12} becomes 
\beq
e^{2i \delta_l} = e ^{2i\phi_l} \frac{1 - (S_l-B) R_l + iP_l R_l}
{1 - (S_l-B) R_l - iP_l R_l}.
\eeqn{34.6}
One obtains an explicit expression for the phase shift, 
\beq
\delta_l  = \phi_l + \arctan \frac{P_l R_l}{1 - (S_l-B) R_l}.
\eeqn{34.7}
The internal wave function \rref{32.19} can be rewritten as
\beq \fl
\uint(r) = \frac{\hbar^2}{\mu a } e^{i(\delta_l - \dem \pi)} C_l 
\frac{(ka P_l)^{1/2}}{|1-(L_l-B)R_l|} \sum_{j=1}^N  \varphi_j(r) 
\sum_{i=1}^N (\ve{C}^{-1})_{ij} \varphi_i(a),
\eeqn{34.7b}
which explicitly shows the phase and modulus of $\uint$ (up to a global sign). 
For an orthonormal basis, this expression can be rewritten using \rref{33.3} and \rref{33.6} as 
\beq \fl
\uint(r) = \frac{\hbar^2}{\mu a } e^{i(\delta_l - \dem \pi)} C_l 
\frac{(ka P_l)^{1/2}}{|1-(L_l-B)R_l|} 
\sum_{n=1}^N  \frac{\phi_{nl} (r) \phi_{nl} (a)}{E_{nl} - E}.
\eeqn{34.7c}
\subsection{On the basis and boundary parameter choices}
\label{sec:bc}
Considerable confusion exists in the literature about the properties that 
basis states $\varphi_i$ should have. 
Improper choices have led to the introduction of corrections and to attempts 
to use the boundary parameter $B$ to correct drawbacks of the basis. 
However we have just seen that the results are independent of the choice of $B$. 
It is thus worthwhile to devote this section to a clarification of this issue 
that has sometimes led to an undeserved reputation of poor convergence 
for the calculable $R$-matrix method. 

In their seminal paper, Wigner and Eisenbud wanted to provide a phenomenological 
description of resonances \cite{WE47}. 
They did not intend to propose a technique of resolution.
To reach their goal they assume that the basis functions all satisfy (for $B = 0$) 
the boundary conditions $\varphi_j(0)=0$ and 
\beq
a\varphi'_j(a) -B \varphi_j(a) = 0.
\eeqn{33.8}
This procedure leads to $R$ matrix \rref{33.7}. 
When used as a technique of resolution, the finite-basis $R$ matrix \rref{32.9} or  \rref{33.4} 
obtained with this procedure does not converge uniformly. 
The reason is simple. 
The first derivative of the wave function \rref{32.19} suffers from a discontinuity at $r=a$ \cite{SH96}. 
The limit of $\uint'$ when $r$ tends towards $a$ to the left is equal to $\uext'(a)$ 
but not to $\uint'(a)$, 
\beq
\lim_{r \rightarrow a^-} \uint'(r) = \uext'(a) \ne \uint'(a).
\eeqn{33.9}
For example, if $B = 0$, $\varphi'_j (a)$ vanishes for all $j$ values 
and one readily sees that $\uint'(a) = 0$ at all energies. 
This property has unfavourable consequences on the convergence of numerical 
methods when the basis is truncated since the logarithmic derivative of the external solution 
depends on the phase shift (and thus on energy) and can not be matched with the internal 
solution (see Figures \ref{fig_c12p2} and \ref{fig_ahe3_2} in section \ref{sec:apcomp}). 
Buttle \cite{Bu67} has proposed a correction to the $R$-matrix truncation. 
His idea is to replace the truncated part by an analytical approximation, 
i.e., in practice, by the same expression for the zero potential. 
Although this correction improves the phase shifts, it does not really solve the problem 
because it does not improve the wave functions. 

This problem received a solution with the works of Lane and Robson \cite{LR66,LR69,Ph75}. 
Their method was successfully applied in nuclear physics where traditional basis functions 
do not satisfy \rref{33.8} and, on the contrary, display a variety of behaviours 
at the channel radius. 
With oscillator basis functions, accurate results for neutron-nucleus scattering could be obtained 
\cite{Ph75,PG74}. 
At the same time, a microscopic extension of the $R$ matrix using a fully antisymmetrized two-centre 
harmonic-oscillator model provided accurate phase shifts for collisions between light nuclei 
with few basis states \cite{BH74,BHL77} (see section \ref{sec:MCM}). 
The success of these calculations relies on the fact that the Bloch operator makes 
condition \rref{33.8} unnecessary. 
Since the results do not depend on $B$, the choice $B=0$ was used. 
A general though economical method for solving coupled-channel problems 
is described in \cite{HSV98}. 

The negative role of condition \rref{33.8} remained long unnoticed in atomic physics 
where in many cases basis states were imposed to satisfy \rref{33.8} 
\cite{Bu74,BR75b,BNS87,BEN95,BNB07}. 
In the literature, the basis functions are often assumed 
to be solutions of a Sturm-Liouville problem satisfying \rref{33.8}.  
As reviewed in \cite{BRT83}, various solutions to the purported convergence problems 
of the calculable $R$ matrix have been proposed, 
such as using two different sizes for the internal region. 
Another type of solution proposed in \cite{Sz98} requires a basis with a boundary condition 
depending on the eigenvalue $E_{nl}$. 
In 1983, Greene \cite{Gr83,Gr85} realized in the context of atomic physics that, 
in place of the traditional choice \rref{33.8} of a common boundary condition to all basis functions at $r = a$, 
a variety of values for the logarithmic derivatives of the basis functions 
should be far more efficient for accurate calculations. 
He also proposed to optimize the boundary parameter $B$ to have a better connection 
between the internal and external logarithmic derivatives. 
With his variational principle \cite{Gr85} applied to potential scattering, 
$B$ is given at each energy $E$ by the generalized eigenvalue problem 
\beq
\ve{C}(E,B) \ve{c} = 0
\eeqn{33.10}
in the present notation. 
One readily sees \cite{LR84} that the only eigenvalue $B$ is $1/R(E)$ 
which, as the logarithmic derivative of the internal wave function, is obviously optimal. 
With this choice, the basis functions allow a perfect matching 
but at the cost of a new calculation of the basis at each energy. 
As we have shown with \rref{32.13}, this complication is unnecessary since 
all physical results are independent of $B$. 

The first accurate calculations in atomic physics with a basis that fully ignores condition \rref{33.8} 
can be found in \cite{VB87,Ma94,va97,PN99}. 
New developments making use of $B$ splines now avoid this condition \cite{ZF00,ZB04,NPT08}. 
So let us emphasize that the calculable $R$ matrix does converge accurately 
when the basis functions are a well chosen part of a complete set displaying 
a variety of logarithmic derivatives at the channel radius $a$. 
The reason of its accuracy is that the Bloch operator imposes a good matching 
at the boundary \cite{BHS98}. 
The choice of a boundary parameter $B$ is irrelevant and the Buttle correction 
is not necessary because the good matching of the internal and external wave functions 
allows an accurate determination of the phase shift. 
Practice has shown that this simple solution allows a much smoother connection 
with the external wave function \cite{BHS98,HSV98}. 
In opposition to the traditional presentation, there is thus no need for a special assumption 
about the behaviour of the basis functions at the boundary. 
\subsection{Resonances}
\label{sec:resbs}
Resonances can be studied in various ways. 
Each of them may be useful, either in calculable applications (section~\ref{subsec:resonances}) 
or in phenomenological applications (section~\ref{sec:phen}). 

In a first approach, the boundary parameter can be chosen as \cite{Bl57,LR69} 
\beq
B = L_l.
\eeqn{34.1}
It thus depends on energy. 
This complex value leads to a complex function $R_l(E,L_l)$ 
which is not an $R$ matrix in the strict sense since $B$ is not a real constant. 
It is the function introduced by Kapur and Peierls \cite{KP38}. 
Nevertheless it is also given by expression \rref{32.9}. 
Equation (\ref{32.12}) then takes the simpler form \cite{LR69,BGS02} 
\beq
U_l = e^{2i \phi_l} \ [1 + (L_l - L_l^*) R_l(E,L_l)].
\eeqn{34.2}
This expression is also valid for complex $k$ values 
if $\phi_l$ is defined as the phase of $I_l (ka)$. 
Since (\ref{34.2}) has no denominator, a direct relation appears between a pole 
of the scattering matrix, i.e.\ a resonance energy, and a pole of the complex $R$ matrix. 
This relation is however valid only when $L_l$ is calculated at the resonance energy. 
This means that the scattering and $R$ matrices have only one common pole 
and only at specific energies. 
Determining $S$-matrix poles in this way thus requires some iterative procedure. 
The choice \rref{34.1} for the boundary parameter can be used to analyze the mathematical 
nature of resonances in the complex plane in coupled-channel cases \cite{HBJ87}. 
However, the same results can also be interpreted with the traditional $R$ matrix 
involving only real parameters \cite{Ba97b}. 
Let us return to real energies for the other approaches. 
We choose $B=0$ to simplify the presentation. 

Another definition of a resonance energy $E_R$ is that it corresponds to 
the value $\pi/2$ of the resonant part $\delta_l - \phi_l$ of the phase shift. 
From \rref{34.7}, it is therefore defined by the equation 
\beq
1-S_l(E_R)R_l(E_R)=0.
\eeqn{eq_res}
In general this equation must be solved numerically. 
To define the resonance width, let us consider the collision matrix (\ref{34.6}) 
for energies close to $E_R$. 
A Taylor expansion of $S_l(E) R_l(E)$ for $E\approx E_R$ provides the Breit-Wigner approximation 
\beq
U_l^{\rm BW} (E)\approx e ^{2i\phi_l} \frac{E_R-E+i\Gamma(E)/2}{E_R-E-i\Gamma(E)/2}.
\eeqn{eq_bw2}
In this expression the (energy-dependent) width of the resonance is given by 
\beq
\Gamma(E) = \frac{2P_l(E)R_l(E)}{[d(S_lR_l)/dE]_{E=E_R}}.
\eeqn{eq_bw3}
Because of the shift of $E_R$ with respect to a pole (see below), 
$R_l(E)$ can be supposed to vary slowly near a narrow resonance (see section~\ref{sec:phen}). 
The total width then reads 
\beq
\Gamma(E) = 2\gamma^2 P_l(E) = \frac{P_l(E)}{P_l(E_R)} \Gamma(E_R),
\eeqn{eq_bw4}
where $\Gamma(E_R)$ is the width at the resonance energy. 
The reduced width $\gamma^2$ defined by \rref{eq_bw4} is given by 
\beq
\gamma^2=R_l(E_R)/[d(S_lR_l)/dE]_{E=E_R}.
\eeqn{eq_bw5}
Let us mention that (\ref{eq_res}) may have solutions which do not correspond to physical states. 
The width of a physical resonance should be small enough to make its lifetime 
longer that the typical collision time. 

In another way of studying a resonance, let us consider an energy very close 
to a pole $E_{nl}$ of the $R$ matrix. 
If all terms with $n' \ne n$ can be neglected, the $R$ matrix 
is approximated as 
$R_l(E,0) \approx \gamma_{nl}^2/(E_{nl} - E)$.
A simple calculation provides 
\beq
\delta_l \approx \phi_l + \arctan \frac{\gamma_{nl}^2 P_l (E)}
{E_{nl} - \gamma_{nl}^2 S_l (E) - E}.
\eeqn{34.11}
This expression resembles the Breit-Wigner form of the phase shift 
\beq
\delta_l^{\rm BW} \approx \phi_l + \arctan \frac{\dem \Gamma(E)}{E_R - E}.
\eeqn{34.12}
By comparison, one defines the resonance energy 
\beq
E_R = E_{nl} - \gamma_{nl}^2 S_l (E_R)
\eeqn{34.13}
and the formal width 
\beq
\Gamma(E) = 2 \gamma_{nl}^2 P_l (E).
\eeqn{34.14}
The resonance energy is defined by an implicit equation which can be solved approximately 
(see section \ref{sec:formobs}). 
The width is an energy-dependent quantity whose asymmetric shape depends on the behaviour of $P_l$. 
Its relation with a measured width is also discussed in section \ref{sec:formobs}. 

With \rref{33.5}, \rref{34.13} and \rref{34.14}, the internal wave function \rref{34.7c} 
can be approximated at the vicinity of a resonance by 
\beq
\uint(r) \approx e^{i(\delta_l - \dem \pi)} C_l 
\left[ \frac{\hbar v \Gamma}{(E_R - E)^2 +(\Gamma/2)^2} \right]^{1/2} \phi_{nl} (r).
\eeqn{34.15}
It is thus proportional to an approximate eigenfunction \rref{33.6} of $H_l + \cL(B)$ 
with a proportionality factor exhibiting the usual Lorentzian energy dependence of a resonance. 
Equation \rref{34.15} is at the basis of the so-called bound-state approximations 
where the resonance is described by a square-integrable wave function \cite{BD85}. 
\subsection{Bound states}
The $R$-matrix formalism can be extended to bound states $(E_B<0)$ \cite{BD83}. In that
case the external wave function $\uext$ is given by 
\beq
\uext (r) = C_l W_l(2\kappa_Br),
\eeqn{eq_witt}
where $W_l(x)$ is a shorthand notation for the Whittaker function \rref{2.8b} 
and where $\kappa_B$ and $\eta_B$ are the wave number and Sommerfeld parameter, 
respectively, of the bound state. 
In (\ref{eq_witt}), $C_l$ is the asymptotic normalization constant (ANC) 
which determines the amplitude of the wave function at large distances.
This quantity plays an important role in some nuclear reactions of
astrophysical interest \cite{MT90} (see section \ref{sec:racar}).

To determine $E_B$, a convenient choice for the boundary parameter $B$ in \rref{32.1} is 
\beq
B = L_l (E_B) = S_l (E_B) = 2\kappa_B a \frac{W_l'(2\kappa_B a)}{W_l(2\kappa_B a)},
\eeqn{eq_bs0}
because it suppresses the right-hand side of the Bloch-Schr\"odinger equation \rref{32.2}. 
Since the wave function is real, $L_l$ is real and identical to the shift factor. 
The penetration factor $P_l$ vanishes.   
The internal wave function is expanded over a basis, as in \rref{32.4}. 
Projecting the Bloch-Schr\"odinger equation \rref{32.2} over a basis function $\varphi_i$ 
provides for $i = 1, N$, 
\beq
\sum_{j=1}^N \la \varphi_i | T_l+\cL (L_l)+V-E_B |\varphi_j \ra c_j = 0.
\eeqn{eq_bs1}
This system of equations is similar to a standard eigenvalue problem, 
but parameter $L_l$ depends on energy $E_B$. 
In practice one starts from $L_l=0$ and iterates until energy $E_B$ has converged. 
At convergence, the $c_j$ can be calculated by solving the system. 
If $\uint$ is normed according to \rref{33.2}, 
the square of the norm of the wave function is given by 
\beq
N_l = 1+(C_l)^2\int_a^{\infty}(W_l(2\kappa_B r))^2 dr.
\eeqn{eq_bs2}
This expression can be rewritten \cite{LT58,BF80} as 
\beq
N_l = 1 + \gamma_{nl}^2 \left[ \frac{dS_l}{dE} \right]_{E=E_B}
\eeqn{eq_bs2a}
where $\gamma_{nl}$ is the reduced width amplitude of the bound state. 

The internal wave function is given by \rref{32.19} multiplied by $N_l^{-1/2}$, 
\beq
\uint (r) = N_l^{-1/2} \sum_{j=1}^N \, v_{nl,j} \varphi_j (r).
\eeqn{eq_bs3}
where coefficients $v_{nl,j}$ correspond for an orthogonal basis to the eigenvector $\ve{v}_{nl}$ 
of $\ve{C}(0,L_l)$ at energy $E_{nl} = E_B$ in \rref{33.1}. 
Using \rref{33.6}, it can be rewritten as 
\beq
\uint (r) = N_l^{-1/2} \phi_{nl} (r).
\eeqn{eq_bs4}
From \rref{32.3} and \rref{eq_witt}, the ANC is given by 
\beq
C_l = N_l^{-1/2} \phi_{nl} (a)/W_l(2\kappa_B a) 
\eeqn{eq_bs5}
or with \rref{33.5} by \cite{MT99} 
\beq
C_l = (2\mu a/\hbar^2 N_l)^{1/2} \gamma_{nl} /W_l(2\kappa_B a).
\eeqn{eq_bs5a}
It should be independent of radius $a$. 
This relation between the ANC and a reduced width amplitude which corresponds to a vanishing width 
can be useful for the phenomenological $R$ matrix. 
A similar formalism can be applied
to resonances \cite{DV90}, and provides widths as well as energies of resonances.
\subsection{Capture cross sections}
\label{subsec:capth}
The determination of capture cross sections requires the calculation of matrix elements of the 
electromagnetic multipole operators $\cM^{\sigma\lambda}_{\mu}$ 
between an initial scattering state and a final bound state. 
One can take into account the division of the configuration space 
in the general case \cite{BD83} but the principle of the calculation can be explained 
more easily in the simple potential model using equations \rref{2.61}, \rref{2.62} and \rref{2.58}. 

According to the $R$-matrix framework, the radial matrix element between an initial scattering 
wave function $u_i (r) \equiv u^{J_i}_{I_fl_i(I l)} (r)$ and a final bound-state 
wave function $u_f (r) \equiv u^{J_f}_{I_fl_f} (r)$ appearing in \rref{2.58} 
can be written for an electric multipole as 
\beq
\int_0^\infty u_f r^\lambda u_i dr 
= \int_0^a u_f^{\rm int} r^\lambda u_i^{\rm int} dr  
+ \int_a^\infty u_f^{\rm ext} r^\lambda u_i^{\rm ext} dr.
\eeqn{eq_cap1}
The internal matrix element is given by 
\beq
\int_0^a u_f^{\rm int} r^\lambda u_i^{\rm int} dr = \sum_{k,k'}  c_{f,k'} c_{i,k} 
\int_0^a  \varphi_{k'} r^\lambda \varphi_k dr,
\eeqn{eq_cap2}
where coefficients $c_{f,k'}$ and $c_{i,k}$ are related to the final and initial wave functions, respectively. 
Notice that the bases could be different for both states. 
In the external region, we have 
\beq \fl
\int_a^\infty u_f^{\rm ext} r^\lambda u_i^{\rm ext} dr
= C_i C_f \int_a^{\infty} W_{l_f}(2\kappa_B r)r^{\lambda}(I_{l_i}(kr)-U^{J_i}_{I_fl_i,Il} O_{l_i}(kr))dr.
\eeqn{eq_cap3}
As for elastic scattering, the total matrix element (\ref{eq_cap1}) should not depend on the channel radius $a$, 
whereas each contribution does depend on $a$. 
For transitions to weakly bound states, the external term is dominant \cite{BD85} 
since the Whittaker function slowly decreases at large distances. 
Neglecting completely the internal contribution leads to the ``external-capture''model \cite{CD61}. 
A typical example is the $^7$Be(p,$\gamma)^8$B reaction where the ground state is bound 
by 137 keV only \cite{MT90}. 
On the contrary, resonant reactions, or transitions to deeply bound states provide a dominant internal term. 
\subsection{Propagation methods}
\label{subsec:propa}
For long-range potentials, the channel radius may become very large. 
This may induce a prohibitive size for the $R$-matrix basis and lead to long computation times. 
This problem was first met in atomic physics because of the long tail of polarization potentials 
and led to the development of propagation methods \cite{LW76,BBM82,BN95}. 
This situation can also occur in nuclear physics, for example, in three-body systems \cite{DTB06} 
or in coupled-channel calculations \cite{CT94}. 

Different methods have been proposed to address this problem. 
The basic idea is, either to propagate the wave function or the $R$-matrix 
over several intervals on which the basis size remains reasonable, 
or to determine ``distorted"  Coulomb functions valid at distances shorter 
than the channel radius. 
We present here a propagation method directly derived for the $R$-matrix 
(see for example \cite{BN95}). 

The idea of propagation methods \cite{LW76} is to divide the internal region $(0,a)$ in $N_s$ subregions 
$(a_{\alpha-1},a_{\alpha})$ for $\alpha = 1, N_s$, with $a_0=0$ and $a_{N_s}=a$. 
The intermediate radii $a_{\alpha}$ can be chosen equidistant, but this is not mandatory. 
The width of the intervals and the basis size in each interval can also depend on the number of 
oscillations of the wave function \cite{BBM82,BN95}. 
With small intervals, the $R$-matrix bases remain limited, but the number $N_s$ of repetitions 
of the calculation may be large. 
Approximations of the potential can often be employed in some intervals. 
In some cases, the size of the intervals is chosen small enough so that the potential 
may be considered as constant which allows an analytical propagation \cite{LW76}. 

We briefly present here the principle of the propagation technique for potential scattering 
with a basis in each interval. 
We refer to references \cite{BBM82,BN95} for a multichannel extension. 
A Bloch operator $\cL_\alpha$ is now  defined at each boundary $a_{\alpha}$ as
\beq
\cL_{\alpha}=\frac{\hbar^2}{2\mu} \delta (r-a_{\alpha}) \frac{d}{dr},
\hspace*{1 cm} \alpha = 1, N_{s}.
\eeqn{eq3.5.2}
The Bloch-Schr\"odinger equation is replaced by a set of equations 
\beq
(H_l + \cL_{\alpha} - \cL_{\alpha-1} - E) \uintm{\alpha} 
= (\cL_{\alpha} - \cL_{\alpha-1}) \uintm{\alpha}, 
\hspace*{1 cm} \alpha = 1, N_s
\eeqn{eq3.5.1}
with $\cL_0 = 0$.  
The boundary conditions are $\uintm{1}(0) = 0$ and 
\beq
\uintm{\alpha}(a_{\alpha})=\uintm{\alpha+1}(a_{\alpha}),
\hspace*{1 cm} \alpha = 1, N_{s}
\eeqn{eq3.5.0}
with $\uintm{N_s+1} \equiv \uext$. 
In each interval, the wave function is expanded over a set of $N_\alpha$ basis functions 
\beq
\uintm{\alpha}(r) = \sum_{j=1}^{N_\alpha} c^{\alpha}_j \varphi^\alpha_j (r).
\eeqn{eq3.5.3}
Equations (\ref{eq3.5.1}) can be solved with approximate Green functions 
as in section \ref{sec:form}. 
By projecting \rref{eq3.5.1} on one of the basis functions, 
the solutions can be approximated as 
\beq
\uintm{\alpha}(r)=\sum_{j j'}\bigl(\ve{C}_{\alpha}^{-1}\bigr)_{jj'}
\la \varphi^\alpha_{j'}|{\cal L_{\alpha}}-\cL_{\alpha-1}|\uintm{\alpha} \ra \varphi^\alpha_{j}(r),
\eeqn{eq3.5.4}
where the symmetric matrix $\ve{C}_{\alpha}$ is defined in each interval by
\beq
C_{\alpha,ii'}=\la \varphi^\alpha_i | H_l + {\cal L}_{\alpha} - \cL_{\alpha-1} -E | \varphi^\alpha_{i'} \ra.
\eeqn{eq3.5.5}
The Dirac notation represents an integration limited to the range $(a_{\alpha-1},a_{\alpha})$. 

For each interval, (\ref{eq3.5.4}) can be used to determine a relation between values of 
the wave function and its first derivative at $a_{\alpha-1}$ and $a_{\alpha}$, 
\beq
\uintm{\alpha}(a_{\alpha-1})
& = & {\cal R}^{\alpha}_{10}\uintmp{\alpha}(a_{\alpha})-{\cal R}^{\alpha}_{11}\uintmp{\alpha}(a_{\alpha-1}), 
\eoln{eq3.5.6b}
\uintm{\alpha}(a_{\alpha})
& = & {\cal R}^{\alpha}_{00}\uintmp{\alpha}(a_{\alpha})-{\cal R}^{\alpha}_{01}\uintmp{\alpha}(a_{\alpha-1}) , 
\eeqn{eq3.5.6}
where various values of the approximate Green functions ${\cal R}^{\alpha}_{\beta\beta'}$ are defined as 
\beq
{\cal R}^{\alpha}_{\beta\beta'}=\frac{\hbar^2}{2\mu} \sum_{jj'}\, \varphi^\alpha_j(a_{\alpha-\beta}) 
\bigl(\ve{C}_{\alpha}^{-1}\bigr)_{jj'} \varphi^\alpha_{j'}(a_{\alpha-\beta'})
\eeqn{eq3.5.7}
with $\beta\beta' = 0,1$. 

An $R$ matrix can be defined at each boundary with an extension of \rref{32.8} as 
\beq
\uintm{\alpha}(a_{\alpha}) = a_{\alpha}R(a_{\alpha})\uintmp{\alpha}(a_{\alpha}),
\hspace*{1 cm} \alpha = 1, N_s.
\eeqn{eq3.5.6a}
Equations (\ref{eq3.5.6b}), (\ref{eq3.5.6}) and (\ref{eq3.5.6a}) provide relationships 
between $R$ matrices at $a_{\alpha-1}$ and $a_{\alpha}$ ($\alpha = 2, N_s$), 
\beq
a_{\alpha-1}R(a_{\alpha-1})=-{\cal R}^{\alpha}_{11}+{\cal R}^{\alpha}_{10}\bigl[
{\cal R}^{\alpha}_{00}-a_{\alpha}R(a_{\alpha})\bigr]^{-1}{\cal R}^{\alpha}_{01},
\eoln{eq3.5.8a}
a_{\alpha}R(a_{\alpha})={\cal R}^{\alpha}_{00}-{\cal R}^{\alpha}_{01}\bigl[
{\cal R}^{\alpha}_{11}+a_{\alpha-1}R(a_{\alpha-1})\bigr]^{-1}{\cal R}^{\alpha}_{10}.
\eeqn{eq3.5.8}
The latter equation provides an outwards propagation [$R(a_{\alpha})$ from $R(a_{\alpha-1})$] 
and the former provides a backwards propagation [$R(a_{\alpha-1})$ from $R(a_{\alpha})$]. 
This technique is quite efficient in multichannel calculations. 
Numerically the main part of the $R$-matrix calculation arises in the inversion of matrices $\ve{C}_\alpha$ 
[see (\ref{eq3.5.7})]. 
It may save computer time to diagonalize them when many energies are needed. 
The size of these matrices is given by the number of basis functions times the number of channels. 
If the calculation involves many channels, 
reducing the number of basis functions may lead to a significant reduction of the computer times.

The collision matrix is obtained from $R(a_{N_s})$ with \rref{32.12}. 
The external wave function $\uext$ is thus known. 
By starting from the last interval, the wave function is determined  by its coefficients 
in each interval. 
If the $\cL_{\alpha-1}$ terms are simplified in \rref{eq3.5.1}, one obtains the system 
\beq
\sum_{j'} \left[ C_{\alpha,jj'} + \la \varphi^\alpha_j|{\cal L}_{\alpha-1}|\varphi^\alpha_{j'}\ra 
\right] c^{\alpha}_{j'} = \la \varphi^\alpha_j|{\cal L}_{\alpha}|\uintm{\alpha}\ra
\eeqn{eq3.5.9}
where \rref{eq3.5.6a} can be used to eliminate $\uintm{\alpha}'(a_{\alpha})$ in the right-hand side. 
The matrix in this system can be inverted with \rref{A.2}. 
Starting with $\alpha = N_s$ and going backwards one obtains the coefficients in all intervals. 

Let us briefly discuss other techniques dealing with long-range potentials. 
In \cite{DTB06}, in the framework of three-body continuum states, 
we propagate the wave functions from $a_{\alpha-1}$ to $a_{\alpha}$ by using the Numerov algorithm. 
This approach avoids the choice of a basis, but requires longer computer times. 
It is also difficult to apply to non-local potentials. 

The Light-Walker propagation consists in approximating the potential by a constant 
in small enough intervals \cite{LW76}. 
In \cite{CT94}, this method is improved by considering linear approximations of the potential. 
In the method suggested by Gailitis \cite{Ga76}, the Coulomb wave functions are modified by $1/r$ expansions, 
which can be used at short distances (see also \cite{NN84,BN95}). 

\subsection{Extension to multichannel collisions}
\label{sec:Rmulti}
Until now the presentation was, for the sake of clarity and simplicity, limited to single-channel 
calculations which also neglect the spins of the colliding particles. 
This does not affect the general properties of the $R$-matrix theory. 
However many problems require a multichannel approach. 

In a many-body problem, the Schr\"odinger equation \rref{2.43} involves Hamiltonian \rref{2.38}. 
The total wave function $\Psi^{JM\pi}$ of the system, with total angular momentum $J$ 
and parity $\pi$, is expanded over a set of channel functions \rref{2.41}, denoted as $|c \ra$. 
If we assume that a single relative coordinate appears in the problem 
or that we approximate all relative coordinates by a single one, 
the wave function is given by \rref{2.50}. 
The Schr\"odinger equation is replaced by a set of differential equations 
\beq
\sum_{c'} \Bigl[(T_{c}+E_c-E)\delta_{c c'} + V_{c c'} \Bigr] u_{c'}=0,
\eeqn{eq_mul3}
where, as before, $T_{c}$ includes the centrifugal term. 
This system is a particular case or a local approximation of \rref{2.45}. 

A typical example is given by coupled-channel calculations where channel functions 
are defined by \rref{2.41}. 
This situation occurs, for example, in coupled-channel calculations with a discretized continuum 
(CDCC) \cite{AIK87}, see section \ref{sec:CDCC}. 
Another example corresponds to three-body scattering, 
where the channel functions $| c\ra$ contain various 
quantum numbers defined in the hyperspherical formalism \cite{ZDF93}, see section \ref{sec:tbcs}. 

In all cases the problem is first to determine the potentials $V_{c c'}$ and 
then to solve (\ref{eq_mul3}) for positive energies. 
Here we essentially deal with the second step. 
We thus assume that the potentials are known and that they present the asymptotic behaviour 
\beq
V_{c c '}\mathop{\longrightarrow}\limits_{r \rightarrow \infty}
\frac{Z_{1c}Z_{2c}e^2}{r} \delta_{c c '}.
\eeqn{eq_mul5}
In these conditions the asymptotic form of the radial wave functions $u_{c}(r)$ 
is given by (\ref{2.44}) and the radial wave functions in the external region 
at energy $E$ are defined as 
\beq
u^{\rm ext}_{c(c_0)}(r)=
\left\{\begin{array}{ll}
v_{c}^{-1/2} \Bigl( I_{c} (k_{c} r)\delta_{c c_0} - U_{c c_0} O_{c} (k_{c} r) \Bigr)
& {\rm \ for\ } E_c < E \\
A_{cc_0} W_{-\eta_{c},l+\frac{1}{2}}(2\kappa_{c}r) & {\rm \ for\ } E_c > E,
\end{array} \right .
\eeqn{eq_mul7}
where $c_0$ is the entrance channel ($E_{c_0} < E$).

The Bloch operator (\ref{32.1}) is defined in the multichannel formalism as 
\beq
\cL = \sum_{c} |c\ra \cL_{c} \la c |, \hspace*{1 cm}
\cL_{c}= \frac{\hbar^2}{2\mu_{c}} \, \delta (r-a) 
\left( \frac{d}{dr} - \frac{B_{c}}{r} \right),
\eeqn{eq_mul8}    
where coefficients $B_{c}$ are chosen as zero or as in (\ref{eq_bs0}) for open and closed
channels, respectively. 
Notice that these coefficients then depend on energy for closed channels. 
This choice may be inefficient in some variants of the $R$ matrix \cite{Ph75} 
but is convenient in \rref{eq_mul12} and \rref{eq_mul13} below. 
The Bloch-Schr\"odinger equation is given by 
\beq
\sum_{c'} \Bigl[(T_{c}+\cL_{c}+E_c-E)\delta_{c c'} + V_{c c'} \Bigr] u^{\rm int}_{c'}(r)
= \cL_{c} u^{\rm ext}_{c}.
\eeqn{eq_mul9}
The internal parts of the radial wave functions are expanded over a basis $\varphi_j(r)$, 
\beq
u^{\rm int}_c (r) = \sum_{j=1}^N c_{c j} \varphi_j(r).
\eeqn{eq_mul10}
The determination of the $R$ matrix and of the collision matrix $\ve{U}$ are direct extensions 
of the formalism developed in sections \ref{sec:form} and \ref{sec:scat}. 

The $R$ matrix is defined as 
\beq
u_c (a) = \sum_{c'} (\mu_{c}/\mu_{c'})^{1/2} R_{c c'} [a u'_{c'} (a) - B_{c'} u_{c'} (a)].
\eeqn{eq_mul10a}
Matrix $\ve{R}$ is symmetric, with elements given by 
\beq
R_{c c'}(E) = \frac{\hbar^2}{2\sqrt{\mu_{c}\mu_{c'}} a} 
\sum_{i,i'=1}^N \varphi_i(a) (\ve{C}^{-1})_{c i, c'i'} \varphi_{i'}(a),
\eeqn{eq_mul11}
where 
\beq
C_{c i, c'i'} = \la \varphi_{i}| T_{c}+\cL_{c}+E_c-E |\varphi_{i'} \ra \delta_{c c'} 
+ \la \varphi_{i}| V_{c c'} |\varphi_{i'} \ra.
\eeqn{eq_mul11a}
Like in section \ref{sec:prop}, the spectral decomposition of the symmetric matrix $\ve{C}$ 
for an orthonormal basis provides the canonical form of the multichannel $R$ matrix as 
\beq
R_{c c'}(E) = \sum_{n} \frac{\gamma_{nc}\gamma_{nc'}}{E_n-E}
\eeqn{eq_mul11b}
where the real poles $E_n$ are the eigenvalues of $\ve{C}$ and the reduced-width amplitude 
of pole $E_n$ in channel $c$ is expressed as a function of the components of the corresponding 
normed eigenvector $\ve{v}_n$ as 
\beq
\gamma_{nc} =  \left( \frac{\hbar^2}{2\mu_{c} a} \right)^{1/2} \sum_{i=1}^N v_{n,ci} \varphi_i(a).
\eeqn{eq_mul11c}
The number of terms in the sum \rref{eq_mul11b} is given by the product of the number of channels by $N$. 

The collision matrix is obtained with 
\beq
\ve{U}=\ve{Z}^{-1} \ve{Z}^*,
\eeqn{eq_mul12}
where an element of matrix $\ve{Z}$ reads 
\beq
Z_{c c'} = (k_{c'} a)^{-1/2} 
\Bigl[ O_{c} (k_{c} a)\delta_{c c '} - k_{c'} a R_{c c'} O'_{c'} (k_{c'} a) \Bigr].
\eeqn{eq_mul13}
For complex potentials as encountered in the optical model, 
expression \rref{eq_mul12} must be modified into 
\beq
\ve{U}=\ve{Z}_O^{-1} \ve{Z}_I,
\eeqn{eq_mul12a}
where $\ve{Z}_O = \ve{Z}$ and $\ve{Z}_I \neq \ve{Z}_O^*$ is given by a similar expression 
with outgoing functions replaced by incoming ones.
Matrix $\ve{U}$ is then not unitary. 

The dimension of the $R$ matrix is equal to the number of channels included 
in the calculation; it does not depend on energy. 
On the contrary the dimension of the collision matrix $\ve{U}$ is given by the number of {\sl open} channels 
and may vary with energy. 
In \rref{eq_mul13}, only open channels contribute thanks to the choice of boundary parameters $B_c$. 
The $R$ matrix can be modified by eliminating the closed channels. 
When not all channels are open, let us denote open channels by $c$ and closed channels by $\bar{c}$. 
Equation (\ref{eq_mul11}) remains valid with matrix $\ve{C}$ replaced by a smaller open-channel matrix 
$\ve{C}^o$ with elements \cite{TW52,LT58,HSV98} 
\beq
C^o_{ci,c'i'} = C_{ci,c'i'} 
- \sum_{\bar{c}j,\bar{c}'j'} \la \varphi_i | V_{c\bar{c}} | \varphi_j \ra (\bar{\ve{C}}^{-1})_{\bar{c}j,\bar{c}'j'} 
\la \varphi_{j'} | V_{\bar{c}'c'} | \varphi_{i'} \ra ,
\eeqn{eq_mul14}
where $\bar{\ve{C}}$ is the restriction of the full matrix to closed channels 
and $\bar{c}$, $\bar{c}'$ are the corresponding indices. 

The internal components of the wave function are given for both open and closed channels by 
\beq \fl
u^{\rm int}_{c (c_0)} (r) & = & \sum_{c'} \frac{\hbar^2 k_{c'}}{2\mu_{c'} a} 
\Bigl[ I'_{c'} (k_{c'} a)\delta_{c' c_0} - U_{c' c_0} O'_{c'} (k_{c'} a) \Bigr] 
\sum_{i,i'=1}^N \varphi_i(r) (\ve{C}^{-1})_{c i, c'i'} \varphi_{i'}(a),
\eeqn{eq_mul15}
where the sum over $c'$ runs over open channels only. 
The coefficients of the external part of the wave function for closed channels 
read 
\beq \fl
A_{\bar{c} c_0} = [W_{-\eta_{\bar{c}},l+\frac{1}{2}}(2\kappa_{\bar{c}}a)]^{-1}
\sum_{c'} (\mu_{c} k_{c'}/\hbar)^{1/2} a R_{\bar{c} c'} 
\Bigl[ I'_{c'} (k_{c'} a)\delta_{c' c_0} - U_{c' c_0} O'_{c'} (k_{c'} a) \Bigr],
\eeqn{eq_mul16}
where the sum over $c'$ also runs over open channels only.

\setcounter{equation}{0}
\renewcommand{\theequation}{4.\arabic{equation}}

\section{Applications of the calculable $R$ matrix}
\label{sec:apcomp}

\subsection{Conditions of the calculations}
\label{subsec:conditions}
In this section we apply the calculable $R$-matrix method to the scattering by a potential. Our goal is not to fit  experimental data, but to illustrate the method under different conditions, and with different types of basis wave functions. In this approach, both colliding particles are assumed to be structureless, and to interact through a potential [see (\ref{31.3})]. In general, this potential $V(r)$ involves
a local term $U(r)$ and a non-local term $W(r,r')$ such that
\beq
V(r)u_l(r)=U(r)u_l(r)+\int_0^{\infty} W(r,r')u_l(r')dr'.
\eeqn{eq3.4.1}
In the following, unless specified otherwise, only the local term is included. In nuclear physics applications, the
local
potential $U(r)$ usually contains a nuclear and a Coulomb contributions, denoted as $V_N(r)$ and $V_C(r)$,
respectively.
The potential may be $l$ dependent, by adapting its parameters to the $l$ value.
The Schr\"odinger equation associated with potential (\ref{eq3.4.1}) can be solved exactly (at the computer precision) with the Numerov method \cite{Ra72,Hu94,Mi09}. The exact phase shifts and wave functions will be of course compared with the $R$-matrix calculations.

In nuclear-physics applications, the reduced mass is expressed in terms of the nucleon mass as
\beq
\mu=\frac{A_1A_2}{A_1+A_2}\,m_N,
\eeqn{eq3.4.4}
where $A_1$ and $A_2$ are the nucleon numbers. The calculations are performed with $\hbar^2/2m_N=20.736$ MeV.fm$^2$ and $e^2=1.44$ MeV.fm.

To cover a broad variety of applications, we have first selected three typical systems:
\begin{enumerate}
	\item The $^{12}$C+p system ($l=0$) which presents a narrow resonance ($\Gamma=37$ keV) at $E=0.42$ MeV.
	\item The $\alpha+\alpha$ system ($l=4$) which presents a broad resonance ($\Gamma=3.5$ MeV) near $E=11.3$ MeV.
	\item The $\alpha+^3$He system ($l=0$) which is non resonant.
\end{enumerate}

These cases are typical examples of nuclear-physics applications, and will be treated by different 
types of basis functions.
Here, we only consider real potentials. This is consistent with the low-energy regime where 
the $R$-matrix method is well adapted. The extension to complex potentials is however trivial. 

Then, the calculation of $e^- -$H phase shifts will provide a typical example
of an atomic-physics application with a non-local potential. Atomic units will be used.

\subsection{Basis functions}
\label{subsec:wf}

We consider different families of basis functions $\varphi_i(r)$ [$i=1,\ldots ,N$, see (\ref{32.4})] vanishing at $r=0$,
commonly used in the literature. The corresponding matrix elements are given in Appendix C.

\begin{enumerate}
\item {{\em Sine functions} \\
The basis functions are given by
\beq
\varphi_i(r)=\sin\frac{\pi r}{a}(i-1/2).
\eeqn{eq3.4.14}
This choice seems natural since it can simulate the oscillating behaviour of the wave function
near the channel radius. However we will show that these basis functions are not suitable
for accurate calculations since the derivative satisfies \rref{33.8} (see section \ref{sec:bc}), 
i.e. it vanishes at $r=a$ for each $i$ value,
\beq
\varphi'_i(a)=0.
\eeqn{eq_sine}
This property means that any finite combination of these basis functions will present a zero derivative
at $r=a$. As discussed before, the matching between the internal and external solutions is therefore expected to be poor,
and the $R$-matrix phase shifts rather inaccurate.}

\item {{\em Gaussian functions} \\
The basis functions have a Gaussian dependence \cite{Ka77} with different size parameters
\beq
\varphi_i(r)=r^{l+1}\exp(-(r/b_i)^2),
\eeqn{eq3.4.5}
where the $b_i$ are chosen as a geometric progression
\beq
b_i=b_1\,x_0^{i-1},
\eeqn{eq3.4.5a}
and are therefore determined by a set of 2 parameters ($b_1,x_0$).
Gaussian-type basis functions (\ref{eq3.4.5}) allow an analytical calculation of the overlap 
and kinetic-energy matrix
elements. For Gaussian potentials, the matrix elements are analytical as well.}

\item {{\em Lagrange functions} \\
The Lagrange functions \cite{HRB02} are defined in the $(0,a)$ interval as
\beq
\varphi_i (r)=(-1)^{N+i}\left( \frac{r}{ax_i}\right) ^n \sqrt{a x_i(1-x_i)} 
\frac{P_N(2r/a-1)}{r-a x_i},
\eeqn{eq3.4.9}
where $P_N$ is the Legendre polynomial of order $N$, and $x_i$ are the zeros of 
\beq
P_N(2x_i-1)=0.
\eeqn{eq3.4.10}
In \rref{eq3.4.9}, the factor $(r/ax_i)^n$ is aimed at regularizing the basis function at the origin. For 
two-body calculations defined in the $(0,a)$ range, we use $n=1$, which ensures that
the wave function vanishes at the origin. 

The basis functions satisfy the Lagrange conditions
\beq
\varphi_i(a x_j)=(a \lambda_i)^{-1/2}\delta_{ij},
\eeqn{eq3.4.11}
where $\lambda_i$ is the weight of the Gauss-Legendre quadrature corresponding to the $(0,1)$ interval.
This basis is exactly equivalent to the Legendre basis defined by
\beq
\varphi_i(r)=rP_{i-1}(2r/a-1).
\eeqn{eq_leg}
However, if the matrix elements with basis functions \rref{eq3.4.9}
are computed at the Gauss approximation of order $N$, consistent with the $N$ mesh
points, their calculation is strongly
simplified.  At this approximation, the overlap is given by
\beq
\langle \varphi_i|\varphi_j\rangle =\int_0^a \varphi_i(r)\varphi_j(r) dr\approx \delta_{ij},
\eeqn{eq3.4.11b}
and the local-potential matrix is diagonal with
elements just given by the value of the potential at the mesh points,
\beq
\langle \varphi_i|U|\varphi_j\rangle =\int_0^a \varphi_i(r)U(r)\varphi_j(r) dr \approx
U(a x_i)\delta_{ij}.
\eeqn{eq3.4.11c}
This calculation is easily extended to non-local potentials $W(r,r')$
\beq
\langle \varphi_i|W|\varphi_j\rangle \approx a \sqrt{\lambda_i \lambda_j} W(a x_i,a x_j).
\eeqn{eq3.4.11d}
These matrix elements do not need any evaluation of integral.
The kinetic energy term is given by analytical expressions \cite{Ba06} (see Appendix C).
This method has been shown to be quite efficient and accurate in various fields for 
bound states \cite{Ba06} as well as for scattering states \cite{HRB02,DTB06}.
In the present applications, the Gauss approximation will be used systematically, except
in the $e^- -$H scattering, where this approximation will be also tested by a Simpson
quadrature for the matrix elements.}
\end{enumerate}

\subsection{Application to a narrow resonance: $^{12}$C+p}
\label{subsec:c12p}
The $^{12}$C+p system is known to present a narrow resonance ($J^{\pi}=1/2^+$, $E_R=0.42$ MeV, 
$\Gamma=37$ keV \cite{Aj91}) at low energies. As mentioned before, our aim is not to fit these 
properties accurately, but to compare the exact solutions of the Schr\"odinger equation (\ref{31.2}) 
with the $R$-matrix approach. For the nuclear and Coulomb potentials we choose
\beq
V_N(r)=-73.8 \exp(-(r/2.70)^2),\nonumber \\
V_C(r)=6e^2/r. 
\eeqn{eq3.4.11a}
In what follows, the units in the nuclear potentials are fm and
MeV for lengths and energies, respectively. 
This potential reproduces the resonant behaviour of the $s$ phase shifts at 0.42 MeV. 
To simulate the Pauli principle, a nucleus-nucleus potential may contain additional (unphysical) bound states
\cite{BDV75,WT77}. These Pauli forbidden states show up
in microscopic calculations, where the internal structure of the colliding nuclei is taken into account,
but their effect can be partly simulated in non-microscopic theories 
by additional bound states in the potential (see \cite{BDV75,WT77} for details).
For the $^{12}$C+p system, the $l=0$ potential contains one forbidden state.

\begin{figure}[ht]
\begin{center}
\includegraphics[width=0.5\textwidth,clip]{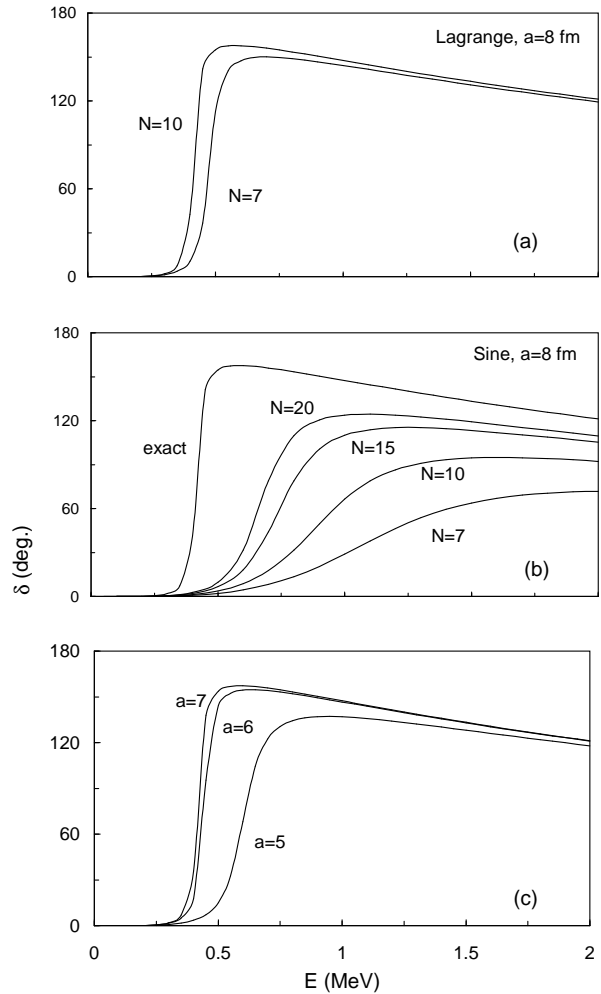} 
\caption{\label{fig_c12p1} $^{12}$C+p $R$-matrix phase shifts (in degrees) for different 
bases and conditions ($l=0$). (a) Lagrange functions
at $a=8$ fm (the exact results are superimposed to the $N=10$ curve), (b) Sine 
functions at $a=8$ fm, (c) Convergence as a function of $a$ for $N=15$; the Lagrange and
Gaussian results are superimposed ($a=7$ fm corresponds to the exact results).}
\end{center}
\end{figure}

In Fig.~\ref{fig_c12p1}, we present the exact phase shifts and $R$-matrix calculations with Lagrange (a) 
and Sine (b) functions. The channel radius is chosen as $a=8$ fm, where the nuclear interaction is 
negligible. Fig.~\ref{fig_c12p1} (a) shows that with the Lagrange functions, the convergence is reached 
with $N\geq10$. We show an example with $N=7$, where $N$ is aimed at being not large enough. Fig.~\ref{fig_c12p1} (b) 
illustrates the Sine functions, poorly adapted to a good matching since the left derivative of the 
wave function at $r=a$ is zero [see \rref{eq_sine}]. Even $N=20$ is far from the exact calculation. In Fig.~\ref{fig_c12p1} (c), 
we present the convergence as a function of the channel radius with the Lagrange functions
($N=15$ is fixed). Results obtained with Gaussian functions (with $b_1=1.4$ fm, $x_0=0.6$) are identical
at the scale of the figure. 
As expected, $a=5$ fm is too small ($\left|V_N(a)/V_C(a)\right|=1.4$). To obtain a good stability, 
radii larger than 6 fm should be used.

The matching problem is illustrated in Fig.~\ref{fig_c12p2}, where we show the wave function at $E=2$ MeV,
with $a=8$ fm, $N=15$. With the Lagrange functions, the matching between internal and external wave functions is quite smooth. For Sine functions, the matching is poor, which has a direct impact on the phase shifts.

\begin{figure}[ht]
\begin{center}
\includegraphics[width=0.5\textwidth,clip]{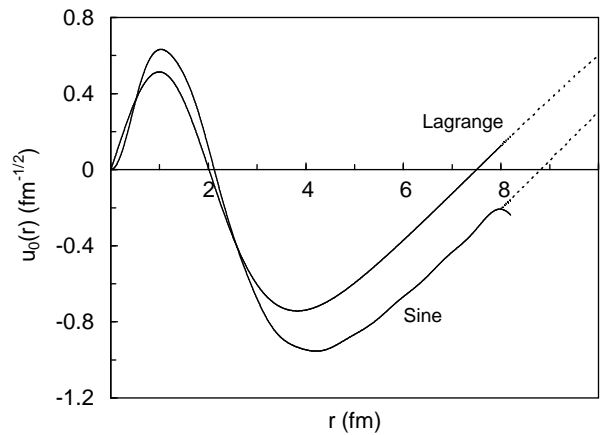} 
\caption{\label{fig_c12p2} $^{12}$C+p $l=0$ wave functions with Lagrange and Sine functions at $E=2$ MeV
($a=8$ fm, $N=15$). Solid lines represent the internal wave function, and dotted lines the external
parts. The Lagrange wave function is superimposed to the exact result.}
\end{center}
\end{figure}

\begin{table}[ht]
\begin{center}
\caption{$^{12}$C+p $l=0$ phase shifts (in degrees) and matching parameters \rref{eq_match} for different bases ($a=8$ fm).
\label{tab_c12p}}
\begin{tabular}{lcccc|ccc}
& \multicolumn{4}{c}{phase shift} & \multicolumn{3}{|c}{matching parameter} \\
\cline{2-5}\cline{6-8}
$E$ (MeV) & exact & $N=7$ & $N=10$ & $N=15$ & $N=7$ & $N=10$ & $N=15$\\
\hline
Lagrange  &  &  &  &  &  &  & \\
    0.5 &   154.66 &   112.90 &   154.94 &   154.59 &     5.96 &     0.65 &     0.01\\
1.0 &   147.48 &   144.22 &   147.55 &   147.48 &     3.68 &     0.28 &     0.00\\
    1.5 &   133.30 &   311.02 &   133.35 &   133.30 &     1.67 &     0.20 &     0.00\\
2.0 &   121.18 &   299.30 &   121.23 &   121.18 &     1.21 &     0.16 &     0.00\\
 &  &  &  &  &  &  & \\
Gaussian&  &  &  &  &  &  & \\
    0.5 &   154.66 &   179.36 &   154.53 &   154.53 &     5.61 &     0.02 &     0.01\\
1.0 &   147.48 &   146.52 &   147.47 &   147.47 &     0.04 &     0.01 &     0.01\\
    1.5 &   133.30 &   130.28 &   133.29 &   133.29 &     1.06 &     0.00 &     0.01\\
2.0 &   121.18 &   112.15 &   121.18 &   121.18 &     1.55 &     0.01 &     0.01\\
 &  &  &  &  &  &  & \\
Sine &  &  &  &  &  &  & \\
    0.5 &   154.66 &     2.03 &     3.74 &     6.68 &     0.12 &     0.66 &     1.54\\
1.0 &   147.48 &    28.90 &    66.57 &   109.98 &     1.30 &     4.66 &    29.35\\
    1.5 &   133.30 &    63.84 &    94.53 &   113.28 &    11.46 &     6.97 &     3.53\\
2.0 &   121.18 &    71.76 &    92.24 &   105.34 &     5.73 &     3.14 &     2.38\\
\end{tabular}
\end{center}
\end{table}

The accuracy of the different basis functions is illustrated in Table~\ref{tab_c12p} where we compare 
various calculations of the phase shifts. The relative difference between left and right 
derivatives provides the matching parameter
$\epsilon$ defined as
\beq
\epsilon=\frac{|\uext'(a)-\uint'(a)|}{[\uext'(a)+\uint'(a)]/2}.
\eeqn{eq_match}
This comparison is done for the different bases and at four typical energies. 
When $\epsilon$ is small, the matching is obviously accurate. However, in some specific cases
where $\uext'(a)+\uint'(a)$ is small, this parameter may be large, although the phase shift is fairly good. This parameter
should therefore be considered as indicative only.
Table~\ref{tab_c12p} confirms that 
Lagrange and Gaussian functions represent a convenient basis, whereas Sine functions do not provide 
satisfactory results. Note that the use of Gaussian functions implies two additional parameters 
($b_1,x_0$) to define the basis. This choice has to be optimized for the different conditions and
is therefore less direct than Lagrange functions.
\subsection{Application to a broad resonance: $\alpha+\alpha$}
\label{subsec:aa}
The $\alpha+\alpha$ system presents a well known rotational band based on the $0^+$
ground state. The $l=0$ narrow resonance 
corresponds to the unstable ground state of $^8$Be. In order to illustrate the 
$R$-matrix formalism applied to a 
broad resonance, we consider the  $l=4$ partial wave. The experimental energy and width 
of the resonance are $E_R=11.35\pm 0.15$ MeV and $\Gamma \approx 3.5$ MeV \cite{TKG04}.
However, for broad resonances these properties may depend on the definition used (see section 3.6). A 
comparison between experiment and theory cannot be done with a high precision.

The $\alpha+\alpha$ potential of Buck {\sl et al.} \cite{BFW77} has a simple 
Gaussian form, is $l$-independent, and reproduces the $l=0,2,4$ experimental $\alpha+\alpha$ phase 
shifts up to about 20 MeV. It is defined by
\beq
V_N(r)=-122.6225 \exp(-(r/2.132)^2), \nonumber \\
V_C(r)=4e^2 {\rm erf}(r/1.33)/r.
\eeqn{eq_potaa}

Fig.~\ref{fig_aa} shows the phase shifts obtained with the Lagrange and Sine functions, 
with $a=8$ fm. With Lagrange functions, the $R$-matrix perfectly reproduces the exact phase shifts with $N\geq10$. With Sine functions, the convergence is, as expected, much slower.
Table~\ref{tab_aa} gives the phase shifts and $\epsilon$ values under different conditions and at different energies. Again, the importance of the matching on the phase shifts is obvious.

\begin{figure}[ht]
\begin{center}
\includegraphics[width=0.6\textwidth,clip]{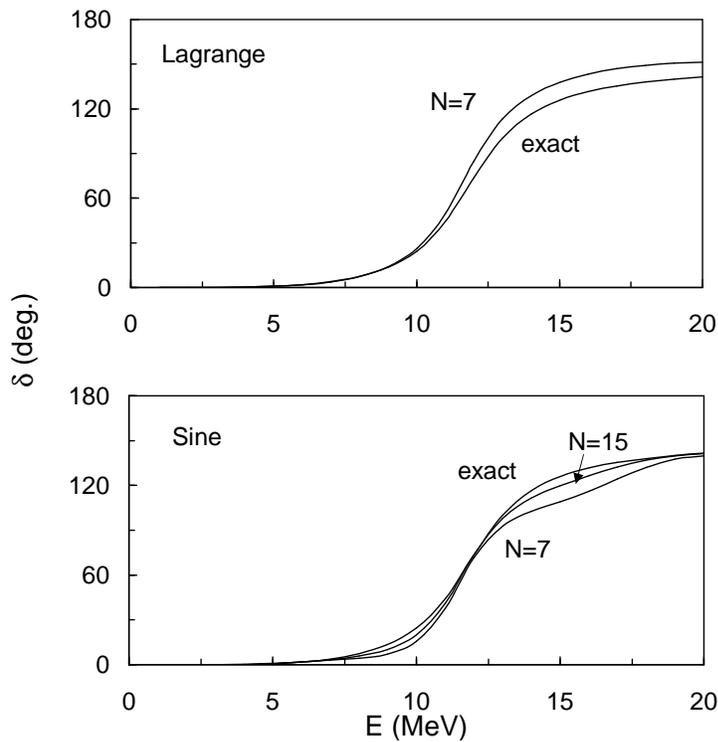} 
\caption{\label{fig_aa} $\alpha+\alpha$ $l=4$ phase shifts with the Lagrange and Sine functions
for different $N$ values. For the Lagrange functions $N\geq 10$ is indistinguishable from the exact result.}
\end{center}
\end{figure}

\begin{table}[ht]
\begin{center}
\caption{$\alpha+\alpha$ $l=4$ phase shifts (in degrees) and matching parameters for Lagrange and Sine bases ($a=8$ fm).
\label{tab_aa}}
\begin{tabular}{lcccc|ccc}
& \multicolumn{4}{c}{phase shift} & \multicolumn{3}{|c}{matching parameter} \\
\cline{2-5}\cline{6-8}
$E$ (MeV) & exact & $N=7$ & $N=10$ & $N=15$ & $N=7$ & $N=10$ & $N=15$\\
\hline
Lagrange  &  &  &  &  &  &  & \\
    5 &     0.79 &     0.81 &     0.80 &     0.80 &     0.05 &     0.02 &     0.00\\
   10 &    24.46 &    26.19 &    24.58 &    24.48 &     0.42 &     0.12 &     0.00\\
   15 &   125.88 &   137.90 &   126.00 &   125.88 &     0.31 &     0.18 &     0.01\\
   20 &   141.55 &   151.24 &   141.53 &   141.54 &     1.26 &     5.67 &     0.13\\
 &  &  &  &  &  &  & \\
Sine  &  &  &  &  &  &  & \\
    5 &     0.79 &     0.53 &     0.62 &     0.68 &     0.66 &     0.65 &     0.65\\
   10 &    24.46 &    15.55 &    17.74 &    19.84 &     2.23 &     2.20 &     2.17\\
   15 &   125.88 &   109.09 &   116.76 &   119.87 &     2.16 &     2.09 &     2.06\\
   20 &   141.55 &   139.82 &   141.24 &   141.37 &     3.27 &     5.84 &     6.21\\
\end{tabular}
\end{center}
\end{table}

\subsection{Application to a non-resonant system: $\alpha+^3$He}
\label{subsec:ahe3}
For $l=0$, the experimental $\alpha+^3$He phase shifts are well reproduced by the potential of 
Buck and Merchant \cite{BM88}, 
\beq
V_N(r)&=&-66.10\exp(-(r/2.52)^2),  \nonumber \\
V_C(r)&=&4e^2(3-(r/r_C)^2)/2r_C  {\rm \ for \ } r\le r_C,\nonumber \\
&=&4e^2/r {\rm \ for \ } r\ge r_C,
\eeqn{eq_potahe3}
with $r_C=3.095$ fm.
In Fig.~\ref{fig_ahe3_1}, we use the Lagrange functions for two radii: 
$a=8$ fm, and $a=5$ fm. For $a=8$ fm, the convergence is reached as soon as $N\geq 10$. On the contrary 
the choice $a=5$ fm is not consistent with one of the $R$-matrix requirements: the nuclear contribution 
is not negligible. In that case, it is impossible to get a good convergence at all energies. 

\begin{figure}[ht]
\begin{center}
\includegraphics[width=0.5\textwidth,clip]{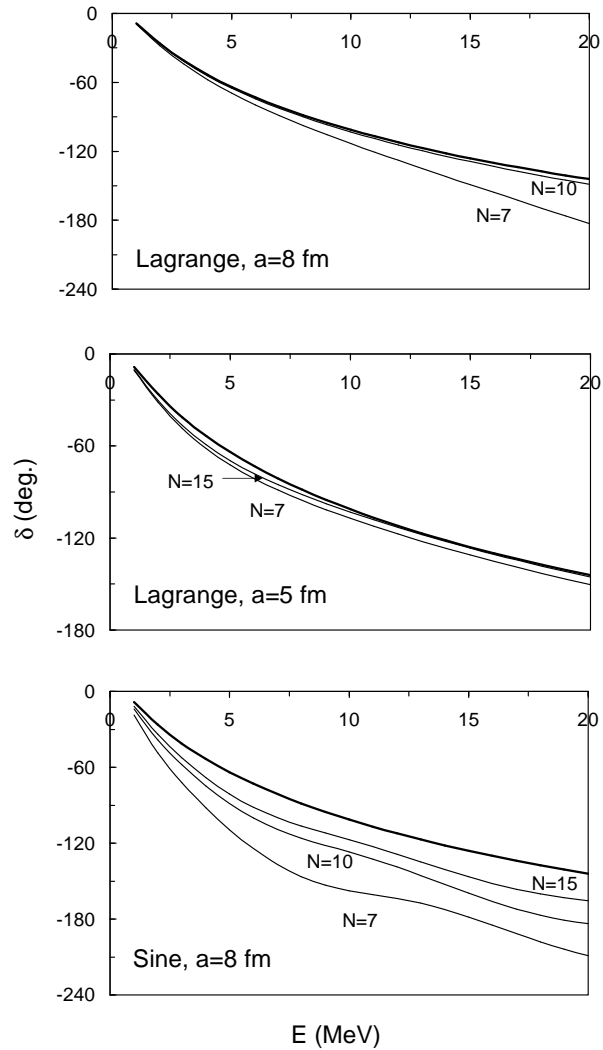} 
\caption{\label{fig_ahe3_1}$\alpha+^3$He $R$-matrix phase shifts with Lagrange ($a=8$ fm and $a=
5$ fm) and Sine ($a=8$ fm) functions for $l=0$. The thick lines represent the exact phase shifts.}
\end{center}
\end{figure}

In Fig.~\ref{fig_ahe3_2}, we compare the wave functions at $E=8$ MeV with the Sine 
and Lagrange functions. In the former case, the matching 
between the internal and external wave functions is poor, which has an impact on the accuracy of 
the phase shifts. Conversely, the $R$-matrix wave function with the Lagrange basis 
is indistinguishable from the exact wave function.

\begin{figure}[ht]
\begin{center}
\includegraphics[width=0.5\textwidth,clip]{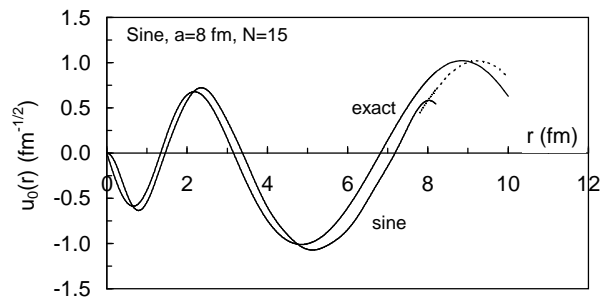} 
\caption{\label{fig_ahe3_2} $\alpha+^3$He $l=0$ wave functions with Sine functions at $E=8$ MeV
($a=8$ fm, $N=15$). Solid lines represent the internal wave function, and the dotted line the external
part with Sine functions. The Lagrange wave function is superimposed to the exact result.}
\end{center}
\end{figure}

Table~\ref{tab_ahe3} gives the phase shifts and matching coefficients under different conditions. For 
$a=8$ fm, $N\geq 10$ provides reasonable phase shifts. If $a$ increases (we choose here $a=10$ fm), 
the $R$-matrix approximation is of course still valid, but the number of basis functions must 
be increased. This is expected as the internal wave function must be reproduced in a wider range. 
In agreement with Fig.~\ref{fig_ahe3_2}, the Sine functions provide a poor approximation of the phase shifts.

\begin{table}[ht]
\begin{center}
\caption{$\alpha+^3$He $l=0$ phase shifts (in degrees) and matching parameters for different bases.
\label{tab_ahe3}}
\begin{tabular}{lccc|cc}
& \multicolumn{3}{c}{phase shift} & \multicolumn{2}{|c}{matching parameter} \\
\cline{2-4}\cline{5-6}
$E$ (MeV) & exact &  $N=10$ & $N=15$ &  $N=10$ & $N=15$\\
\hline
Lagrange, $a=8$ fm &    &  &  &  &   \\
    5  & -63.97 &  -65.02 & -63.96 &        0.16 &     0.01\\
   10  & -101.2 &  -103.1 & -101.2 &        0.89 &     0.22\\
   15  & -125.9 &  -128.9 & -125.9 &        0.13 &     0.01\\
   20  & -144.2 &  -148.5 & -144.2 &        0.14 &     0.02\\
  &  &  &  &  &   \\
Lagrange, $a=10$ fm &    &  &  &  &   \\
    5  & -63.97 &  -70.25 & -63.97 &        0.30 &     0.04\\
   10  & -101.2 &  -109.5 & -101.3 &        0.17 &     0.02\\
   15  & -125.9 &  -134.5 & -126.0 &        5.19 &     0.07\\
   20  & -144.2 &  -149.4 & -144.3 &        0.28 &     0.02\\
 &  &  &  &    &   \\
Sine, $a=8$ fm &    &  &  &  &   \\
    5  & -63.97 &  -88.74 & -81.18 &        2.66 &     2.47\\
   10  & -101.2 &  -127.0 & -117.2 &        1.16 &     0.81\\
   15  & -125.9 &  -159.4 & -146.5 &        2.66 &     2.40\\
   20  & -144.2 &  -183.7 & -165.4 &        1.95 &     1.82\\
\end{tabular}
\end{center}
\end{table}

\subsection{Application to a deep potential}
\label{subsec:deep}
As mentioned before, the potential may contain additional (unphysical) bound states
to simulate the Pauli principle.
In the applications considered so far, the potentials involved a small number of forbidden states. 
The situation is different in heavy-ion reactions, where the number $m_{l}$ of 
forbidden states may be rather large. 
In such conditions the internal wave function presents several oscillations which must be reproduced 
by the basis functions to provide accurate phase shifts.

This problem is illustrated here with the $^{13}$C+$^{12}$C system ($l=0,m_l=12$). 
We take the deep Gaussian potential of 
\cite{BD90} (depth of $-273.7$ MeV and range of 3.06 fm) for the nuclear interaction.
The Coulomb potential has a point-sphere shape [see \rref{eq_potahe3}], with a radius of 8.2 fm. The $R$-matrix phase shifts 
are determined with the Lagrange basis, and compared in Fig.~\ref{fig_c12c13} (upper panel) 
with the exact phase shifts. 
The channel radius is $a=10$ fm. It is clear that small $N$ values do not reproduce the phase shifts. Values 
larger than $N\approx35$ are necessary to obtain accurate results.

We show in Fig.~\ref{fig_c12c13} (lower panel) the wave function for $N=20,30,40$ at $E=10$ MeV. 
Clearly, $N=20$ is not able to give a good description of the oscillations of the internal wave function. The situation 
is of course improved with $N=30$, but only $N=40$ is superimposed to the exact wave function. 
This problem is common to all potentials presenting many bound states.

\begin{figure}[ht]
\begin{center}
\includegraphics[width=0.5\textwidth,clip]{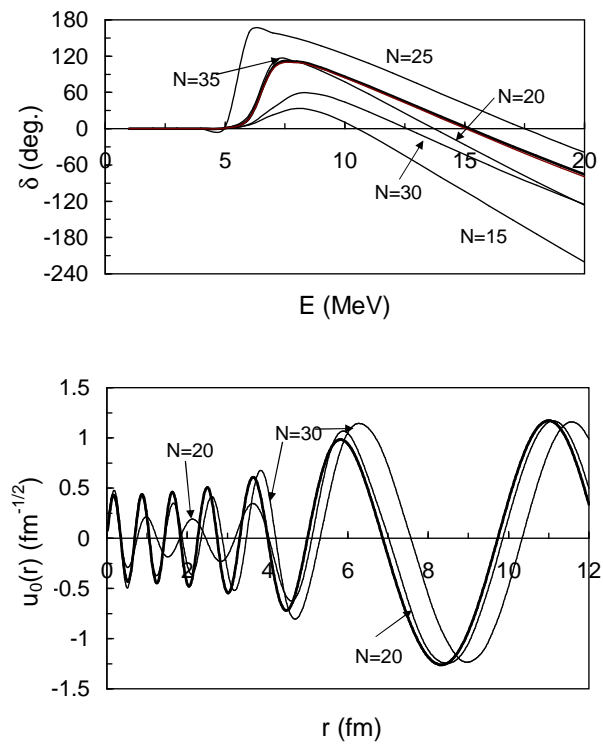} 
\caption{\label{fig_c12c13} Upper panel: $^{13}$C+$^{12}$C $l=0$ phase shifts with the Lagrange functions ($a=10$ fm).
Lower panel: corresponding wave functions at $E=10$ MeV. The exact wave function (superimposed on
the $N=40$ result) is shown as a thick line.}
\end{center}
\end{figure}

\subsection{Application to a non-local potential: $e^--$H scattering in the static-exchange approximation}
\label{subsec:nonlocal}
The electron + hydrogen-atom scattering has been studied with various methods \cite{BNS87}. In the simplest version,
referred to as the static-exchange approximation \cite{MA33,Jo60}, the problem is limited to a single-channel calculation,
but involves a non-local potential. The local term $U(r)$ reads
\beq
U(r)=-E_0\left(\frac{a_0}{r}+1\right)\exp(-2r/a_0).
\eeqn{eq_nl2}
We use here the traditional atomic units, involving the Bohr radius $a_0$ and the Hartree 
energy $E_0=e^2/a_0$ 

The non-local contribution  is defined as
\beq
W(r,r')=(-1)^{S+1}\frac{4r\,r'}{a_0^3}\exp\left(-\frac{r+r'}{a_0}\right)\left(E-E_0(1+\frac{a_0}{r_>})\right)
\eeqn{eq_nl4}
where $E$ is the total energy, $S$ is the total spin of the electrons ($S=0$ or 1), 
and $r_>=\max(r,r')$. The $1/r_>$ singularity is responsible for numerical 
problems \cite{ALL88}, and
deserves special attention in the literature.

This example is used to illustrate the applicability of the Lagrange functions for non-local potentials
(notice that the singlet phase shifts have been already determined in the $R$-matrix framework \cite{BNS87}).
Here the Gauss approximation associated with the Lagrange basis is expected to present convergence problems,
owing to the singularity mentioned above. This problem will be addressed by performing, in parallel
with the consistent Gauss approximation of the potential matrix elements [see \rref{eq3.4.11c} 
and \rref{eq3.4.11d}], a ``traditional" calculation using the Simpson method
(with typically 1000 integration points) for the numerical quadrature.

Results are shown in Tables \ref{tab_nl1} and \ref{tab_nl2} for $S=0$ and $S=1$, respectively,
as a function of the wave number $k$. For
the Lagrange-mesh method, we select a channel radius $a=14\ a_0$, and consider various basis sizes.
Comparing with the alternative method (Simpson integration) which is less sensitive to
singularities in the potential, we immediately see that a reasonable accuracy 
(typically 0.02 rad.) can be
achieved with small $N$ values at the Gauss approximation. However, when $N$ increases the convergence is rather slow. If we do not use the
Gauss approximation for the potential matrix elements, at least 5 digits are exact with $N=20$
and for a given $a$ value. The
sensitivity with respect to the channel radius is about $10^{-4}$ rad. when going from $a=14\ a_0$ to
$a=16\ a_0$. For comparison we give in Table \ref{tab_nl1} the results of Burke {\sl et al.} \cite{BNS87},
obtained for $S=0$, up to $k=0.7\ a_0^{-1}$. 

\begin{table}[ht]
\begin{center}
\caption{$R$-matrix phase shifts (in rad.) for $e^--$H scattering ($S=0$). The potential matrix
elements are computed, either with the associated Gauss approximation \rref{eq3.4.11c}
and \rref{eq3.4.11d}, or with the Simpson quadrature. 
The results are compared with
those of Burke {\sl et al.} \cite{BNS87}.
\label{tab_nl1}}
\begin{tabular}{c|ccccc|cc|c}
& \multicolumn{5}{c}{Gauss approximation, $a=14\ a_0$} & \multicolumn{2}{c}{$N=20$, Simpson} & \cite{BNS87}\\
\hline
  $k\ (a_0^{-1})$& $N=20$ & $N=40$ & $N=60$ & $N=80$ & $N=100$ & $a=14\ a_0$ & $a=16\ a_0$ & \\
  0.1 &   2.3790 &   2.3917 &   2.3940 &   2.3949 &   2.3952 &   2.3959 &   2.3958 &   2.3960\\
  0.2 &   1.8509 &   1.8654 &   1.8681 &   1.8691 &   1.8695 &   1.8703 &   1.8702 &   1.8704\\
  0.3 &   1.4902 &   1.5036 &   1.5061 &   1.5070 &   1.5074 &   1.5081 &   1.5081 &   1.5081\\
  0.4 &   1.2228 &   1.2352 &   1.2375 &   1.2384 &   1.2388 &   1.2394 &   1.2395 &   1.2391\\
  0.5 &   1.0155 &   1.0274 &   1.0296 &   1.0304 &   1.0308 &   1.0314 &   1.0315 &   1.0309\\
  0.6 &   0.8532 &   0.8650 &   0.8673 &   0.8681 &   0.8684 &   0.8691 &   0.8690 &   0.8685\\
  0.7 &   0.7281 &   0.7401 &   0.7424 &   0.7432 &   0.7435 &   0.7442 &   0.7441 &   0.7438\\
  0.8 &   0.6347 &   0.6470 &   0.6494 &   0.6502 &   0.6506 &   0.6513 &   0.6512 & \\
  0.9 &   0.5685 &   0.5813 &   0.5837 &   0.5846 &   0.5850 &   0.5857 &   0.5857 & \\
  1.0 &   0.5251 &   0.5384 &   0.5409 &   0.5417 &   0.5422 &   0.5429 &   0.5429 & \\
\end{tabular}
\end{center}
\end{table}

Similar conclusions can be drawn for the $S=1$ phase shifts, but the Gauss approximation 
is slightly more accurate because the antisymmetry of the spatial part of the wave function decreases
the effect of the $1/r_>$ singularity.
The $S=0$ and $S=1$ phase shifts are also in good agreement with those
of Apagyi {\sl et al.} \cite{ALL88} who use the Schwinger variational method \cite{Sc66}, much
more sensitive to numerical problems than the $R$-matrix method.

\begin{table}[ht]
\begin{center}
\caption{See caption to Table \ref{tab_nl1} for $S=1$.
\label{tab_nl2}}
\begin{tabular}{c|ccccc|cc}
 & \multicolumn{5}{c}{Gauss approximation, $a=14\ a_0$} & \multicolumn{2}{c}{$N=20$, Simpson} \\
\hline
 $k\ (a_0^{-1})$ & $N=20$ & $N=40$ & $N=60$ & $N=80$ & $N=100$ & $a=14\ a_0$ & $a=16\ a_0$  \\
  0.1 &   2.9080 &   2.9077 &   2.9077 &   2.9077 &   2.9077 &   2.9075 &   2.9076\\
  0.2 &   2.6799 &   2.6794 &   2.6793 &   2.6792 &   2.6792 &   2.6791 &   2.6792\\
  0.3 &   2.4622 &   2.4614 &   2.4612 &   2.4612 &   2.4611 &   2.4611 &   2.4611\\
  0.4 &   2.2588 &   2.2576 &   2.2574 &   2.2574 &   2.2573 &   2.2572 &   2.2573\\
  0.5 &   2.0721 &   2.0706 &   2.0703 &   2.0702 &   2.0702 &   2.0701 &   2.0701\\
  0.6 &   1.9032 &   1.9013 &   1.9009 &   1.9008 &   1.9007 &   1.9006 &   1.9006\\
  0.7 &   1.7521 &   1.7497 &   1.7492 &   1.7490 &   1.7490 &   1.7488 &   1.7488\\
  0.8 &   1.6180 &   1.6150 &   1.6145 &   1.6143 &   1.6142 &   1.6140 &   1.6141\\
  0.9 &   1.4998 &   1.4963 &   1.4957 &   1.4954 &   1.4953 &   1.4951 &   1.4952\\
  1.0 &   1.3959 &   1.3919 &   1.3911 &   1.3909 &   1.3908 &   1.3905 &   1.3905\\
\end{tabular}
\end{center}
\end{table}

\subsection{Discussion of resonances}
\label{subsec:resonances}
In this section we determine the resonance energies and widths of the $^{12}$C+p ($l=0$) and $\alpha+\alpha\ (l=4)$
systems, as an illustration of section~\ref{sec:resbs}. In Fig.~\ref{fig_reso}, we show the functions $R(E)$ and $1/S(E)$
for both systems. The resonances energies are calculated from the roots of equation (\ref{eq_res}), i.e.\ from the crossing points
of both curves. The numerical values, as well as the widths (\ref{eq_bw4}) are given in Table \ref{tab_reso}.

\begin{figure}[ht]
\begin{center}
\includegraphics[width=0.5\textwidth,clip]{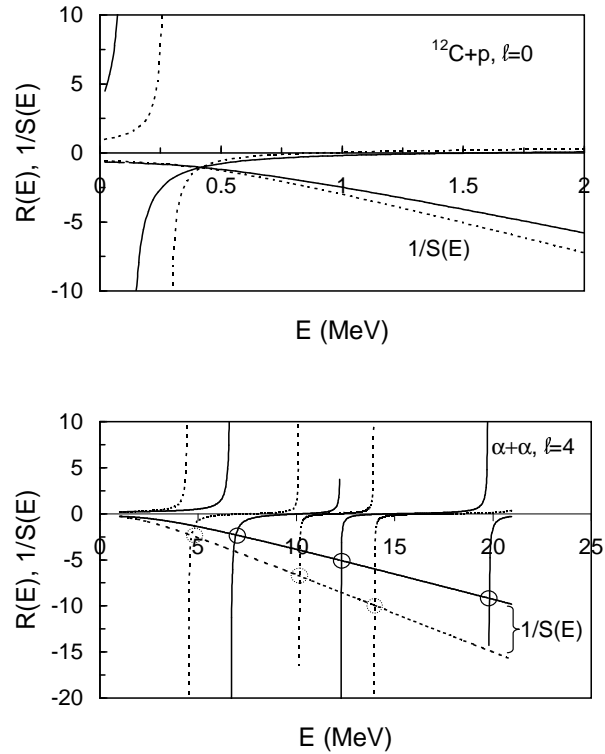} 
\caption{\label{fig_reso} $R$-matrix $R_l(E)$ and inverse of the shift function  $S_l(E)$ for the $^{12}$C+p ($l=0$) and $\alpha+\alpha\ (l=4)$ collisions. Solid lines correspond to $a=8$ fm and dotted lines to $a=10$ fm. The circles show the roots of
(\ref{eq_res}).}
\end{center}
\end{figure}

The $^{12}$C+p ($l=0$) state is a typical example of a narrow resonance. The phase shifts near 0.42 MeV are 
well described by the Breit-Wigner approximation. Although the $R$ matrices are rather different for $a=8$ fm
and $a=10$ fm (see Fig.~\ref{fig_reso}), the energy and width are weakly sensitive to the channel radius
(see Table \ref{tab_reso}). This
result is typical of narrow and isolated resonances. 

The situation is different for the $\alpha+\alpha\ (l=4)$ system, which is typical of a broad resonance
(around 12.5 MeV, see Fig.~\ref{fig_aa}).
The resonance properties are rather sensitive to the channel radius, whereas the phase shifts are
almost identical. Table \ref{tab_reso} shows that the second eigenvalue might be related
to the $4^+$ resonance. However the width is not significantly smaller than the energy difference
with neighbouring states. In that case, a single-level approximation is not well adapted to reproduce
the phase shifts in a wide energy range.

\begin{table}[ht]
\begin{center}
\caption{Resonance energies and widths (in MeV) of the $^{12}$C+p ($l=0$) and $\alpha+\alpha\ (l=4)$ systems.
The Lagrange basis is used with $N=15$.
\label{tab_reso}}
\begin{tabular}{c|cc|cc}
& \multicolumn{2}{c}{$a=8$ fm} & \multicolumn{2}{|c}{$a=10$ fm} \\
\hline
Eigenvalue & $E_R$ & $\Gamma$ & $E_R$ & $\Gamma$\\
\hline
$^{12}$C+p ($l=0$) & & & & \\
1 &  0.418 & 0.0375 & 0.417 &  0.0369\\
 &  &  &  & \\
 $\alpha+\alpha\ (l=4)$ & & & & \\
1 &   6.92   &   5.14   &   4.67   &   3.68  \\
2 &   12.3  &   2.47   &   10.2  &   2.74  \\
3 &   19.8  &   7.08   &   14.0  &   3.33  \\
\end{tabular}
\end{center}
\end{table}

\subsection{Application to bound states}
\label{subsec:bs}
The $R$ matrix is applied to the $^{13}$N ground state ($J^{\pi}=1/2^-,l=1$) described by a
$^{12}$C+p potential model. In order to simulate the experimental binding energy ($-1.94$
MeV), the depth of the Gaussian potential used in section~\ref{subsec:c12p} for $l=0$
has been modified to $-55.3$ MeV.

The calculation is performed with the Lagrange basis, and the results are presented in Table
\ref{tab_bs}. With $a=8$ fm, $N=7$ does not provide the exact energy. This situation
is of course improved by increasing $N$. For $N=15$, we have 5 exact digits,
and the matching parameter is close to 0. In each case, 4 iterations are sufficient in system \rref{eq_bs1}
to provide an energy stable by better than 0.1 keV. The ANC value $C_l$ [see (\ref{eq_witt})]
is also very stable. 

As in section\ 4.3 for the phase shifts, the choice $a=5$ fm is too small to make
the nuclear interaction negligible. Although the matching parameter is quite acceptable, the
convergence is slower and the final result does not converge to the exact energy. This means that a
small matching parameter is not sufficient to ensure that the calculation is accurate. Testing
the stability against the channel radius is a more severe test.

\begin{table}[ht]
\begin{center}
\caption{Energy of $^{13}$N ground state (in MeV) with a Lagrange basis and different $a$ and $N$ values.
The exact binding energy and ANC are $-1.942$ MeV and 2.063 fm$^{-1/2}$, respectively.
\label{tab_bs}}
\begin{tabular}{ccccc}
Iteration & $a=8,N=7$ & $a=8,N=10$ & $a=8,N=15$ & $a=5,N=10$\\
\hline
1 & $-2.012 $ & $-2.052 $ & $-2.053 $ & $-3.236 $\\
2 & $-1.897 $ & $-1.940 $ & $-1.941 $ & $-1.763 $\\
3 & $-1.899 $ & $-1.941 $ & $-1.942 $ & $-1.894 $\\
4 & $-1.899 $ & $-1.941 $ & $-1.942 $ & $-1.881 $\\
5 & $-1.899 $ & $-1.941 $ & $-1.942 $ & $-1.882 $\\
6 &  &  &  & $-1.882 $\\
$C_l$ (fm$^{-1/2}$) & $ 2.060 $ & $ 2.071 $ & $ 2.072 $ & $ 2.034 $\\
$\epsilon $ &  9.1  &  0.19 &  0.005 &  0.002\\
\end{tabular}
\end{center}
\end{table}

\subsection{Application to a multichannel problem: $\alpha$+d}
\label{subsec:multi}
Here we use the $\alpha$+deuteron potential of Dubovichenko \cite{Du98} to investigate the $R$-matrix formalism
in a multichannel problem. This nuclear potential $V_N(r)$ contains central $V_c(r)$ and tensor $V_t(r)$ forces, defined as
\beq
V_N(r)&=&V_c(r)+V_t(r)S_{12}\nonumber \\
V_c(r)&=&V_0\exp(-\alpha r^2) \nonumber \\
V_t(r)&=&V_1\exp(-\beta r^2) ,
\eeqn{eqad1}
where
\beq
S_{12}=\frac{6}{r^2}(\ve{S\cdot r})^2-2S^2
\eeqn{eqad2}
is the usual tensor operator. The Coulomb potential is the bare potential.
With $V_0=-91.979$ MeV, $V_1=-25.0$ MeV, $\alpha=0.2$ fm$^{-2}$,
$\beta=1.12$ fm$^{-2}$, this potential reproduces most of the $^6$Li ground-state properties, in particular the
binding energy, the quadrupole moment, and the $d$-wave admixture amplitude. The spin and parity of the deuteron
being $1^+$, non-natural parity states (i.e. $\pi=(-1)^{J+1}$) involve two $l$ values: $l=|J-1|$
and $l=J+1$. The system to be solved (\ref{eq_mul3}) involves the potentials  (we use the notation of section~\ref{sec:Rmulti})
\beq
V_{11}=V_c+V_C-2(J-1)V_t/(2J+1) \nonumber \\
V_{12}=V_{21}= 6\sqrt{J(J+1)} V_t/(2J+1)\nonumber \\
V_{22}=V_c+V_C-2(J+2)V_t/(2J+1)
\eeqn{eqad3}

As in previous sections the $R$-matrix phase shifts are computed in various conditions, and compared
to the ``exact" results obtained by the Numerov algorithm. The $R$-matrix calculations
are performed with Lagrange functions. To compare with experiment \cite{KKK91},
the collision matrix is diagonalized [see \rref{2.44c}] and parametrized as
\beq
\fl \ve{U}=
\left(
\begin{array}{cc}
\cos \epsilon & -\sin \epsilon\\
\sin \epsilon & \cos \epsilon
\end{array}
\right)
\left(
\begin{array}{cc}
\exp(2i\delta_1) & 0\\
0 & \exp(2i\delta_2)
\end{array}
\right)
\left(
\begin{array}{cc}
\cos \epsilon & \sin \epsilon\\
-\sin \epsilon & \cos \epsilon
\end{array}
\right),
\eeqn{eqad4}
where $\delta_1$ and $\delta_2$ are the (real) eigenphases, and $\epsilon$ is the
mixing angle (removing the tensor force provides $\epsilon=0$).
These values are presented in Fig.~\ref{fig_alphad} for $J^{\pi}=1^+$
and compared with the exact solutions (also shown in
Figs.~4 and 5 of \cite{Du98}). We choose $a=8$ fm and show the result for $N=7$. With this small number
of basis functions, slight deviations can be observed. As soon as $N>7$, the $R$-matrix phase
shifts cannot be distinguished from the exact values.

Other applications of the $R$-matrix theory to multichannel problems can be found in \cite{HSV98}.

\begin{figure}[ht]
\begin{center}
\includegraphics[width=0.4\textwidth,clip]{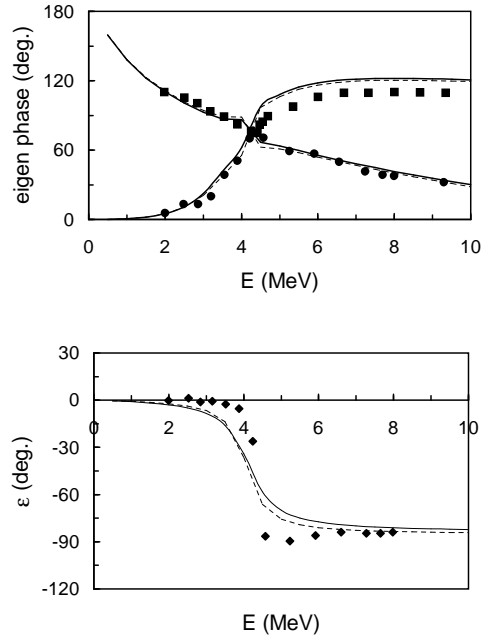} 
\caption{\label{fig_alphad} $\alpha$+d eigenphases (upper panel) and mixing angles (lower panel)
 for $J^{\pi}=1^+$. The solid lines represent
the exact calculation and the dotted lines correspond to the $R$-matrix calculation with $a=8$ fm,
$N=7$. The experimental data are taken from \cite{KKK91}.}
\end{center}
\end{figure}

\subsection{Application to propagation methods}
\label{subsec:propa2}
To illustrate the propagation of the $R$ matrix,
we present here a numerical example, with the $\alpha+\alpha$ system. 
We choose $a=80$ fm, which is of course unnecessarily large for this system, but typical of
long range potentials, where propagation methods should be used.
The goal of this example is just to provide a numerical illustration of the method. The $R$ matrix at 80 fm has been computed in 3 ways: (i) exactly with the Numerov algorithm, (ii) with 100 basis functions without propagation $(N_S=1)$, (iii) with 100 basis functions split in several intervals. The $R$-matrix basis functions are the Lagrange functions, but any other choice would provide similar conclusions. Matrix elements over an interval $(a_1,a_2)$ are given in Appendix C [see \rref{eq_lag5b}-\rref{eq_lag6}].

Table \ref{tab_propa} gives the $l=0$ $R$-matrix at $a=80$ fm for typical energies. Without propagation ($N_S=1$), we reproduce at least 4 significant digits. This is still true for $N_S=2$, but the computer time is reduced by more than a factor of two. Essentially the difference is that, in the former case we have to invert one $100\times 100$ matrix, whereas the latter calculation requires two inversions of $50\times 50$ matrices. The computer time is still reduced with $N_S=4$ (four inversions of $25\times 25$ matrices). The
precision is however reduced since the first interval $(0,20 \ {\rm fm})$ only contains 25 basis functions.
Increasing this number to 35 (and hence the full basis size to 110 functions) provides the exact
$R$-matrix (with at least 4 significant digits) with similar computer times.

\begin{table}[ht]
\begin{center}
\caption{$R$-matrix for the $\alpha+\alpha$ collision ($l=0$) and for different $N_S$ values ($N=100,a=80$ fm).
For each interval the number of basis functions is $100/N_S$.
\label{tab_propa}}
\begin{tabular}{lllll}
$E$ (MeV) &   exact & $N_S=1$ & $N_S=2$ & $N_S=4$\\
\hline
5 & $ 2.648\times 10^{-2}$ & $ 2.648\times 10^{-2}$ & $ 2.648\times 10^{-2}$ & $ 2.643\times 10^{-2}$\\
10 & $-3.729\times 10^{-1}$ & $-3.729\times 10^{-1}$ & $-3.730\times 10^{-1}$ & $-3.957\times 10^{-1}$\\
15 & $ 1.195\times 10^{-3}$ & $ 1.195\times 10^{-3}$ & $ 1.195\times 10^{-3}$ & $ 1.157\times 10^{-3}$\\
20 & $-1.246\times 10^{-2}$ & $-1.246\times 10^{-2}$ & $-1.246\times 10^{-2}$ & $-1.262\times 10^{-2}$\\
time (ms)   &  &  1.2 & 0.50 & 0.25\\
\end{tabular}
\end{center}
\end{table}

\subsection{Application to capture reactions: $\cpg$}
\label{subsec:capture}
The $\cpg$ reaction is the first reaction of the CNO cycle. It represents 
an ideal candidate for an $R$-matrix treatment since the $1/2^+$ ($l=0$) resonance at 
$E_{R}=0.42$ MeV determines the $S$-factor \rref{2.59} in a wide energy range. As in previous sections,
our goal here is not to find the best fit to the data (see \cite{BF80}). 
Rather, we want to illustrate the different contributions to the capture matrix elements (\ref{eq_cap1}), and to discuss various approximations.
Our procedure is as follows:
\begin{enumerate}
	\item The initial states are determined as in section~\ref{subsec:c12p}.
	\item As our goal is to illustrate the different contributions to the capture matrix elements, 
	the bound-state wave function (and the corresponding ANC) is computed exactly with the Numerov
	algorithm \cite{Ra72}. This avoids the optimization of the basis for the bound state. 
	\item The basis functions for the initial state
	are chosen as Lagrange functions ($N=15$), and we determine the internal and external contributions 
	from (\ref{eq_cap2}) and (\ref{eq_cap3}), respectively.
\end{enumerate}

The matrix elements are computed as in section~\ref{subsec:capth}, but the $R$-matrix expansion of the final
state is replaced by the exact wave function. Table \ref{tab_cap} gives, at typical energies, the exact values of the matrix elements (i.e.\ with scattering wave functions obtained from the Numerov method), as 
well as their values
in the $R$-matrix theory with $a=8$ fm and $a=10$ fm. 
Large values must be used to ensure that the nuclear interaction is negligible.
Several comments can be made:
\begin{enumerate}
	\item As expected the matrix element presents a maximum at $E=0.42$ MeV, as the initial potential has been fitted to provide a resonance at this energy.
	\item Since the external scattering wave function (\ref{eq_cap3}) involves the phase shift, the external contribution also presents a resonant behaviour.
	\item Each term in (\ref{eq_cap1}) depends on the channel radius. Their sum, however, should be insensitive to its choice, if the conditions of the calculation are properly defined. The relative difference
	is maximum near the resonance but is always less than 1\%.
\end{enumerate}

\begin{table}[ht]
\begin{center}
\caption{$\cpg$ matrix elements [see \rref{eq_cap2} and \rref{eq_cap3}]
with Lagrange functions ($N=15$).
\label{tab_cap}}
\begin{tabular}{c|ccc|ccc}
 & \multicolumn{3}{c}{exact} & \multicolumn{3}{c}{$R$-matrix} \\
\cline{2-4}\cline{5-7}
$E$ (MeV)& internal & external & total & internal & external & total\\
\hline
$a=8$ fm & & & & & & \\
    0.20 &   220.78 &    76.56 &   297.34 &   219.24 &    76.51 &   295.74\\
    0.40 &   2034.4 &   414.38 &   2448.8 &   2002.1 &   410.58 &   2412.7\\
    0.60 &   247.93 &    19.26 &   267.18 &   246.76 &    19.32 &   266.08\\
    0.80 &   116.02 &  $-  3.83$ &   112.19 &   115.35 &  $-  3.81$ &   111.54\\
    1.00 &    74.60 &  $-  9.69$ &    64.92 &    74.14 &  $-  9.68$ &    64.46\\
    & & & & & &\\
$a=10$ fm & & & & & &\\
    0.20 &   248.93 &    48.41 &   297.34 &   247.97 &    48.42 &   296.40\\
    0.40 &   2240.6 &   208.22 &   2448.8 &   2228.7 &   208.03 &   2436.7\\
    0.60 &   266.55 &     0.63 &   267.18 &   265.39 &     0.64 &   266.03\\
    0.80 &   121.69 &  $-  9.49$ &   112.19 &   121.13 &  $-  9.49$ &   111.64\\
    1.00 &    76.28 & $-  11.37$ &    64.92 &    75.92 & $-  11.36$ &    64.56\\
\end{tabular}
\end{center}
\end{table}

For the sake of completeness we show in Fig.~\ref{fig_c12pg} the $\cpg$ $S$-factor compared to the
available data sets \cite{Vo63,RA74}. As usual in the potential model, a spectroscopic factor ${\cal S}$
should be introduced for a more realistic comparison with the data. The spectroscopic factor scales
the total cross section. Values lower than unity mean that the final ground state is more 
complicated than a simple $^{12}$C+p structure. With ${\cal S}=1$ the theoretical $S$-factor
is larger than the data. A reasonable agreement can be obtained with ${\cal S}=0.45$, although
the high energy part is slightly overestimated by the model.

\begin{figure}[ht]
\begin{center}
\includegraphics[width=0.7\textwidth,clip]{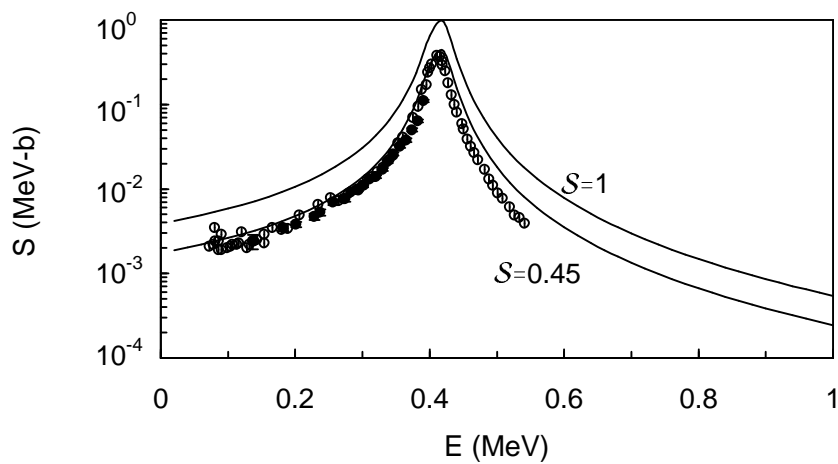} 
\caption{\label{fig_c12pg} $\cpg$ $S$-factors with different spectroscopic factors ${\cal S}$. The experimental data
are from \cite{Vo63} (open circles) and \cite{RA74} (closed circles).}
\end{center}
\end{figure}

\setcounter{equation}{0}
\renewcommand{\theequation}{5.\arabic{equation}}
\section{The phenomenological $R$ matrix}
\label{sec:phen}
\subsection{Introduction}
The goal of the phenomenological $R$-matrix method is to use a parametrization based on 
expression (\ref{33.7}) of the $R$ matrix or its multichannel generalization \rref{eq_mul11b} 
with a finite number of poles.  
The properties of these poles are adjusted to some data, 
in place of being derived from some Hamiltonian, as in the calculable approach. 
We present here various applications in nuclear physics. In particular, this technique
is very successful in nuclear astrophysics \cite{Cl83}, where the main issue is to fit 
cross-section data, and to extrapolate them down to stellar energies at which direct
measurements are in general impossible (see, for example, \cite{BK91}). 
Another recent application is the analysis of low-energy
scattering data in experiments involving radioactive beams. Those experiments usually
probe the nuclear structure at low level density, well adapted to the $R$-matrix formalism
(see, for example references \cite{RKV07,CAW06,ATC03} for $^{8}$B+p, $^{11}$C+p 
and $^{18}$Ne+p, respectively). 
The method is of course
not limited to elastic scattering, but can be extended to inelastic \cite{PAA08} as well as to
transfer \cite{DAC09} reactions.

One of the main drawbacks of the phenomenological $R$-matrix formalism is that, though the pole
energies and reduced widths are associated with physical properties, they cannot be directly
compared with experiment. 
In particular, the $R$-matrix parameters significantly depend on the channel radius. 
This is in contrast with the so-called ``observed" data, such as resonance energies and widths 
directly fitted to experimental data which are by definition independent of such a radius. 
However, these ``observed" values of the energy and width of a resonance depend on the 
assumptions made about the theoretical description of the resonance (see section \ref{sec:resbs} 
for a part of the possible definitions). 
Within the phenomenological $R$-matrix approach, the ``observed" quantities have a rather 
precise definition from which the $R$-matrix parameters can be determined. 
For an isolated resonance in some partial wave $l$, the observed energy $E_R$ and width $\Gamma_R$ 
are obtained in the single-channel approach by fitting the elastic cross section \rref{2.55} 
where the phase shift is parametrized by the generalized Breit-Wigner expression \rref{34.12},
\beq
\delta_l^{\rm BW} \approx \phi_l + \arctan \frac{\dem \Gamma_R}{E_R - E}.
\eeqn{eq5.0}
This expression may be simplified by neglecting $\phi_l$ or made more realistic by multiplying 
$\Gamma_R$ by $P_l(E)/P_l(E_R)$. 
Notice that, except for the simplest Breit-Wigner approximation, 
these expressions slightly depend on $a$ through $\phi_l$ or $P_l$. 
The ``observed" reduced width $\gamma_{obs}^2$ is then extracted from $\Gamma_R$ with the relation 
similar to  (\ref{34.14}), 
\beq
\Gamma_R = 2\gamma_{obs}^2 P_l(E_R).
\eeqn{eq5_4}
The corresponding ``formal" parameters, i.e.\ the pole location $E_{nl}$ and the reduced width 
$\gamma_{nl}^2$ can then be deduced for a given channel radius $a$ but their determination 
is not immediate because of the shift factor $S_l$ and its energy dependence. 
The problem is more complicated if several resonances or several channels 
must be taken into account. 

In the next sections, we present different methods to link the ``formal" $R$-matrix 
parameters with the ``observed" values. 
The simplest and most common  case deals with a single isolated resonance, described by one pole 
(section \ref{sec:formobs}). 
However some applications require to include several poles in the same partial wave or, 
in other words, to consider interference effects between these poles (section \ref{sec:multires}). 
Other applications require to take several channels into account (section \ref{sec:multicha}). 
As a first application, we consider the $^{12}$C+p elastic scattering from 0.3 MeV to 1.8 MeV, 
where accurate data exist for many years (section \ref{sec:CPelas}). 
Then more recent multichannel application to $^1$H($^{18}$Ne,p)$^{18}$Ne(g.s.) and 
$^1$H($^{18}$Ne,p')$^{18}$Ne*($2^+$, 1.887 MeV) 
will be shown as an example of radioactive-beam experiments (section \ref{sec:HNe}). 
Finally, the $R$-matrix method will be applied to radiative 
capture reactions, with $\cpg$ as a typical example (section \ref{sec:racar}).

\subsection{Single-pole approximation of elastic scattering\\}
\label{sec:formobs}
In section~\ref{sec:resbs}, we have presented a general procedure to define the resonance energy and width
from the $R$-matrix expression. 
Let us particularize (\ref{33.7}) to a single pole with energy $E_1$ and reduced width
$\gamma_1^2$ (we drop the index $l$ for the sake of clarity); the $R$ matrix is then given by
\beq
R_{l}(E)=\frac{\gamma_1^2}{E_1-E}.
\eeqn{eq5_1}
This approximation is frequently used at low energy, where single isolated resonances are
present. The $R$-matrix phase shift associated with (\ref{eq5_1}) is given by (\ref{34.11}). To
go further, let us use the Thomas approximation \cite{LT58}, which consists in a linearization
of the shift function $S_l(E)$. Near the pole energy $E_1$, it reads
\beq
S_l(E)\approx S_l(E_1)+(E-E_1)S'_l(E_1),
\eeqn{eq5_2}
where $S'_l$ is the derivative of the shift factor with respect to energy 
(which appears both in the wave number $k$ and in the Sommerfeld parameter $\eta$). 
The validity of this approximation is supported by Figs.~\ref{fig3.1}, \ref{fig3.2} and
\ref{fig3.3}, where it is clear that,
in a limited energy range, the linearization of the shift function is quite appropriate.

Comparing expression (\ref{34.11}) of the phase shift at the Thomas approximation 
with the Breit-Wigner expression \rref{eq5.0}, 
one derives the observed properties $(E_R,\gamma_{obs})$ from the formal parameters 
$(E_1,\gamma_1)$ of an isolated pole. 
The observed energy $E_R$ reads 
\beq
E_R=E_1-\gamma_1^2 \frac{S_l(E_1)}{1+\gamma_1^2 S'_l(E_1)},
\eeqn{eq5_3}
where the shift between both energies depends on $E_1$. 
This shift also depends on $a$. 
It is in general non-negligible, unless $\gamma_1^2$ is very small. 
The observed reduced width reads 
\beq
\gamma_{obs}^2= \frac{\gamma_1^2}{1+\gamma_1^2 S'_l(E_1)}.
\eeqn{eq5_5}
It also depends on $E_1$.

In practice, however, the reversed relationships are needed. 
Indeed, in many cases, observed values $(E_R,\gamma_{obs})$ are known from experiments 
(or from other theoretical works) and one wants to derive the corresponding 
$R$-matrix parameters $(E_1,\gamma_1)$. 
The inverses of (\ref{eq5_3}) and (\ref{eq5_5}) are obtained by linearizing 
the shift factor $S_l(E)$ around $E_R$ as 
\beq
&&\gamma_{1}^2= \frac{\gamma_{obs}^2}{1-\gamma_{obs}^2 S'_l(E_R)},
\eeqn{eq5_5b}
\beq
&&E_1=E_R+\gamma_{1}^2 S_l(E_R).
\eeqn{eq5_5c}
These results are well known \cite{LT58}. 

In the literature, the reduced width is frequently expressed in units of the Wigner limit 
\cite{TW52}
\beq
\gamma_W^2=\frac{3\hbar^2}{2\mu  a^2},
\eeqn{eq_wl}
which provides the dimensionless reduced widths
\beq
\theta^2_{obs}=\gamma^2_{obs}/\gamma_W^2,\nonumber \\
\theta^2_{1}=\gamma^2_{1}/\gamma_W^2.
\eeqn{eq_th}
Notice that Lane and Thomas \cite{LT58} define the dimensionless reduced width as 
$\theta^2=\gamma^2_{obs} \mu  a^2 /\hbar^2$ without an explicit reference to the Wigner limit.
In general, a value of $\theta^2$ not far from unity indicates the occurrence of a cluster structure, 
i.e.\ the colliding nuclei partly conserve their identity within the resonance. 
Conversely, a plausible guess on $\theta^2$ (possibly inspired by a model) can be used 
to estimate the width of a resonance. 
This concept of dimensionless reduced width, as defined by \rref{eq_th}, was first used by
various authors \cite{LS54,MP58,JPM60} in the interpretation of low-energy scattering data.

The difference between observed and formal parameters is illustrated by numerical applications 
in Table \ref{tab_phen1}
for the narrow $1/2^+$ resonance in $^{12}$C+p ($E_R=0.42$ MeV, $\Gamma_R=32$ keV) and for
the broad $1^-$ resonance in $^{12}$C+$\alpha$ ($E_R=2.42$ MeV, $\Gamma_R=420$ keV). 
The formal parameters ($E_1,\gamma_1)$ are calculated for various channel radii 
and are then used to determine the phase shifts shown in Fig.~\ref{fig_phen_delta}. 
As usual in applications of the phenomenological 
$R$-matrix method, the channel radii are somewhat smaller than in the ``calculable'' variant. 

\begin{table}[h]
\begin{center}
\caption{$R$-matrix parameters (\ref{eq5_5b}) and (\ref{eq5_5c}) for resonances in $^{12}$C+p and 
$^{12}$C+$\alpha$ (in MeV). The observed reduced widths are obtained from (\ref{eq5_4}).
\label{tab_phen1}}
\begin{tabular}{ccccc}
\multicolumn{5}{l}{$^{12}$C+p ($J^{\pi}=1/2^+, l=0, E_R=0.42$ MeV, $\Gamma_R=32$ keV)}\\
\hline
 & $a=4$ fm & $a=5$ fm & $a=6$ fm & $a=7$ fm\\
 \hline
$\gamma_{obs}^2$ &    1.089 &    0.592 &    0.353 &    0.227\\
$\theta_{obs}^2$ &   0.258 &    0.220 &    0.189 &    0.165\\
$\gamma_1^2$ &    3.083 &    1.157 &    0.569 &    0.323\\
$E_1$ & $-2.152$ & $-0.614$ & $-0.110$ &    0.113\\
 &  &  &  & \\
\multicolumn{5}{l}{$^{12}$C+$\alpha$ ($J^{\pi}=1^-, l=1, E_R=2.42$ MeV, $\Gamma_R=420$ keV)}\\
\hline
 & & $a=5$ fm & $a=6$ fm & $a=7$ fm\\
 \hline
$\gamma_{obs}^2$ &    &    0.574 &    0.277 &    0.165\\
$\theta_{obs}^2$ &   &    0.6920 &    0.481 &    0.389\\
$\gamma_1^2$ &    &    1.172 &    0.374 &    0.191\\
$E_1$ &     &    0.491 &    1.921 &    2.219\\
\end{tabular}
\end{center}
\end{table}

\begin{figure}[ht]
\begin{center}
\includegraphics[width=0.5\textwidth,clip]{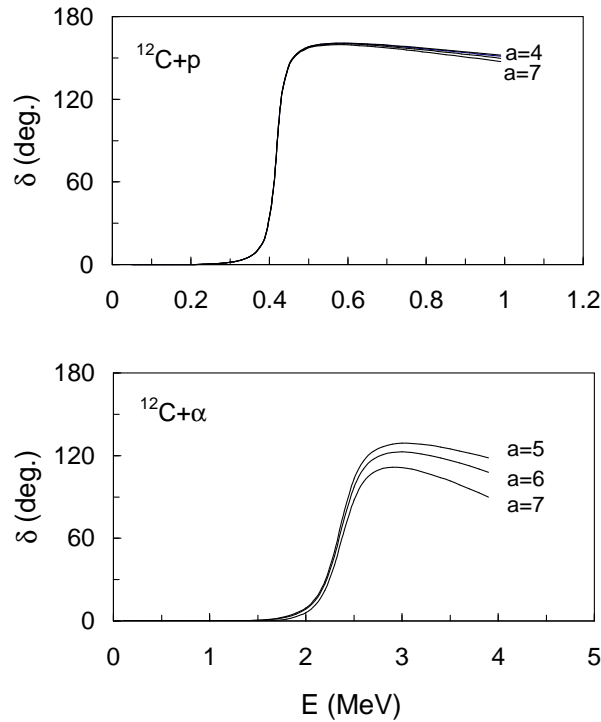} 
\caption{\label{fig_phen_delta} $^{12}$C+p $(l=0)$ and $^{12}$C+$\alpha$ $(l=1)$ phase shifts
in the single-pole $R$-matrix approximation computed with the parameters of Table
\ref{tab_phen1} for different channel radii.}
\end{center}
\end{figure}

From Table \ref{tab_phen1}, it is clear that the dimensionless reduced widths $\theta^2_{obs}$ 
are less dependent on the channel radius than $\gamma_{obs}^2$. For both systems, and in particular for
$^{12}$C+$\alpha$, the $\theta_{obs}^2$ values are rather large. These states can therefore be considered
as cluster states. As expected, the formal parameters (\ref{eq5_5b}) are strongly
dependent on the channel radius.  
However, the corresponding phase shifts 
in Fig.~\ref{fig_phen_delta} are very close to each other, in particular for the narrow
resonance in $^{12}$C+p. 
For the broad resonance in $^{12}$C+$\alpha$, the phase shift is
rather stable near the resonance energy, but more significant differences appear at higher energies.
In such a case, the validity of the Breit-Wigner approximation is more limited. In other
words, the single-pole approximation (\ref{eq5_1}) should be replaced by 
an $R$ matrix containing several terms.

\subsection{Multiresonance elastic scattering\\}
\label{sec:multires}
As mentioned above, the single-pole approximation is often valid at low energies. 
However, for nuclei with a high level density, several resonances may appear with the same angular momentum and parity. 
Then the interferences between different resonances and/or bound states in the same partial wave may be important.
Typical examples are $^{12}$C+$\alpha$ or $^{14}$N+p where several partial waves involve more than one state,
even in a limited energy range.

In this case, the link between formal and observed parameters is more complicated, and was addressed for
many years in a rather indirect way (see, for example, \cite{Ba72,ABB94}). More recently, an iterative
procedure was proposed to determine the formal $R$-matrix parameters from observed values \cite{AD00}. This
method is valid for single-channel systems only. A generalization to multichannel problems was then
developed by Brune~\cite{Br02}. 
The idea is to propose an alternative parametrization of the $R$ matrix, where the input parameters
are the observed data. It is based on the invariance of these values when the boundary parameters $B_c$
[see \rref{eq_mul8}] are changed. 
We briefly summarize Brune's method here, by assuming elastic scattering only. 
A more general presentation can be found in \cite{Br02}.

Let us assume $N$ resonances in a given partial wave, with observed energies $E_{Ri}$ and widths $\Gamma_{Ri}$. 
For each resonance, an observed reduced width is defined, according to (\ref{eq5_4}), as 
\beq
\gamma_{obs,i}^2=\Gamma_{Ri}/2P_l(E_{Ri}).
\eeqn{eq5_6}
In the notations of \cite{Br02}, \rref{eq5_5b} is written as
\beq
\tilde{\gamma}_i^2=\frac{\gamma_{obs,i}^2}{1-\gamma_{obs,i}^2 S'_l(E_{Ri})}.
\eeqn{eq5_8}

From these expressions, the formal pole energies $E_n$, used in the $N$-pole $R$-matrix
expansion, are obtained from 
\beq
\ve{N}\, \ve{b}_n=E_n\, \ve{M}\, \ve{b}_n
\eeqn{eq5_9}
where matrix elements $N_{ij}$ and $M_{ij}$ are given by
\beq
N_{ij}=
\left\{\begin{array}{ll}
E_{Ri}+ \tilde{\gamma}_i^2 S_l(E_{Ri})& {\rm for \ } i=j,\\
\tilde{\gamma}_i \tilde{\gamma}_j
\frac{E_{Ri}S_l(E_{Rj})-E_{Rj}S_l(E_{Ri})}{E_{Ri}-E_{Rj}} & {\rm for \ } i \neq j,
\end{array} \right .
\eeqn{eq5_10}
and
\beq
M_{ij}=
\left\{\begin{array}{ll}
1& {\rm for \ } i=j,\\
-\tilde{\gamma}_i \tilde{\gamma}_j
\frac{S_l(E_{Ri})-S_l(E_{Rj})}{E_{Ri}-E_{Rj}} & {\rm for \ } i \neq j.
\end{array} \right .
\eeqn{eq5_11}
The generalized eigenvalue problem (\ref{eq5_9}) provides the formal energies $E_n$. The formal reduced-width
amplitudes $\gamma_n$ are derived from the eigenvectors $\ve{b}_n$ as
\beq
\gamma_n=\sum_{j=1}^N b_{n,j} \tilde{\gamma}_j.
\eeqn{eq5_12}
When $N=1$, it is easy to see that \rref{eq5_5b} and \rref{eq5_5c} are recovered. 
This method provides an efficient way to derive the $R$-matrix parameters. 
It represents the starting point of an alternative parametrization 
of the $R$ matrix, proposed by Brune \cite{Br02}.

This formalism is illustrated in Table \ref{tab_phen2} and Fig.~\ref{fig_phen_delta2} with the $^{14}$N+p
system. In the $J^{\pi}=3/2^+$ partial wave $(l=0)$, we take three resonances into account. 
The observed values $E_{Ri}$ and
$\Gamma_{Ri}$  \cite{Aj91} are given in Table \ref{tab_phen2}. As for the
single-pole approximation, the formal parameters ($E_n,\gamma_n$) do depend on the channel radius.
Fig.~\ref{fig_phen_delta2} shows that the influence of $a$ on the corresponding phase shifts is 
weak near the narrow resonance at 0.987 MeV (this resonance presents a small $\theta^2$ value 
and has thus a complicated structure). 
As expected, it is more important in the vicinity of the
broader states at 2.2 and 3.2 MeV.

\begin{table}[h]
\begin{center}
\caption{$R$-matrix parameters (in MeV) with 3 poles for the $J^{\pi}=3/2^+,l=0$ 
partial wave in $^{14}$N+p.
\label{tab_phen2}}
\begin{tabular}{l|cccc}
observed values& & $a=5$ fm & $a=6$ fm & $a=7$ fm\\
\hline
$E_{R1}=0.987$ & $\gamma_{obs,1}^2$ &        0.0082 &   0.0054 &   0.0039\\  
$\Gamma_{R1}=0.00367$ &  $\theta^2_{obs,1}$ &      0.0031 &   0.0029 &   0.0028\\ 
 & $\gamma_1^2$ &      0.0097 &   0.0061 &   0.0042\\ 
                           & $E_1$ &       0.981 &   0.983 &   0.985\\  
$E_{R2}=2.187$ & $\gamma_{obs,2}^2$ &        0.114 &   0.086 &   0.069\\  
$\Gamma_{R2}=0.2$ & $\theta^2_{obs,2}$ &      0.043 &   0.047 &   0.050\\ 
 &$\gamma_2^2$ &           0.119 &   0.089 &   0.070\\ 
                            & $E_2$ &       2.153 &   2.165 &   2.172\\ 
$E_{R3}=3.209$ & $\gamma_{obs,3}^2$ &        0.053 &   0.042 &   0.034\\  
$\Gamma_{R3}=0.14$ & $\theta^2_{obs,3}$ &      0.020 &   0.023 &   0.025\\ 
 & $\gamma_3^2$&           0.051 &   0.041 &   0.034\\ 
                           & $E_3$&          3.199 &   3.202 &   3.204\\
\end{tabular}
\end{center}
\end{table}

\begin{figure}[ht]
\begin{center}
\includegraphics[width=0.7\textwidth,clip]{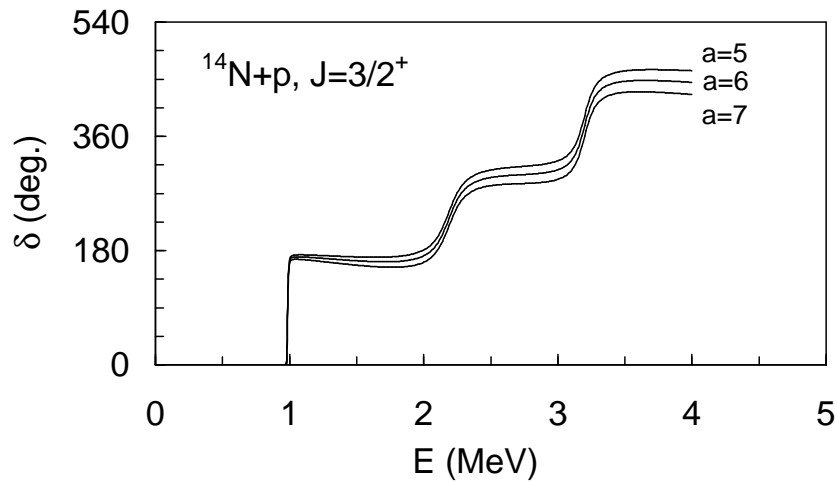} 
\caption{\label{fig_phen_delta2} $^{14}$N+p $R$-matrix phase shifts ($J^{\pi}=3/2^+$) computed with 
the parameters of Table \ref{tab_phen2}.}
\end{center}
\end{figure}

\subsection{Phenomenological parametrization of multichannel collisions\\}
\label{sec:multicha}
To deal with inelastic or transfer cross sections, the $R$-matrix formalism must be extended
to several channels. Rigorously this is even true in elastic scattering involving
particles with non-zero spins. When the spins are different from zero, several $(lI)$ values
[see (\ref{2.41})] are to be considered, each of them being characterized by a partial reduced width. 

Let us consider the multichannel $R$ matrix \rref{eq_mul11b}.
In the calculable variant, the number of
poles is equal to the product of the number of basis functions $N$ by the number of channels.
In the phenomenological approach, including all partial waves is often not realistic, since too many
parameters may be involved. At low energies, the choice of the relative angular momenta $l$ is guided
by the penetration factor, i.e.\ the minimum $l$ value for given $J$ and $\pi$ is often adopted. 
Such a selection mechanism does not exist for the channel spin $I$. 
In general, all possible values should be considered. 
The choice of the relevant partial waves is made according to the quality of the fit.
For these reasons, the multichannel approach in the phenomenological variant of the $R$ matrix is most 
often restricted to two channels. 

Here we limit the presentation to a single pole at energy $E_1$, the most frequently used approximation. 
In that case, the $R$ matrix (\ref{eq_mul11b}) is reduced to
\beq
R_{c c'}(E)=\frac{\gamma_{c}\gamma_{c'}}{E_1-E},
\eeqn{eq5_13}
where $\gamma_{c}$ is the reduced-width amplitude of the pole in channel $c$ (for simplicity we drop the
index $1$ of the reduced widths). These parameters are real. In this approximation, the property
\beq
R_{c c'}^2(E)=R_{c c}(E) R_{c' c'}(E),
\eeqn{eq5_13b}
provides the multichannel $\ve{Z}$ matrix \rref{eq_mul13} as
\beq
Z_{c c'}  =  (ak_{c'})^{-1/2}O_{c'}\biggl[\delta_{cc'}-L_{c'}\sqrt{R_{cc}R_{c'c'}}\biggr]. 
\eeqn{eq5_14a}
With the help of \rref{A.2}, we obtain the inverse as
\beq
\bigl(\ve{Z}^{-1}\bigr)_{c c'}  =  (ak_{c})^{1/2}O_{c}^{-1}\biggl[\delta_{cc'}+
\frac{L_{c'}\sqrt{R_{cc}R_{c'c'}}}{1-\sum_c R_{cc}L_c}\biggr], 
\eeqn{eq5_14b}
where the logarithmic derivatives $L_{c}$ \rref{34.3} and outgoing functions $O_{c}$ are defined for each channel. 
Using definition \rref{eq_mul12} for a real $R$ matrix, one obtains 
\beq
U_{c c'}=e^{i(\phi_c+\phi_{c'})} \left[ \delta_{cc'}+
2i\frac{\sqrt{P_c P_{c'} R_{cc}R_{c'c'}}}{1-\sum_c R_{cc}L_c}
\right], 
\eeqn{eq5_14}
where  $\phi_{c}$ is the hard-sphere phase shift \rref{32.11} in channel $c$. 

From the denominator in \rref{eq5_14}, the resonance energy and the reduced width 
are direct generalizations of (\ref{eq5_3}) and (\ref{eq5_5}),
\beq
&& E_R=E_1-\frac{\sum_{c}\gamma_{c}^2S_{c}(E_1)}
{1+\sum_{c}\gamma_{c}^2S'_{c}(E_1)},
\eoln{eq5_15}
&& \gamma_{obs,c}^2= \frac{\gamma_{c}^2}{1+\sum_{c}\gamma_{c}^2 S'_{c}(E_1)},
\eeqn{eq5_17a}
where index $l$ in the penetration and shift factors is replaced by the more general channel index $c$. 
With \rref{eq5_13}, the collision matrix is parametrized at the generalized Breit-Wigner approximation 
\cite{WT77} as 
\beq
U_{c c'}^{\rm BW}=e^{i(\phi_c+\phi_{c'})} 
\left[ \delta_{cc'}+\frac{i\sqrt{\Gamma_{c}(E)\Gamma_{c'}(E)}}{E_R-E-i\Gamma(E)/2} \right],
\eeqn{eq5_16}
where $\Gamma_{c}$ is the observed partial width in channel $c$ and $\Gamma=\sum_c \Gamma_c$. A simple
calculation gives
\beq
\Gamma_{c}(E)=2\gamma_{obs,c}^2P_{c}(E)
\eeqn{eq5_17}
and $\Gamma_{cR}=\Gamma_{c}(E_R)$.
Again, the $R$-matrix parameters $(E_1,\gamma_c)$ can be deduced from the observed
values as
\beq
\gamma_{c}^2= \frac{\gamma_{obs,c}^2}{1-\sum_{c}\gamma_{obs,c}^2 S'_{c}(E_R)},
\eeqn{eq5_18a}
\beq
E_1=E_R+\sum_{c}\gamma_{c}^2 S_{c}(E_R).
\eeqn{eq5_18}
These analytical formulas are direct extensions of (\ref{eq5_5b}) and (\ref{eq5_5c}) 
obtained for elastic scattering in the single-pole approximation.

\subsection{Application to the $^{12}$C+p elastic scattering}
\label{sec:CPelas}
Data on $^{12}$C+p exist for many years. In particular, elastic scattering cross sections
have been measured with a high precision \cite{MPS76}. This system
is well adapted to the phenomenological $R$-matrix approach, since the level density near
threshold is quite low. This example should be considered as a typical application of the method.
Similar fits have been done on the $^{16}$O+p $l=0$ elastic phase shifts \cite{Br96} and
on the $^{14}$O+p cross section \cite{BDL05}.

From \cite{Aj91}, three resonances are expected in the energy range covered by the data:
$1/2^+$ at 0.421 MeV, $3/2^-$ at 1.558 MeV and $5/2^+$ at 1.603 MeV. 
The pole corresponding to the bound state is neglected.
Data sets at three c.m. angles
($\theta=89.1^{\circ},118.7^{\circ},146.9^{\circ}$)  are available. The smallest and largest angles
are fitted simultaneously by using
the single-pole approximation (\ref{eq5_1}) for the resonant partial waves. For other
partial waves, the hard-sphere phase shift is used. This is consistent with the absence
of resonance ($R_l=0$), but plays a minor role in the cross sections. Replacing the hard-sphere
phase shifts by zero provides essentially the same fits.

The observed resonance properties are given in Table \ref{tab_phen3} for different channel radii.
Clearly the results are almost independent of $a$, as expected from physical arguments. The
fitted values are consistent with the literature \cite{MPS76}, and the corresponding
cross sections are shown in Fig.~\ref{fig_phen_c12p} for both scattering angles. The three
channel radii provide fits which are indistinguishable at the scale of the figure.
As expected \cite{MPS76}, the $R$-matrix parametrization reproduces the data very well, not
only in the vicinity of the resonances, but also between them, where the process is mostly
non-resonant. This technique is very successful in the analysis of recent data involving
radioactive beams (see, for example, \cite{RKV07,CAW06,ATC03}).

\begin{table}[ht]
\begin{center}
\caption{$R$-matrix parameters from a simultaneous fit of $^{12}$C+p scattering data \cite{MPS76} at $\theta=89.1^{\circ}$ and $146.9^{\circ}$. Resonance energies $E_R$ are expressed in MeV and widths $\Gamma_R$ in keV.
\label{tab_phen3}}
\begin{tabular}{ccccccc}
& \multicolumn{2}{c}{$J^{\pi}=1/2^+$} & \multicolumn{2}{c}{$J^{\pi}=3/2^-$} &\multicolumn{2}{c}{$J^{\pi}=5/2^+$} \\
\cline{2-7}
& $E_R$ & $\Gamma_R$ & $E_R$ & $\Gamma_R$ &$E_R$ & $\Gamma_R$ \\
\hline
$a=4$ fm &    0.427 &   33.8 &    1.560 &   51.4 &    1.603 &   48.1\\
$a=5$ fm &    0.427 &   32.9 &    1.559 &   51.4 &    1.604 &   48.1\\
$a=6$ fm &    0.427 &   30.9 &    1.558 &   51.3 &    1.606 &   47.8\\
Exp. \cite{MPS76} &    0.424 &   33 &    1.558 &   55 &    1.604 &   50\\
\end{tabular}
\end{center}
\end{table}

\begin{figure}[ht]
\begin{center}
\includegraphics[width=0.5\textwidth,clip]{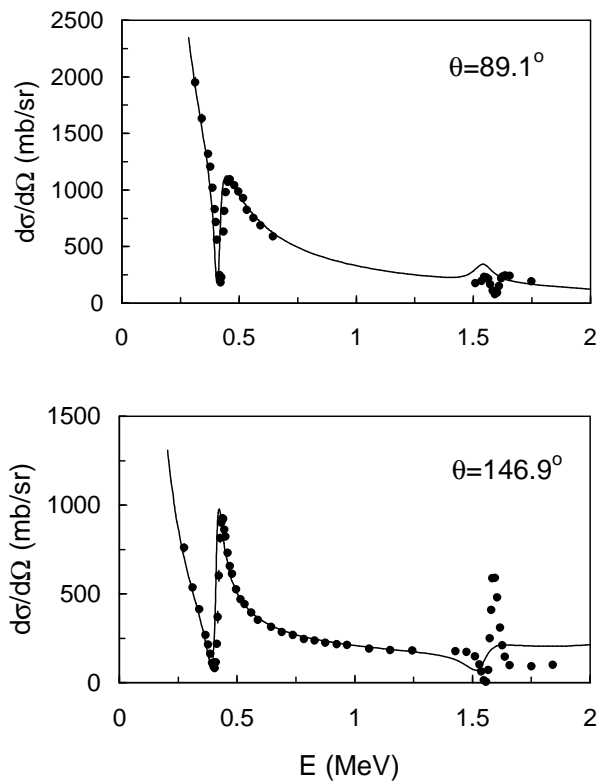} 
\caption{\label{fig_phen_c12p} $R$-matrix fits of $^{12}$C+p experimental excitation functions 
at two c.m. angles \cite{MPS76} with the parameters of Table \ref{tab_phen3}.}
\end{center}
\end{figure}

\subsection{Application to the $^{18}$Ne(p,p')$^{18}$Ne$(2^+)$ inelastic scattering}
\label{sec:HNe}
We present here an application of the phenomenological $R$-matrix method to inelastic scattering.
The $^{18}$Ne(p,p')$^{18}$Ne$(2^+,1.887{\rm \ MeV})$ cross section has been measured in parallel
with elastic cross sections \cite{PAA08}. These data were obtained at various angles and 
complemented a previous
data set, obtained at lower energies, and aimed at investigating elastic scattering only \cite{AAD03}.

Here our goal is not to repeat the analysis of \cite{PAA08}, where several angles were
simultaneously included, but where previous elastic data were not considered. Instead, we select a single angle
but cover a broader energy range by including data sets of \cite{AAD03,PAA08} in a global fit. 
Both experiments measured the elastic cross sections in different energy ranges, but also at
slightly different angles. We select
the elastic data sets of \cite{PAA08} at $\theta_{lab}=6.2^{\circ}$ and of \cite{AAD03}
at $\theta_{lab}=4.9^{\circ}$. As the angular dependence is weak, we combine these both data 
sets at a common angle, taken as the average ($\theta_{lab}=5.6^{\circ}$).
For the inelastic cross section, the experimental angle \cite{PAA08}
$\theta_{lab}=6.2^{\circ}$ is used.

\begin{figure}[htb]
\begin{center}
\includegraphics[width=0.5\textwidth,clip]{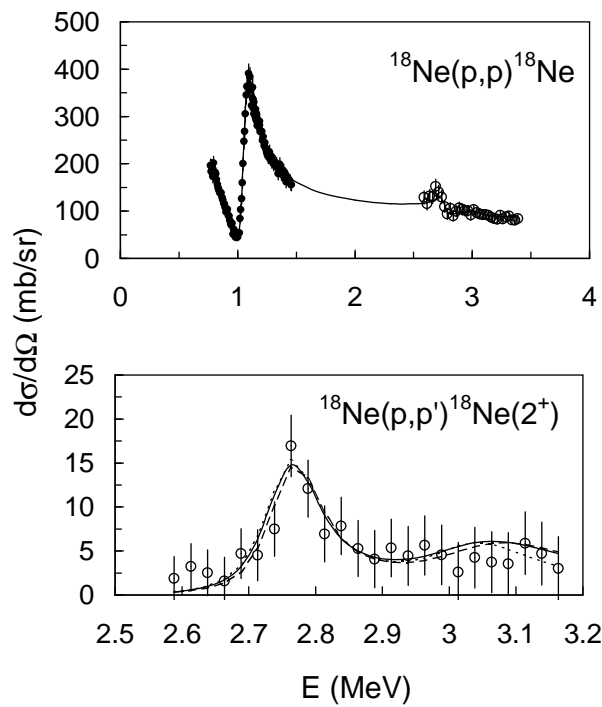} 
\caption{\label{fig_phen_ne18p} $R$-matrix fits of $^{18}$Ne(p,p)$^{18}$Ne elastic (upper panel) and
$^{18}$Ne(p,p')$^{18}$Ne$(2^+)$ inelastic (lower panel) cross sections with the parameters of
Table \ref{tab_phen4}. The data are from \cite{AAD03} (full circles) and \cite{PAA08} (open
circles). The fits are done with $a=4.5$ fm (dashed lines), $a=5.0$ fm (solid lines), and $a=5.5$ fm 
(dotted lines).}
\end{center}
\end{figure}

The experimental data and the corresponding $R$-matrix fits are presented in Fig.~\ref{fig_phen_ne18p}.
As suggested in \cite{PAA08}, the fits are performed by including three resonances ($J^{\pi}=1/2^+,
5/2^+,3/2^+$), which are characterized by their energy $E_R$ and their partial widths
$\Gamma_1$ and $\Gamma_2$ corresponding to the p+$^{18}$Ne$(0^+)$ and p+$^{18}$Ne$(2^+)$ channels,
respectively. The fitted parameters are given in Table \ref{tab_phen4} for different channel radii.
As expected we essentially reproduce the results of \cite{AAD03,PAA08}. The $1/2^+$ resonance
at 1.06 MeV is below the inelastic threshold ($\Gamma_2=0$), and corresponds to a single-particle state, with a
large reduced width. The higher-lying resonances $(5/2^+,3/2^+)$ correspond to $s$ states in the
p+$^{18}$Ne$(2^+)$ channel. They present a dominant width in that channel $(\Gamma_2 \gg \Gamma_1)$
and correspond to a significant fraction of the Wigner limit \rref{eq_wl}. 
These resonances are hardly
visible in the elastic data, and could not be properly analyzed without the inelastic cross
sections. 

Owing to the use of a radioactive beam, the error bars are rather large in the inelastic cross
sections, and the sensitivity of the
$R$-matrix parameters to the channel radius is slightly stronger than in $^{12}$C+p. In $R$-matrix
analyses, this sensitivity should be taken into account in the evaluation of the recommended
error bars.

\begin{table}[h]
\begin{center}
\caption{$R$-matrix parameters from a simultaneous fit of elastic and inelastic $^{18}$Ne+p cross sections
(see text). Resonance energies are given in MeV and widths in keV. The bracketed values represent
the dimensionless reduced widths.
\label{tab_phen4}}
\begin{tabular}{llccc}
$J^{\pi}$& & $a=4.5$ fm & $a=5$ fm & $a=5.5$ fm\\
\hline 
$1/2^+$ & $E_R$&   1.064 &    1.063 &    1.062\\
 & $\Gamma_1$ &   101 (0.26)&    98 (0.23)&    94 (0.20)\\
$5/2^+$ &  $E_R$&  2.773 &    2.768 &    2.766\\
 & $\Gamma_1$ &   7 (0.01) &    6 (0.01) &    7 (0.01)\\
 & $\Gamma_2$ &   85 (0.46) &    85 (0.40)&    80 (0.34)\\
$3/2^+$ &  $E_R$&  3.108 &    3.080 &    3.036\\
 &  $\Gamma_1$ &  11 (0.01) &    11 (0.01) &    9 (0.01)\\
 &  $\Gamma_2$ &  300 (0.47)&    379 (0.57) &    306 (0.48)\\
$\chi^2$&   &  0.43 &    0.44 &    0.54\\
\end{tabular}
\end{center}
\end{table}

\subsection{Radiative capture reactions}
\label{sec:racar}
\subsubsection{Extension of $R$-matrix formalism\\}
The general formalism of radiative-capture reactions has been given in section~\ref{subsec:rc}. In the $R$-matrix
theory, the matrix element $\tilde{U}$ \rref{2.62} is split in internal and external contributions
as~\cite{BK91}
\beq
\tilde{U}=\tilde{U}_{\rm int}+\tilde{U}_{\rm ext},
\eeqn{eq5_19}
where we have dropped all indices for the sake of clarity. 
The terms $\tilde{U}_{\rm int}$ and $\tilde{U}_{\rm ext}$ of the r.h.s. involve the internal
and external wave functions, respectively. We assume here single-channel calculations, but the
spins may be different from zero.

By using expansion \rref{34.7c} for the initial radial wave function $u^{\rm int}_{l_i}$ and $C_{l_i}=1$, 
the internal part of \rref{2.62} becomes 
\beq
\tilde{U}_{\rm int}=
\frac{e^{i(\delta_{l_i}^{J\pi}-\frac{\pi}{2})}}{|1-L_{l_i}R_{l_i}^{J\pi}|}\sum_n \epsilon_n
\frac{\sqrt{\Gamma_{\gamma,n}(E) \Gamma_n(E)}}{E_n-E},
\eeqn{eq5_21}
where we have introduced the formal gamma width of pole $n$ as
\beq
\Gamma_{\gamma,n}&=&
\frac{2J_f+1}{2J+1} 
\frac{8\pi(\lambda+1)k_{\gamma,n}^{2\lambda+1}}{ \lambda(2\lambda+1)!!^2} 
 \left|\la \Psi^{J_f \pi_f}||{\cal M}^{\sigma \lambda}||\Phi^{J\pi}_n \ra \right|^2
\eeqn{eq5_20}
with $k_{\gamma,n} = (E_n - E_f)/\hbar c$. 
The matrix element is calculated over the internal region only. 
Its definition involves a state $\Phi^{J\pi}_n$ corresponding to the pole $E_n$, 
whose radial part $r^{-1} \phi^{J\pi}_n (r)$ is defined like in the orthogonal basis \rref{33.6} 
but for partial wave $J \pi$. 
Matrix element \rref{eq5_21} depends on the particle width $\Gamma_n(E)$  \rref{eq_bw4} and on
the gamma width at collision energy $E$,
\beq
\Gamma_{\gamma,n}(E)=\left(\frac{E-E_f}{E_n-E_f}\right) ^{2\lambda+1} \Gamma_{\gamma,n}(E_n)
\eeqn{eq5_22}
(for simplicity, $\Gamma_{\gamma,n}(E_n)$ is often denoted as $\Gamma_{\gamma,n}$). 
In \rref{eq5_21},
$\epsilon_n=\pm 1$ is the product of the signs of the matrix element in \rref{eq5_20} and of
the reduced width amplitude. For a single-pole approximation, 
this sign does not play a role but determines interference effects in multi-pole calculations. 

The external contribution is determined as in \rref{eq_cap3}. A calculation similar to the one leading to 
\rref{2.58} provides
\beq
\fl \tilde{U}_{\rm ext}&=&
eZ_{\rm eff}^{({\rm E}\lambda)} C_{l_f}^{J_f \pi_f} \frac{1}{(2J+1)^{1/2}}
(-1)^{I_f - J_f+\lambda} 
\left(\frac{2(\lambda+1)k_{\gamma}^{2\lambda+1}}{\hbar v \lambda(2\lambda+1)!!^2} 
\right)^{1/2}\nonumber \\
\fl &&\times Z(l_i J l_f J_f,I_f \lambda) 
\int_a^{\infty} W_{l_f}(2\kappa_B r)r^{\lambda}(I_{l_i}(kr)-U^{J\pi}_{l_i} O_{l_i}(kr))dr,
\eeqn{eq5_23}
where $C_{l_f}^{J_f \pi_f}$ is the ANC of the final bound state and $U^{J\pi}_{l_i}$ the collision matrix at energy $E$. 
This contribution is often referred to as ``direct capture". In fact it is closely related to the
internal term through the collision matrix.  A resonant behaviour of the collision matrix 
affects the external term \rref{eq5_23} (see also Table \ref{tab_cap}).
As usual, in the calculable $R$-matrix theory, the total matrix element \rref{eq5_19} should not depend
on the channel radius $a$, although each individual term does depend on $a$. 
In addition, one easily shows that both terms present an identical phase factor. 
The calculation can therefore be reduced to real expressions.

In the calculable $R$-matrix, the gamma width \rref{eq5_20} is computed from basis functions. 
In the phenomenological variant, the constant $\Gamma_{\gamma,n}(E_n)$ appearing in \rref{eq5_22} becomes a parameter. 
The treatment of radiative-capture reactions therefore requires one additional parameter for each pole, the 
gamma width $\Gamma_{\gamma,n}(E_n)$  (and the associated interference sign), 
and a global parameter, the ANC $C_{l_f}^{J_f \pi_f}$ of the final bound state. 
The latter parameter is sometimes available independently. 
As for elastic widths, the fitted values of the gamma widths may depend on the channel radius. 
The importance of this dependence will vary with the amplitude of the external contribution. 

\subsubsection{Isolated resonance approximation\\}
Let us consider the single-pole approximation \rref{eq5_1}. 
Starting from expression \rref{eq5_21} of the internal matrix element $\tilde{U}_{\rm int}$, 
a simple calculation using definition \rref{34.7} of the phase shift provides 
near the resonance energy $E_R$ the approximation 
\beq
\tilde{U}_{\rm int}(E)\approx e^{i(\phi_{l_i}-\frac{\pi}{2})}
\frac{\sqrt{\Gamma_{\gamma,R}(E)\Gamma_R(E)}}
{E_R-E-i\Gamma_R(E)/2},
\eeqn{eq5_24}
where the observed particle width $\Gamma_R$ is defined from \rref{eq5_17} 
and where the observed gamma width $\Gamma_{\gamma,R}$ is given by
\beq
\Gamma_{\gamma,R}(E)=\frac{\Gamma_{\gamma,1}(E)}
{1+\gamma_1^2S'_{l_i}(E_1)}.
\eeqn{eq5_25}
The correction factor is thus identical for the particle and gamma widths [see \rref{eq5_5}]. For a resonant
process, the main part of the wave function is located at short distances, and the external term
\rref{eq5_23} can often be neglected to a good approximation. In that case, the capture cross section \rref{2.61}
takes the usual Breit-Wigner form
\beq
\sigma^{\sigma \lambda}_{J_f\pi_f,J\pi}(E)\approx
\frac{\pi}{k^2}\frac{2J+1}{(2I_1+1)(2I_2+1)}
\frac{\Gamma_{\gamma,R}(E)\Gamma_R(E)}
{(E_R-E)^2+(\Gamma_R(E)/2)^2}.
\eeqn{eq5_26}
As the electromagnetic interaction is weak, it is implicitly 
assumed that $\Gamma_{\gamma,R}(E)\ll \Gamma_R(E)$.
This formula is of course an approximation which assumes that (i) there is no background or other 
resonances interfering and (ii) the external contribution is negligible. Going beyond these two
approximations can be done by using the more general formulas \rref{eq5_21} and \rref{eq5_23}. Notice
that the relative roles of $\tilde{U}_{\rm int}$ and $\tilde{U}_{\rm ext}$ depend on energy. When $E$ becomes small,
the internal part of the initial state becomes smaller and smaller, and the importance of the
external contribution increases. The contribution of the external term also depends on the binding energy of the
final state. If $E_f$ is small, the asymptotic decrease of the bound-state wave function is slow and the
external matrix element \rref{eq5_23} may be important.

In the single-channel approximation, the internal and external components can be combined, which yields a 
slightly different definition for the gamma width \cite{BF80}. In \cite{HJL78}, the term
involving the collision matrix in \rref{eq5_23} is recast with the internal contribution. This
provides modified electromagnetic matrix elements. As long as the external capture is negligible,
all definitions of the gamma width are equivalent. 

\subsubsection{Application to $\cpg$ \\}
As discussed in section 4.12, the $\cpg$ $S$-factor at low energies is essentially determined by the properties of
the $1/2^+\ (l=0)$ resonance at $E_R=0.42$ MeV in $^{13}$N \cite{RA74,BF80}. As usual
in nuclear astrophysics the main issue is to extrapolate the available data down to
stellar energies (around 24 keV at the typical temperature $1.5\times 10^7$ K). We use the phenomenological
$R$-matrix approach with the single-pole approximation. In that case, four parameters
are to be considered: the energy, proton widtf $\Gamma_R$, gamma widths $\Gamma_{\gamma,R}(E_R)$ of the resonance, 
and the ANC $C_{l_f}$ of the $^{13}$N ground state ($J_f=1/2^-,l_f=1$). 
Since we are dealing with very low scattering energies and since $^{13}$N is not strongly bound 
($E_f=-1.94$ MeV), the ANC should be included. 
The fit is performed using \rref{2.61}, \rref{eq5_23}, \rref{eq5_24} at a channel radius of $a=5$ fm, and the resulting
observed parameters are given in Table \ref{tab_phen5}.

\begin{table}[ht]
\begin{center}
\caption{Observed $R$-matrix parameters at the resonance energy $E_R$ for the $\cpg$ reaction.
\label{tab_phen5}}
\begin{tabular}{cccc}
\hline
$E_R$ (MeV) & $\Gamma_R$ (keV)& $\Gamma_{\gamma,R}$ (eV) &$C_1$ (fm$^{-1/2}$)  \\
0.415& 31& 0.4 & 1.1 \\
\end{tabular}
\end{center}
\end{table}

Figure \ref{fig_phen_c12pg} presents the $R$-matrix $S$ factor compared with experimental data. 
With the single-pole approximation the fit is not perfect, in particular above the resonance,
where the $R$-matrix calculation slightly overestimates the data. This was already observed
in Fig.~\ref{fig_c12pg}, with the calculable approach. This problem has been addressed by Barker and Ferdous
\cite{BF80} who showed that an excellent fit of the data requires at least two poles in
the $R$-matrix expansion.
In
addition to the total $S$ factor, we also present the internal and external contributions independently.
This analysis is done for $a=5$ fm, but also for $a=6$ fm, where the same observed parameters
are used.
The internal part (dashed line) corresponds to the Breit-Wigner approximation \rref{eq5_26}; it is
almost insensitive to the choice of the channel radius. On the contrary, the external part (dotted lines)
does depend on $a$. Its influence near the resonance energy is weak but, as expected, it
increases at low energies. Neglecting the external term when extrapolating down to
stellar energies would provide a strong underestimation of the $S$ factor. 

\begin{figure}[ht]
\begin{center}
\includegraphics[width=0.5\textwidth,clip]{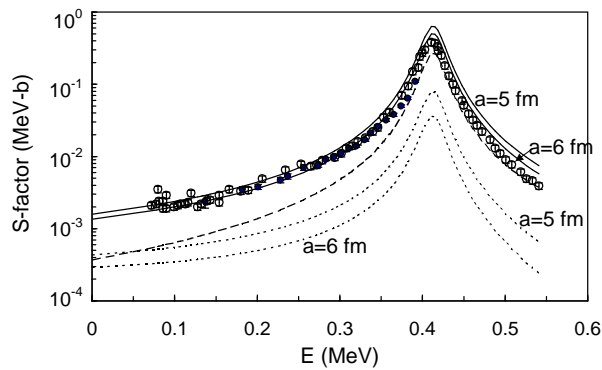} 
\caption{\label{fig_phen_c12pg} $\cpg$ $S$-factor computed with 
the parameters of Table \ref{tab_phen5} at $a=5$ and 6 fm. The dashed line corresponds to the internal
contribution ($a$=5 fm and $a=6$ fm are indistinguishable), and the dotted lines to the external term.}
\end{center}
\end{figure}

\setcounter{equation}{0}
\renewcommand{\theequation}{6.\arabic{equation}}
\section{Recent applications of the $R$-matrix method}
\label{sec:recent}
\subsection{Introduction}
In section~4 we gave simple examples of the calculable $R$-matrix method, in order
to illustrate the theoretical framework with applications which can 
be easily reproduced by the reader. However, in most cases, alternative methods, sometimes  
simpler, are available.

In this section, we present more ambitious applications of the $R$-matrix theory in nuclear physics. 
The first deals 
with microscopic cluster models \cite{WT77}, where the relative motion between the 
colliding particles is not given by a potential, but by a nucleon-nucleon interaction. 
The $R$-matrix theory is also very efficient to solve coupled-channel problems \cite{Th88}. 
In various models, the Schr\"odinger equation is reduced to a system of coupled
differential equations. This can be solved, for example, with the generalized Numerov 
algorithm, but this method looses stability when the size of the system increases \cite{NT99}. 
In that case, the $R$-matrix theory provides an efficient alternative, in particular when it 
is associated with the Lagrange-mesh method \cite{Ba06}. Two applications concerning the three-body
continuum are presented: (i) the 
Continuum Discretized Coupled Channel (CDCC) method \cite{Ra74b,AIK87}, and (ii) the
three-body hyperspherical formalism \cite{MF53,TDE00,DTB06}.
Recent applications of the $R$-matrix theory in atomic physics can be found, for example, 
in \cite{ZB04,NPT08,TGR03,NB08}.

\subsection{Microscopic cluster models}
\label{sec:MCM}
\subsubsection{General presentation\\}
In nuclear physics, a microscopic theory is based 
on a description of all nucleons taking full account of antisymmetrization and derived 
from interactions between nucleons.  
The Hamiltonian \rref{2.38} reads
\beq
H=\sum_{i=1}^A \frac{\ve{p}_i^2}{2m_N}+\sum_{i>j=1}^A V_{ij}+\sum_{i>j>k=1}^A V_{ijk}-T_{cm},
\eeqn{eq6.1}
where $A$ is the nucleon number, ${\ve p}_i$ is the momentum of nucleon $i$ 
and $V_{ij}$ and $V_{ijk}$ are two- and three-nucleon interactions. 
In \rref{eq6.1}, subtracting the center-of-mass energy $T_{cm}$ guarantees that the
wave function is free of spurious c.m.\ components \cite{GS57}. 
The two-body interactions $V_{ij}$ involve a nuclear term with spin-orbit, tensor and other 
components, and the Coulomb interaction. 
The three-body interactions $V_{ijk}$ are necessary to explain the binding energies 
of the $^3$H, $^3$He and $^4$He nuclei. 
Realistic interactions are derived from field theories and partly fitted on properties 
of the nucleon-nucleon system. 
For small nucleon numbers (i.e.\ $A \leq 4$), different techniques exist to find numerically exact 
solutions of the Schr\"odinger equation \cite{Fo08}. 
Few-body calculations can be performed by using various realistic interactions 
and compare well with experiment. 
For heavier systems, {\sl ab initio} calculations \cite{Pi02,NQS09} become available. 
Different variants are being developed, but are currently limited to $A \approx 12$, and 
their application to continuum states is quite difficult \cite{NPW07,QN08}. 

The microscopic cluster approach is based on an assumed cluster structure, i.e.\ on the existence of 
correlated subsystems in the fully antisymmetric wave function of the $A$-nucleon system \cite{WT77}. 
The microscopic cluster model provides a unified framework 
for the description of nuclear spectroscopy and of nuclear reactions.
The cluster assumption allows the 
application of a microscopic theory to heavier systems (typically up to $A \sim 20-24$), 
but requires the use of effective nucleon-nucleon two-body interactions (see, for example, 
\cite{Vo65,TLT77}) which are adapted to the cluster approximation. 
We use here the Minnesota effective interaction \cite{TLT77} which does not include tensor forces,
but simulates their contribution in the binding energy of the deuteron by the central term.
Three-body effects are in general also approximately simulated in these effective interactions. 

\subsubsection{The Resonating Group Method\\}
The Resonating Group Method (RGM) assumes that the wave function can be expressed in terms of 
normed cluster wave functions $\phi_i$ involving $A_i$ among the $A$ nucleons and depending 
on translation-invariant internal coordinates, 
and of an unknown wave function $g_l(r)$ for their relative motion. 
Using the isospin formalism, the $A$-nucleon approximate wave function reads 
\beq
\Psi_{lm}=\frac{A!}{A_1!A_2!}{\mathcal A} \phi_{1} \phi_{2} g_l(r) Y_l^m(\Omega),
\eeqn{eq6.2}
where $\ve{r} = (r,\Omega)$ is the relative coordinate between the c.m.\ of the clusters, 
and ${\mathcal A}$ is the $A$-nucleon antisymmetrization projector
\beq
{\mathcal A}=\frac{1}{A!}\sum_{p=1}^{A!}(-1)^{p}P_p,
\eeqn{eq6.2b}
where the operator $P_p$ performs the permutation $p$ among $A$ particles.
For simplicity, we assume a two-cluster structure, 
with zero-spin clusters [see \rref{2.42} for a multichannel definition]. 
More general presentations can be found in \cite{WT77,BHL77,BD83,Ta81,La94}. 
The antisymmetric internal wave functions $\phi_1$ and $\phi_2$ are normalized to unity and 
defined in the harmonic-oscillator shell model, with a common oscillator parameter $b$ \cite{WT77}. 
First applications were limited to $s$-shell clusters \cite{BH74}, 
but progressively, heavier systems were considered \cite{MKF75,WT87}.

The unknown relative function $g_l(r)$ is obtained by solving the Schr\"odinger equation
\beq
H\Psi_{lm}=E\Psi_{lm},
\eeqn{eq6.3}
with Hamiltonian (\ref{eq6.1}) and approximate wave function (\ref{eq6.2}). 
By projecting the Schr\"odinger equation over 
$\la \phi_1 \phi_2 Y_l^m(\Omega)|$, this technique 
provides an integro-differential equation involving local and non-local potentials \cite{Ta81} 
\beq
(T_r + V_D(r) - E) g_l(r) + \int_0^\infty \biggl( {\mathcal K}_{Hl}(r,r')  
-E {\mathcal K}_{Nl}(r,r') \biggr) g_l(r') dr'= 0,
\eeqn{eq6.3b}
where $T_{r}$ is the relative kinetic energy operator \rref{2.46}, 
$V_D(r)$ is the direct potential \rref{2.46a}, and 
${\mathcal K}_{Hl}(r,r')$ and ${\mathcal K}_{Nl}(r,r')$ are the Hamiltonian and overlap 
exchange kernels, respectively. 

At large distances, the Hamiltonian (\ref{eq6.1}) can be written as 
\beq
H \arrow{r}{\infty} H_1 + H_2+T_{r}+V_C(r),
\eeqn{eq6.4}
where $H_1$ and $H_2$ are the internal Hamiltonians of the clusters, similar to \rref{eq6.1}, 
and $V_C(r)$ is the Coulomb potential between charges $Z_1e$ and $Z_2e$. 
The internal energies $E_i$ of clusters $i=1,2$ are defined by the variational expressions
\beq
E_i=\la \phi_i| H_i | \phi_i\ra .
\eeqn{eq6.4b}
In parallel, the wave function (\ref{eq6.2}) tends to
\beq
\Psi_{lm}\arrow{r}{\infty} \phi_{1} \phi_{2} g_l(r) Y_l^m(\Omega),
\eeqn{eq6.5}
since the antisymmetrization operator ${\mathcal A}$ acts at short distances only. 
For large $r$ values, the radial wave function $r g_l(r)$ is given by the Coulomb equation 
(\ref{2.4}). 
Although it is also suitable for bound states,
the ansatz (\ref{eq6.2}) of the total wave function is therefore well 
adapted to the treatment of scattering states. 

The main problem of the RGM is not to solve (\ref{eq6.3b}). This can be done by 
standard techniques, or by the $R$-matrix method \cite{HRB02}. 
Recent calculations, using realistic nucleon-nucleon interactions,
have been performed on the $^3$He+p and $^3$H+n scattering \cite{PHH01} and 
on the $\alpha+$nucleon scattering \cite{QN08}. In these references, the RGM equation
\rref{eq6.3b} is extended to a multichannel generalization. In \cite{QN08}, it
is solved on a Lagrange mesh.
The drawback of the RGM is that 
the calculation of the overlap and Hamiltonian kernels is not systematic. 
The reason is that the relative coordinate $r$ and the cluster internal coordinates 
in \rref{eq6.2} are modified in different ways by the terms of the antisymmetrizer ${\mathcal A}$. 
This method requires heavy analytical calculations \cite{WT77,Ta81} 
(see also \cite{HRB02,TMO07} for examples of kernels). 
This problem is simplified by using the Generator Coordinate Method (GCM), described in the next subsection.

\subsubsection{The Generator Coordinate Method\\}
In the GCM, the relative wave function $g_l(r)$ is expanded over projected Gaussian functions \cite{Ho77} as
\beq
g_l(r)=\int f_l(R) \Gamma_l(r,R) dR,
\eeqn{eq6.6}
where $R$ is the generator coordinate and $\Gamma_l(r,R)$ is defined as 
\beq
\Gamma_l(r,R)=\left( \frac{\mu_0}{\pi b^2} \right) ^{3/4} \exp\left( -\mu_0\frac{r^2+R^2}{2b^2} \right)
i_l \left( \frac{\mu_0 r R}{b^2}\right) .
\eeqn{eq6.7}
In this equation, $\mu_0$ is the reduced mass in units of the nucleon mass, and
$i_l(x)=\sqrt{\pi/2x} I_{l+1/2}(x)$, $I_{l+1/2}(x)$ being the modified spherical Bessel
function of the first kind \cite{AS72}. The calculation of $g_l(r)$ is therefore 
replaced by the calculation of the generator function $f_l(R)$.

Inserting (\ref{eq6.6}) in the RGM definition (\ref{eq6.2}) provides
\beq
\Psi_{lm}=\int f_l(R) \Phi_{lm}(R) dR,
\eeqn{eq6.8}
where
\beq
\Phi_{lm}(R)=\frac{A!}{A_1! A_2!}{\mathcal A} \phi_{1} \phi_{2} \Gamma_l(r,R) Y_l^m(\Omega).
\eeqn{eq6.9}
After multiplication by an appropriate factor depending on the c.m.\ coordinate of the $A$ nucleons, 
the basis function $\Phi_{lm}(R)$ can be expressed as a projected Slater determinant 
provided that the oscillator parameters of the clusters are identical \cite{Ho77}. 
The Slater determinant is defined from $A_1$ and $A_2$ shell-model orbitals 
centered at $-A_2R/A$ and $A_1R/A$ for the first and second cluster, respectively. 
This property is well adapted to systematic and numerical calculations since it involves 
matrix elements of single-particle orbitals only. 
Well-known techniques exist to determine matrix elements between Slater determinants from 
single-particle matrix elements \cite{Br66,Lo55b}. 
They allow a rather simple extension of the cluster model to heavy systems \cite{BS79} and 
to multichannel problems \cite{DD07}. The projection over angular momentum $l$ can be 
performed numerically \cite{Br66}. 

In practice, the integral in (\ref{eq6.6}) 
is replaced by a finite sum over a set of values $R_n$ of the generator coordinate. 
This means that, at large distances $r$, 
the radial wave function $g_l(r)$ presents a Gaussian behaviour, 
not consistent with the physical asymptotic behaviour. 
This problem can be addressed by using the microscopic $R$-matrix method \cite{BHL77}. 
The wave function is approximated in the internal region by a discretized version of \rref{eq6.8} as 
\beq
\Psi_{lm}^{\rm int}&=&\sum_{n=1}^N f_l(R_n) \Phi_{lm}(R_n). 
\eeqn{eq6.10}
In the external region, it is approximated by the asymptotic expression \rref{eq6.5} as 
\beq
\Psi_{lm}^{\rm ext}&=&\phi_{1} \phi_{2} g_l^{\rm ext}(r) Y_l^m(\Omega)
\eeqn{eq6.10b}
where the external radial function $r g_l^{\rm ext}(r)$ is a linear combination of 
Coulomb functions, as in \rref{32.1a}.

The application of the $R$-matrix method to the GCM is straightforward and follows the 
method detailed in section~\ref{sec:form}. The generalized matrix $\ve{C}$ defined in (\ref{32.6}) 
involves matrix elements of the Hamiltonian 
in the internal region only. This is achieved by subtracting the 
external contributions~\cite{BHL77}. 
By definition of the channel radius $a$, antisymmetrization effects and the nuclear interaction are
negligible in the external region.
The relevant matrix elements are therefore given by 
\beq
\fl \langle \Phi_{l}(R_n) | \Phi_{l}(R_{n'}) \rangle _{\rm int}=
\langle \Phi_{l}(R_n) | \Phi_{l}(R_{n'}) \rangle -
\int_a^{\infty} \Gamma_l(r,R_n) \Gamma_l(r,R_{n'})r^2 dr, \nonumber \\
\fl \langle \Phi_{l}(R_n) |H| \Phi_{l}(R_{n'}) \rangle _{\rm int}=
\langle \Phi_{l}(R_n) |H| \Phi_{l}(R_{n'}) \rangle \nonumber \\
 -\int_a^{\infty} \Gamma_l(r,R_n) (T_{r}+V_C(r)+E_1+E_2)
\Gamma_l(r,R_{n'})r^2 dr, 
\eeqn{eq6.11}
where the first terms in the r.h.s. are matrix elements over the whole space, involving 
Slater determinants. 
The second terms represent the external contributions of the basis functions (\ref{eq6.7}) 
and can easily be computed numerically. Then the $R$-matrix and the associated collision matrix 
are obtained as in section 3.2. Similarly, the collision matrix should not depend on the choice 
of the channel radius $a$, provided it is large enough to make the nuclear interaction and 
the antisymmetrization effects negligible in the external region. 
A generalization to multichannel systems can be found in \cite{BHL77,BD83}.

\subsubsection{Applications: $\alpha + \alpha$ and $^{12}$C+p\\}
We present three typical applications of the GCM associated with the microscopic 
$R$-matrix method.
The first deals with the well known $\alpha + \alpha$ phase shifts (see section~\ref{subsec:aa}). 
Then we compute the $^{12}$C+p elastic
cross section, as well as the $\cpg$ $S$ factor at low energies. In all cases we use the 
Minnesota (MN) effective interaction \cite{TLT77} as central nucleon-nucleon force. 
The Minnesota potential provides the correct binding energy of the deuteron (without tensor
force) and reproduces fairly well some properties of nucleon-nucleon scattering. It
involves the admixture parameter $u$
whose standard value is $u=1$, but which can be slightly modified to fit important physical
quantities, such as the energy of a bound state or of a resonance. 
For the $^{12}$C+p system, a zero-range
spin-orbit force (with amplitude $S_0$) is added \cite{BP81}.

The $\alpha + \alpha$ system is described by cluster wave functions $\phi_i$ defined from 
four $0s$ harmonic-oscillator orbitals with an oscillator parameter $b=1.36$ fm 
for the two protons and the two neutrons. With the MN force, the binding energy of the $\alpha$
particle is 24.28 MeV, which is smaller than the experimental value 28.30 MeV. This
difference does not play an important role, since all theoretical energies are defined with
respect to the $\alpha + \alpha$ threshold.
The admixture parameter $u$ is taken
as $u=0.94687$, as recommended in \cite{TMO07}. This value provides an excellent description
of the $\alpha + \alpha$ phase shifts in a wide energy range.
We use $N=10$ basis functions in (\ref{eq6.10}) with $R_n$ values ranging from 0.8 fm to 8 fm by steps
of 0.8 fm. In Table \ref{tab_gcm_aa}, we give the
$0^+,2^+,4^+$ phase shifts at typical energies, and for various conditions of the calculation. 
The channel radius $a$ is taken as $a=6.4$ fm or $a=7.2$ fm, and $N=9$ or 10 are considered. 
In all cases, the phase shifts are very stable when the conditions are changed. 
They can be obtained with an accuracy better than $0.1^{\circ}$. Fig.~\ref{fig_gcm1} 
shows the phase shifts as a function of energy. It is known that a cluster
model is well adapted to the $\alpha + \alpha$ system since the $\alpha$ particle has a large binding energy
and since the first open threshold ($^7$Li+p) is near 17 MeV. The GCM phase shifts are therefore in very good
agreement with experiment. 
The same results can be obtained from the RGM equation \rref{eq6.3b} \cite{TMO07}. 

\begin{table}[h]
\begin{center}
\caption{Microscopic $\alpha + \alpha$ phase shifts (in degrees) for different conditions of calculation.
\label{tab_gcm_aa}}
\begin{tabular}{l|ccccc}
& & \multicolumn{2}{c}{$N=9$} & \multicolumn{2}{c}{$N=10$}\\
\cline{3-4}\cline{5-6}
 & $E$ (MeV) & $a=6.4$ fm & $a=7.2$ fm & $a=6.4$ fm & $a=7.2$ fm\\
\hline
$l=0$ & 1  & 146.00 &   145.93  & 146.00 & 146.00           \\
      & 5  &   47.48 &   47.42  &   47.49 &   47.48       \\
      & 10  &  $-5.67$ & $-5.79$ & $-5.67$ & $-5.67$       \\
      & 15 & $-38.47$ & $-38.52$ & $-38.46$ & $-38.46$ \\
 &  &  &  &  & \\
$l=2$ & 1  &     0.56 &   0.53 &   0.56 &   0.56     \\
      & 5  &    112.00 & 111.90& 112.00 & 112.00    \\
      & 10 &   94.97 &   94.94  &   94.98 &   94.97 \\
      & 15 &   77.50 &   77.38  &   77.50 &   77.50\\
 &  &  &  &  & \\
$l=4$ & 1  &   0.00 & 0.00      &   0.00 &   0.00            \\
      & 5 &   1.00 &   0.84     &   1.01&   1.00              \\
      & 10&   26.66 &   26.64    &   26.66 &   26.65      \\
      & 15 &   118.40 &   118.40 &   118.40 &   118.40    \\
\end{tabular}
\end{center}
\end{table}

\begin{figure}[htb]
\begin{center}
\includegraphics[width=0.5\textwidth,clip]{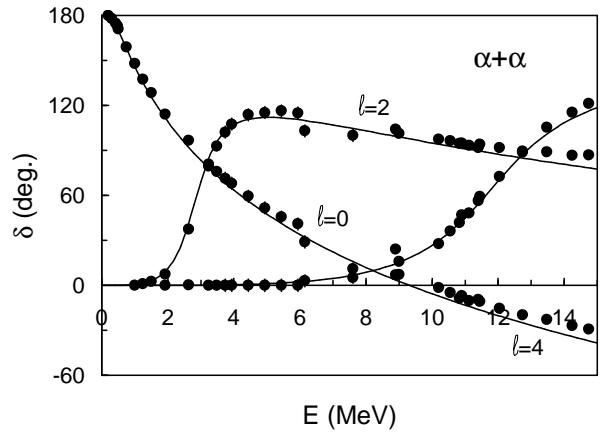} 
\caption{\label{fig_gcm1} $\alpha + \alpha$ phase shifts with the microscopic cluster
model. Experimental data are taken from \cite{AAA69}.}
\end{center}
\end{figure}

The second application deals with the $^{12}$C+p elastic scattering. The $^{12}$C internal wave 
function ($b=1.65$ fm) is described in the $p$ shell limited to zero angular momentum and isospin. 
This provides four independent shell-model states defined from a linear combination of 
Slater determinants \cite{De96}.
The calculation is therefore performed with four channels, obtained from the 
diagonalization of the $^{12}$C basis. The first eigenstate corresponds to the ground state, 
and the three additional eigenvalues are considered
as virtual excitations, which simulate the distortion of $^{12}$C in the $^{12}$C+p system.

The spin-orbit amplitude is fixed as $S_0=36.3$ MeV.fm$^5$, which allows to reproduce the $1/2^-$ and
$3/2^-$ binding energies of $^{13}$N with $u=0.77$. This value is used for negative-parity partial waves. In 
positive parity, $u=0.998$ reproduces the experimental energy of the $1/2^+$ resonance. The GCM widths
of the $1/2^+$ and $3/2^-$ resonances are 43 keV and 99 keV, somewhat larger than the experimental values
($31.7\pm 0.8$ keV and $62\pm 4$ keV, respectively \cite{Aj91}).

\begin{figure}[htb]
\begin{center}
\includegraphics[width=0.5\textwidth,clip]{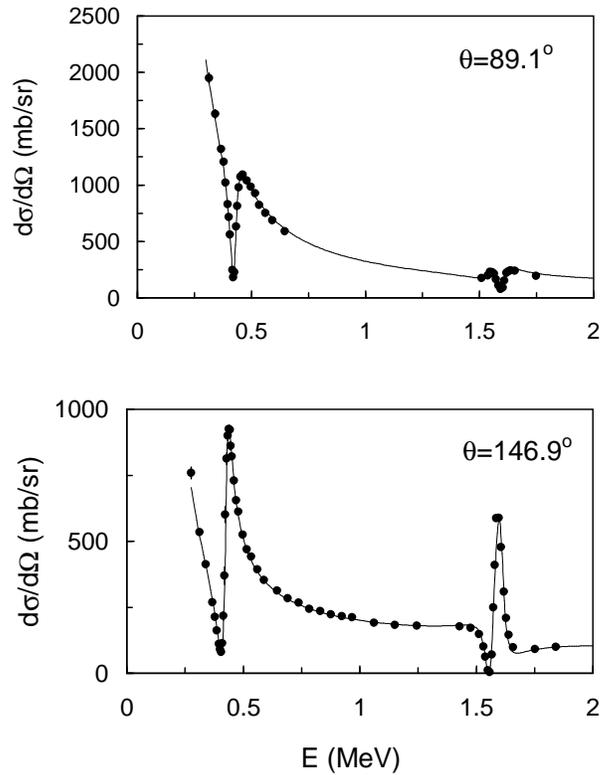} 
\caption{\label{fig_gcm2}$^{12}$C+p excitation functions at two c.m. angles. The data are from \cite{MPS76}.}
\end{center}
\end{figure}

The $^{12}$C+p excitation functions at $\theta=89.1^{\circ}$ and $\theta=146.9^{\circ}$ are 
shown in Fig.~\ref{fig_gcm2}. Below 1.5 MeV, the agreement with experiment is quite good. 
However the resonant structure near 1.6 MeV is not well described since the 
experimental data involve a $5/2^+$ resonance, which cannot be described with a simple cluster structure. 
This problem often occurs in microscopic cluster models, where some experimental states do not
have a cluster structure, and are therefore not present in the model (see, for example, \cite{BDH98}
for the $^{16}$O+p system). Of course the phenomenological approach (see Fig.~\ref{fig_phen_c12p})
provides a better description, but the parameters are adjusted to the data, and all resonances
are included from the very beginning.

Our third application of the GCM is the $\cpg$ cross section at astrophysical energies. The matrix
elements of the $E1$ operator are determined as explained in section~\ref{sec:racar} (see \cite{BD83} for detail).
The capture cross section requires the $^{13}$N ground state wave function, as well as $^{12}$C+p 
scattering states. Only $s$ waves, corresponding to the $1/2^+$ resonant partial wave are included 
in the scattering state.
Calculations with $d$ waves show that this component is of the order of 15\% at 1 MeV, but becomes
negligible as soon as the energy decreases, because of the higher centrifugal barrier.

The resulting $S$ factor \rref{2.59} is shown in Fig.~\ref{fig_gcm3}. As expected from the overestimation of
the proton width, the peak near the $1/2^+$ resonance at 0.42 MeV is slightly too broad. This type 
of calculation cannot be expected to perfectly reproduce the data. 
However the $S$ factor is quite satisfactory considering the fact that no parameter is fitted to 
capture data. 
The model has thus a predictive power which is very useful for capture reactions 
that have not been measured yet or can not be measured.
The GCM has been used to study many reactions of astrophysical interest, 
to compute either the cross section or the properties
of low-energy resonances (see for example \cite{De08}).

\begin{figure}[htb]
\begin{center}
\includegraphics[width=0.5\textwidth,clip]{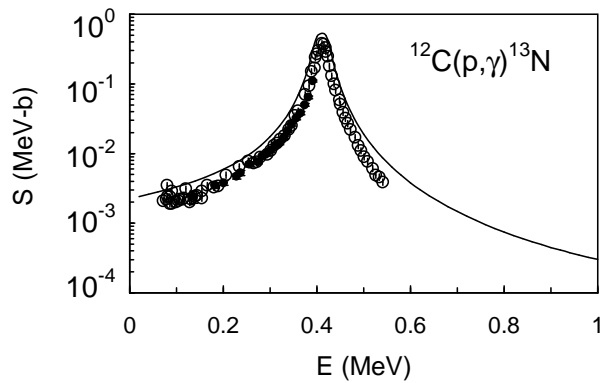} 
\caption{\label{fig_gcm3} $\cpg$ GCM $S$-factor. The experimental data
are from \cite{Vo63} (open circles) and \cite{RA74} (closed circles).}
\end{center}
\end{figure}

\subsection{The Continuum Discretized Coupled Channel (CDCC) method}
\label{sec:CDCC}
\subsubsection{Formalism\\}
The purpose of the CDCC method is to determine, as accurately as possible, the 
scattering and dissociation cross sections 
of a nucleus which can be easily broken up in the nuclear or Coulomb field of a target. 
The final states may thus involve three particles: the target and the fragments of the projectile. 
The relative motion of these fragments is described by approximate continuum wave functions 
at discrete energies.
The CDCC method was suggested by Rawitscher \cite{Ra74b} and first applied to deuteron + nucleus 
elastic scattering and breakup reactions. 
It was then extensively developed and used by several groups \cite{AIK87,NT99}. 
Its interest has been still revived by the availability of radioactive beams of weakly 
bound nuclei dissociating into three fragments, such as $^6$He 
whose first dissociation channel is $\alpha$+n+n \cite{MHO04,RAG08}.
Although variants of the CDCC method also exist in atomic physics \cite{BHS96}, we focus here on applications
in nuclear physics with two-body projectiles. To simplify the presentation, we assume a spin
zero for the constituents of the projectile, and for the target $t$.
As usual in CDCC calculations, the internal structure of the three particles is neglected.

Let us consider the coordinate system of Fig.~\ref{fig_cdcc1}: 
\ve{R} is the internal coordinate of the projectile and 
$\ve{r}$ is the coordinate of the relative motion between projectile and target. 
The three-body Hamiltonian is given by 
\beq
H=H_0+T_r+V_{t1}\biggl(\ve{r}+\frac{A_2}{A}\ve{R}\biggr)
+V_{t2}\biggl(\ve{r}-\frac{A_1}{A}\ve{R}\biggr),
\eeqn{eq6.20}
where $H_0$ is the two-body Hamiltonian of the projectile 
\beq
H_0=T_R+V_{12}(R).
\eeqn{eq6.21}
In general, potential $V_{12}(R)$ associated with the projectile is real, 
whereas the interactions $V_{ti}$ between the fragments $i$
and the target $t$ are derived from the optical model, and
thus complex. In a schematic notation, the wave function associated with (\ref{eq6.20}) 
can be expanded as 
\beq
\Psi(\ve{r},\ve{R})=\sum_B \Phi_B(\ve{R}) u_B(\ve{r})
+\int \Phi_k(\ve{R}) u_k(\ve{r}) dk,
\eeqn{eq6.22}
where $B$ denotes the bound states of the projectile, and ${\phi}_k(\ve{R})$ are 
two-body scattering wave functions with wave number $k$. The first term represents the elastic
and inelastic channels, and the second term is associated with the breakup contribution.

In practice, two methods are
available to perform the continuum discretization, i.e.\ discretize the integral over $k$.
In the ``pseudo-state'' approach, it is replaced  
by a sum over square-integrable positive-energy eigenstates of Hamiltonian \rref{eq6.21}. 
The projectile Hamiltonian $H_0$ is diagonalized over a finite basis, yielding the square-integrable
radial functions $\Phi_i^L(R)$ at energies $E_i^L$,
\beq
H_0 \Phi_i^L(R)Y_L^M(\Omega_R)=E_i^L\Phi_i^L(R)Y_L^M(\Omega_R).
\eeqn{eq6.23}
These functions are associated with bound states ($i=B, E_i<0$), or represent 
square-integrable approximations of continuum wave functions ($E_i>0$).

The alternative is to replace the integral over $k$
by averages of exact scattering states over a range 
of energies (``bin" method) \cite{AIK87}.  This approach also provides square-integrable
basis functions. As far as the applicability of the $R$-matrix is
concerned, both methods are treated in the same way. We use here the pseudo-state  method.

\begin{figure}[htb]
\begin{center}
\includegraphics[width=0.5\textwidth,clip]{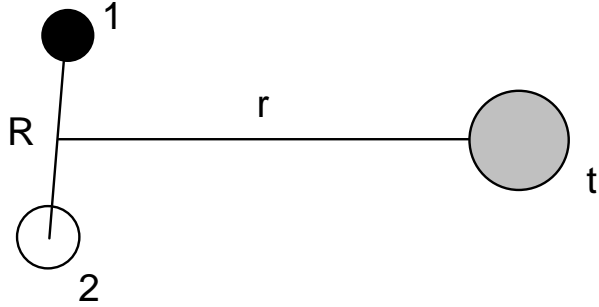} 
\caption{\label{fig_cdcc1} Coordinate system used for CDCC three-body calculations.}
\end{center}
\end{figure}
 
The total wave function $(\ref{eq6.22})$ is then rewritten, 
for an angular momentum $J$ and parity $\pi=(-1)^{l+L}$, as 
\beq
\Psi^{JM\pi}(\ve{r},\ve{R})=\sum_{lLi} 
Y_{lL}^{JM}(\Omega_r,\Omega_R)
\Phi_i^L(R) u^{J\pi}_{lLi}(r),
\eeqn{eq6.24}
where $J$ results from the coupling of orbital momenta $l$ and $L$, and
\beq
Y_{lL}^{JM}(\Omega_r,\Omega_R) =i^{l+L}
\left[ Y_l(\Omega_r)\otimes Y_L(\Omega_R) \right]^{JM}.
\eeqn{eq6.24b}
The relative wave functions $u^{J\pi}_{lLi}(r)$ are given by a set of coupled equations
\beq
\fl \left[ -\frac{\hbar^2}{2\mu}\left( \frac{d^2}{dr^2}-\frac{l(l+1)}{r^2}\right) +E^L_i-E \right] 
u^{J\pi}_{c}(r)
+\sum_{c'}V^{J\pi}_{cc'}(r) u^{J\pi}_{c'}(r)=0,
\eeqn{eq6.25}
where the channel index $c$ stands for $(lLi)$. Of course, the sum over $L$ must be truncated 
at some value $L_{\rm max}$. The sum over the pseudo-states $i$ is limited by the number of 
basis states and can be reduced further by eliminating states above a maximum energy $E_{\rm max}$. 
The CDCC problem is therefore equivalent to a system of coupled equations where the 
potentials $V^{J\pi}_{cc'}(r)$ are given by
\beq
\fl V^{J\pi}_{cc'}(r)= \nonumber \\
\fl  \langle
Y_{lL}^{JM}(\Omega_r,\Omega_R) \Phi_i^L(R) |
V_{t1}\biggl(\ve{r}+\frac{A_2}{A}\ve{R}\biggr)+V_{t2}\biggl(\ve{r}-\frac{A_1}{A}\ve{R}\biggr)
|Y_{l'L'}^{JM}(\Omega_r,\Omega_R) \Phi_{i'}^{L'}(R) \rangle.
\eeqn{eq6.26}
This matrix element represents a 5-dimensional integral over $(\Omega_r,\Omega_R,R)$. 
In practice the potentials are expanded into multipoles as
\beq
V_{t1}\biggl(\ve{r}+\frac{A_2}{A}\ve{R}\biggr)+V_{t2}\biggl(\ve{r}-\frac{A_1}{A}\ve{R}\biggr)=
\sum_{\lambda}V_{\lambda}(r, R)P_{\lambda}(\cos \theta_{Rr}),
\eeqn{eq6.26b}
where $\theta_{Rr}$ is the angle between $\ve{R}$ and $\ve{r}$, and $P_{\lambda}(x)$ a Legendre
polynomial. In practice the number of $\lambda$ values
is limited by angular-momentum couplings. The four angular integrals 
in (\ref{eq6.26}) are performed analytically, whereas the integration over $R$ may
require a numerical approach.

The system (\ref{eq6.25}) can be solved by various methods \cite{Th88}, in particular with
the $R$-matrix formalism. We expand the radial wave functions $u^{J\pi}_{c}(r)$ over
Lagrange functions \rref{eq3.4.9}, and the calculation of potential (\ref{eq6.26}) is therefore limited 
to the mesh points as in \rref{eq3.4.11c}. 
Further simplifications are possible by describing the projectile 
wave functions $\Phi^L_i(R)$ [see (\ref{eq6.23})] in a Lagrange basis as well. 
In this well known two-body problem \cite{Ba06}, the wave function is expanded in a 
basis differing from \rref{eq3.4.9} by the fact that it is constructed from Laguerre polynomials. 
The matrix elements involving $\Phi^L_i (R)$ [as, for example, in the potentials 
(\ref{eq6.26})] are determined in a fast and accurate way. The extension to three-body 
projectiles \cite{MHO04,RAG08} can also be considered in this approach. The calculations are 
much more time-consuming since the projectile wave functions depend on two radial coordinates, 
and are more difficult to handle. Consequently, even though the formulation is similar, 
the calculation of the potential matrix elements (\ref{eq6.26}) raises important numerical
difficulties, which could be addressed by the present technique. 

\subsubsection{Application to the d+$^{58}$Ni elastic scattering\\}
The CDCC theory, associated with the $R$-matrix method, is applied to the 
elastic scattering of deuterons on $^{58}$Ni at $E_{\rm lab}=80$ MeV. This collision 
has been widely covered in the 
literature \cite{AIK87,PKK99}. The $p+n$ and nucleon-$^{58}$Ni interactions are chosen as in these references.

The first step is to determine the deuteron ground  state, and the $p+n$ pseudo-states, 
from (\ref{eq6.23}). These wave functions are expanded over a Lagrange-Laguerre basis, involving 
a scaling parameter $h$ which is adapted to the size of the system \cite{Ba06}. Typical values 
are $h\sim 0.3-0.4$ fm with $15\sim 20$ basis functions. We include 
partial waves $L=0,2,4$. In a second step, the potentials $V_{cc'}(r)$ 
[see (\ref{eq6.25})] are computed at the mesh points of the d+$^{58}$Ni relative motion.
These mesh points are defined from the channel radius $a$ and from the size of the basis $N$
[see \rref{eq3.4.10}].

\begin{table}[h]
\begin{center}
\caption{Amplitude $\eta^{J\pi}$ and phase shift $\delta^{J\pi}$ for elastic d+$^{58}$Ni elastic
scattering $(L_{\rm max}=4, E_{\rm max}=40$ MeV) for $J^{\pi}=0^+$ and $17^-$.
\label{tab_cdcc1}}
\begin{tabular}{cc|cc|cc}
$a$ (fm) & $N$ & $\eta^{0+}$ &  $\delta^{0+}$ & $\eta^{17-}$ & $\delta^{17-}$  \\
 \hline
15 & 30 &   0.117 &   31.5 &   0.5958 &   20.454\\
17 & 30 &   0.105 &   29.7 &   0.5958 &   20.454\\
15 & 40 &   0.116 &   31.2 &   0.5958 &   20.454\\
17 & 40 &   0.116 &   31.1 &   0.5959 &   20.454\\
\multicolumn{2}{c|}{\cite{PKK99}} &  &  &   0.5956 &   19.9\\
\end{tabular}
\end{center}
\end{table}

In Table \ref{tab_cdcc1} and Fig.~\ref{fig_cdcc2}, we present the elastic part of the 
collision matrix 
\beq
U^{J\pi}_{11}=\eta^{J\pi}\exp(2i\delta^{J\pi}) 
\eeq
for $J^{\pi}=0^+$ and $17^-$. Values for $J^{\pi}=17^-$ can be 
compared with the literature \cite{PKK99}. The amplitude $\eta$ and phase shift $\delta$ 
are computed for different channel radii and numbers of wave functions. For the pseudo-states, 
we truncate at $L_{\rm max}=4$ and $E_{\rm max}=50$ MeV. The number of pseudo-states depends on the size
of the basis (here, typically their number is about 20). We have checked that changing the size
of the basis does not affect the phase shifts.
Table \ref{tab_cdcc1} shows that, 
for high partial waves $(J^{\pi}={17}^-)$, the stability with the numerical conditions is 
virtually perfect. The amplitude is in excellent agreement with \cite{PKK99}, but 
the phase shift is slightly different $(0.5^{\circ})$. For low partial waves $(J^{\pi}=0^+)$, 
the stability is still acceptable but higher $N$ values are necessary. This can be understood 
by the fact that the internal part is more and more important as $J$ decreases. 

\begin{figure}[ht]
\begin{center}
\includegraphics[width=0.7\textwidth,clip]{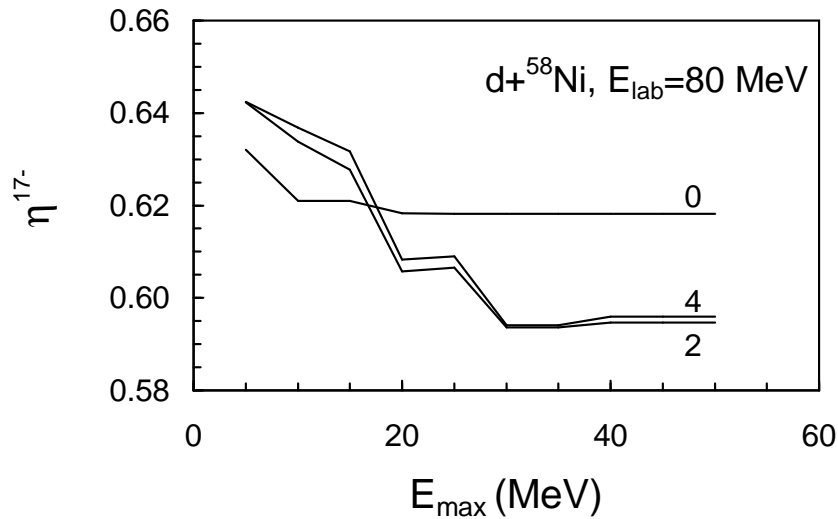} 
\caption{\label{fig_cdcc2} Amplitude $\eta^{17-}$ for different $L_{\rm max}$ values (labels) as a function
of the truncation energy $E_{\rm max}$.}
\end{center}
\end{figure}

Fig.~\ref{fig_cdcc2} displays the amplitude ${\eta}^{17^-}$ as a function of the 
$p+n$ truncation energy $E_{\rm max}$, and for different $L_{\rm max}$ values. The convergence 
with respect to $E_{\rm max}$ is reached near $E_{\rm max}\approx 40$ MeV, which corresponds 
to $k_{\rm max}\approx 1$ fm$^{-1}$. This result agrees with the conclusion of \cite{PKK99}. 
From the figure, it is clear that $L=0$ pseudo-states are not sufficient to provide 
accurate values. However, a very good convergence is already obtained with $L_{\rm max}=2$.

The elastic cross section is presented in Fig.~\ref{fig_cdcc3}, and compared with 
experimental data (quoted in \cite{YNI82}). 
A first calculation, referred to as ``no-breakup" approximation, is done by including only the ground state
of the deuteron (dotted line). This approximation clearly overestimates the data above
$30^{\circ}$. With $L=0$ deuteron pseudo-states, the agreement in the region $30^{\circ}-50^{\circ}$
is significantly improved.
The present result is very close to the 
CDCC calculation of \cite{YNI82}. The curves with $L_{\rm max}=2$ and $L_{\rm max}=4$ 
are indistinguishable at the scale of the figure.

\begin{figure}[htb]
\begin{center}
\includegraphics[width=0.7\textwidth,clip]{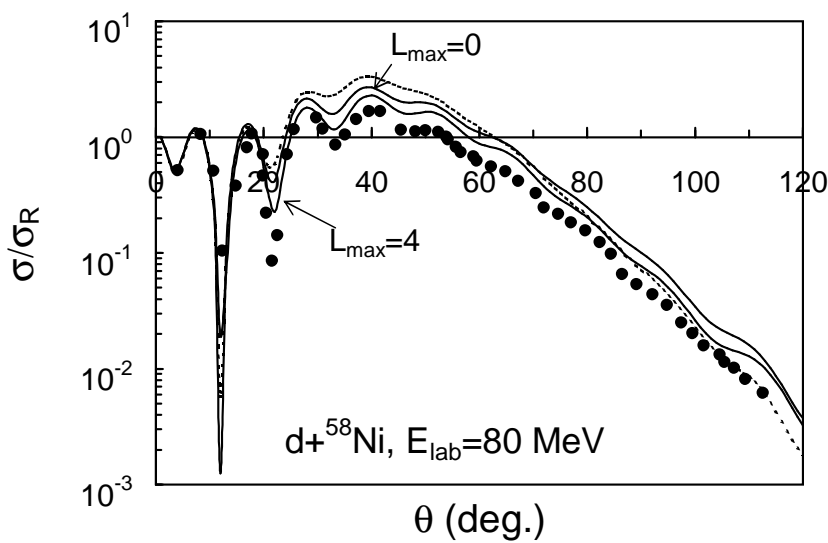} 
\caption{\label{fig_cdcc3} d+$^{58}$Ni elastic cross section relative to the Rutherford cross
section at $E_{\rm lab}=80$ MeV. Solid lines correspond to $E_{\rm max}=40$ MeV and to $L_{\rm max}=0,4$. 
The dotted line corresponds to the no-breakup approximation. Experimental data are
taken from \cite{YNI82}.}
\end{center}
\end{figure}

From this example, it turns out that the $R$-matrix method is a very efficient tool to solve 
the CDCC equations. By using Lagrange functions, the calculation of the coupling 
potentials $V_{cc'}(r)$ is very fast, since only values at the mesh points are 
required (typically $N \sim 30-40$). Consequently, the main part of the computer time is 
devoted to the inversion of the (complex) matrix $\ve{C}$ [see (\ref{32.9})]. Fast techniques are 
available for matrix inversion, and the total computer time is therefore very short 
(typically a few seconds for the angular distribution of Fig.~\ref{fig_cdcc3}). 
This opens encouraging perspectives for CDCC calculations involving three-body projectiles 
which require large computer times with standard techniques \cite{MHO04,RAG08}.

\subsection{Three-body continuum states}
\label{sec:tbcs}
\subsubsection{Hyperspherical formalism\\}
The separation energy being very low in exotic nuclei, a 
precise treatment of the continuum is necessary for the description of reactions 
leading to a three-body dissociation. 
The $R$-matrix method provides an efficient way to treat three-body continuum 
states \cite{TDE00,DTB06}. We use the hyperspherical formalism, 
well adapted to three-body systems \cite{MF53,ZDF93,TND04}. This model is well known, and is 
just briefly outlined here.

In a three-body system, the Hamiltonian is defined, after removal of the c.m. motion, as 
\beq
H=\sum_{i=1}^3 T_i -T_{cm}+\sum_{i>j=1}^3 V_{ij}(\ve{r}_j-\ve{r}_i),
\eeqn{eq6.27}
where $\ve{r}_i$ are the space coordinates of the particles, $T_i$ their kinetic energy 
and $V_{ij}$ the interaction between particles $i$ and $j$. The scaled Jacobi coordinates 
[$\ve{x}=(x,\Omega_x)$ and $\ve{y}=(y,\Omega_y)$] 
are defined from the coordinates $\ve{r}_i$, and provide the hyperradius $\rho$ and 
hyperangle $\alpha$ as
\beq
&&\rho=\sqrt{x^2+y^2}, \nonumber \\
&&\alpha=\arctan(y/x).
\eeqn{eq6.28} 
In this coordinate system, the 3-body kinetic energy involves the operator $\ve{K}^2$ which generalizes
the concept of angular momentum in 2-body systems. It commutes with $\ve{l}_x^2$ and $\ve{l}_y^2$ and
their common eigenfunctions 
${\cal Y}^{JM}_{\gamma K}(\Omega_{5})$ are known analytically (see \cite{ZDF93} for details). 
The eigenvalue of $\ve{K}^2$ is $K(K+4)$ where the integer $K$ is the hypermomentum quantum number. 
In these expressions, $\Omega_5= (\Omega_x,\Omega_y,\alpha)$ and $\gamma$ stands for 
$\gamma=(l_x,l_y,L,S)$ where $(l_x,l_y)$ are the angular momenta associated with $(x,y)$,
$L$ and $S$ are the total orbital momentum and spin, respectively. The total angular momentum
$J$ results from the coupling of $L$ and $S$, and the parity is given by $\pi=(-1)^K$. 

The wave function of partial wave $J\pi$ associated with Hamiltonian (\ref{eq6.27}) is then expanded over 
hyperspherical harmonics  as
\beq
\Psi^{JM\pi}(\rho,\Omega_5)=\rho^{-5/2}\sum_{\gamma K} \chi^{J\pi}_{\gamma K}(\rho)\ {\cal Y}^{JM}_{\gamma K}(\Omega_5),
\eeqn{eq6.29}
where the radial functions $\chi^{J\pi}_{\gamma K}(\rho)$ have to be determined. 
The Schr\"odinger equation is then reduced to a system of coupled 
differential equations
\beq
&&\left[-\frac{\hbar^2}{2m_N} \left( \frac{d^2}{d\rho^2} - \frac{(K+3/2)(K+5/2)}{\rho^2}\right) -E \right]
{\chi}^{J\pi}_{\gamma K}(\rho)
 \nonumber \\
&& +\sum_{K' \gamma'} V^{J\pi}_{K' \gamma',K \gamma}(\rho)\, {\chi}^{J\pi}_{\gamma' K'}(\rho)=0,
\eeqn{eq6.30} 
where the potentials matrix elements are defined as
\beq
V^{J\pi}_{K' \gamma',K \gamma}(\rho)=
\langle {\cal Y}^{JM}_{\gamma' K'}(\Omega_5) | \sum_{i>j=1}^3 V_{ij}(\ve{r}_j-\ve{r}_i) |
{\cal Y}^{JM}_{\gamma K}(\Omega_5) \rangle.
\eeqn{eq6.31} 
The integral over $\Omega_x$ and $\Omega_y$ are performed analytically, whereas a numerical
quadrature is used for the integral over the hyperangle $\alpha$.
With the Raynal-Revai coefficients \cite{RR70,Ra76} the evaluation of 
(\ref{eq6.31}) is rather easy. 
A truncation must be done in the summation over $K$; this provides a maximum $K$ 
value, denoted as $K_{\rm max}$. The number of components in (\ref{eq6.29}) increases rapidly when 
$K_{\rm max}$ increases. 

This formalism has been extensively applied to three-body bound states \cite{ZDF93}.
In \cite{DDB03}, we have addressed this problem by expanding the radial wave functions 
over Lagrange functions. This approach provides a very fast method to evaluate matrix 
elements (\ref{eq6.31}).

One of the main issues associated with nuclear three-body problems is the presence of 
forbidden states in the nucleus-nucleus interaction \cite{WT77,BFW77}. These two-body 
forbidden states introduce spurious states in the three-body problem and should be removed. 
This problem has been discussed in detail by Thompson {\sl et al.} \cite{TDE00}, and is 
usually solved, either by introducing a projector over the forbidden states \cite{KP78}, 
or by using supersymmetric potentials \cite{Ba87}.

\subsubsection{The $R$-matrix method for three-body states\\}
The treatment of three-body slates, with exact three-body asymptotic conditions, is more 
recent \cite{TDE00,DTB06}. 
In the $R$-matrix method, the solutions of the
system (\ref{eq6.30}) are split in two regions, 
\beq
\chi^{J\pi}_{\gamma K, {\rm int}}(\rho)=\sum_{i=1}^N\, c^{J\pi}_{\gamma Ki}\, \varphi_i(\rho),
\eeqn{eq6.32}
for $\rho < a$ where $\varphi_i(\rho)$ are basis functions and 
\beq
\chi^{J\pi}_{\gamma K, {\rm ext}}(\rho)=C^{J\pi}_{\gamma K}
\left[ H^-_{\gamma K}(k\rho)\delta_{\gamma \gamma'}\delta_{KK'}
-U^{J\pi}_{\gamma K,\gamma' K'}H^+_{\gamma K}(k\rho) \right] 
\eeqn{eq6.32a}
for $\rho \geq a$. 
In this equation, $C^{J\pi}_{\gamma K}$ is a normalization coefficient, 
$\ve{U}^{J\pi}$ is the three-body collision matrix 
and the incoming and outgoing functions $H^{\pm}_{\gamma K} (x)$ are defined as
\beq
H^{\pm}_{\gamma K} (x)=\pm i \left( \frac{\pi x}{2} \right) ^{1/2}
\left[ J_{K+2}(x)\pm i Y_{K+2}(x) \right] ,
\label{eq6.32b}
\eeq
where $J_n (x)$ and $Y_n (x)$ are Bessel functions of first and second kind, respectively.
We assume here systems without two-body Coulomb interaction.

As in previous applications, we choose Lagrange functions for the basis states $\varphi_i(\rho)$ \cite{DTB06}. 
The $R$-matrix method is then used to solve the coupled system (\ref{eq6.30}). Formally 
this is equivalent to the CDCC system [see (\ref{eq6.25})], 
although the external wave function involves Bessel functions. Another difference arises 
from the long range of the three-body potential (\ref{eq6.31}). As shown in \cite{DTB06}, 
this potential behaves as 
\beq
V^{J\pi}_{K' \gamma',K \gamma}(\rho) \rightarrow \frac{V^{J\pi}_{0,K' \gamma',K \gamma}}{\rho^3},
\eeqn{eq6.34}
even with short-range two-body interactions. 
This arises from the definition of the hyperspherical coordinates. Even for large $\rho$ values, two 
particles can still be close to each other and interact strongly. 
Constants $V^{J\pi}_{0,K' \gamma',K \gamma}$ can take rather large values (examples are given in \cite{DTB06}). For this reason, 
the $R$-matrix radius should take large values (typically $a \sim 200-300$ fm) to ensure 
that potential (\ref{eq6.34}) is negligible compared with the centrifugal term in \rref{eq6.30}. 
In those conditions, the propagation techniques described in section~\ref{subsec:propa} are necessary 
to avoid huge basis sizes. 

\subsubsection{Application to $\alpha+n+n$ three-body scattering\\}
The $^6$He nucleus is an ideal test case for three-body continuum states and has been 
considered in previous studies \cite{TDE00,DTB06}. Accurate $\alpha+n$ and $n+n$ 
interactions exist in the literature \cite{KKN79, TLT77}. The $\alpha+n$ subsystem 
presents one forbidden slate for $l=0$, which is removed by using a pair of supersymmetric 
transformations \cite{Ba87}. Details can be found in \cite{DTB06}.

Figure \ref{fig_3b1} illustrates the need for propagation techniques. Dotted lines correspond 
to channel radii $a=20$ fm and $a=30$ fm, and are compared with the exact values (solid 
line, $a=250$ fm) obtained by propagation. As expected from the long range of the 
three-body potential (\ref{eq6.31}), large values for the channel radius are necessary. 
As soon as this condition is satisfied, the $R$-matrix phase shifts are very stable against variations of $a$.

\begin{figure}[htb]
\begin{center}
\includegraphics[width=0.5\textwidth,clip]{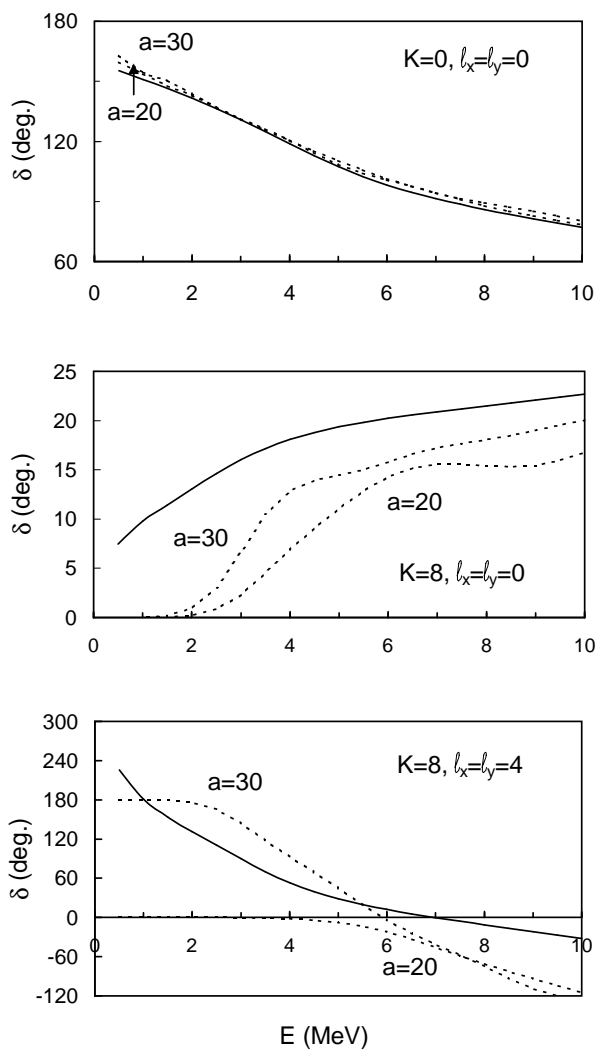} 
\caption{\label{fig_3b1} $\alpha$+n+n phase shifts ($J^{\pi}=0^+$) for channel radii $a=20$ fm 
$(N=20)$ and $a=30$ fm $(N=30)$ without propagation (dashed lines). Solid lines are obtained 
with propagation up to $a=250$ fm (from \cite{DTB06}).}
\end{center}
\end{figure}

In Fig.~\ref{fig_3b2}, we display the $0^+$ eigenphases as a function of $K_{\rm max}$. The eigenphases are 
obtained after diagonalization of the collision matrix [see \rref{2.44c}]. 
In each case we select the eigenphase with the dominant resonant structure.
Above 4 MeV, a fair convergence is obtained, 
but high hypermomenta are necessary near 2 MeV. The $0^+,1^-,2^+$ eigenphases are displayed in 
Fig.~\ref{fig_3b3} with $K_{\rm max}=24,19,{\rm\ and \ }16$, respectively. 
The $2^+$ phase shift presents an experimentally well known narrow resonance at low energies. 
For the $0^+$ and $1^-$ partial waves, the calculation shows a broad structure near 1.5 MeV. The existence 
of three-body resonances at low energies, and in particular for $J^{\pi}=1^-$, is still an open debate 
(see, for example, the discussion in \cite{BCD09}), from the experimental as well as from 
the theoretical viewpoints.

\begin{figure}[htb]
\begin{center}
\includegraphics[width=0.5\textwidth,clip]{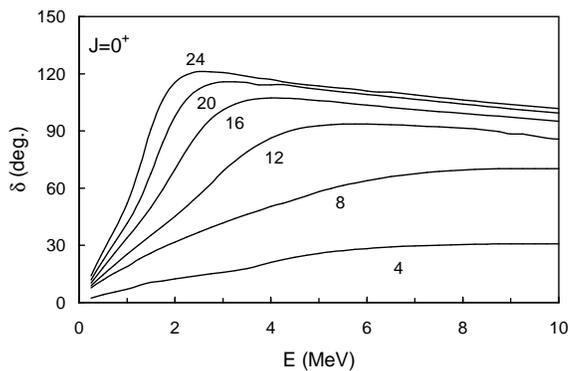} 
\caption{\label{fig_3b2} Energy dependence of $\alpha$+n+n eigenphases ($J^{\pi}=0^+$) 
for different $K_{\rm max}$ values (from \cite{DTB06}).}
\end{center}
\end{figure}

\begin{figure}[htb]
\begin{center}
\includegraphics[width=0.5\textwidth,clip]{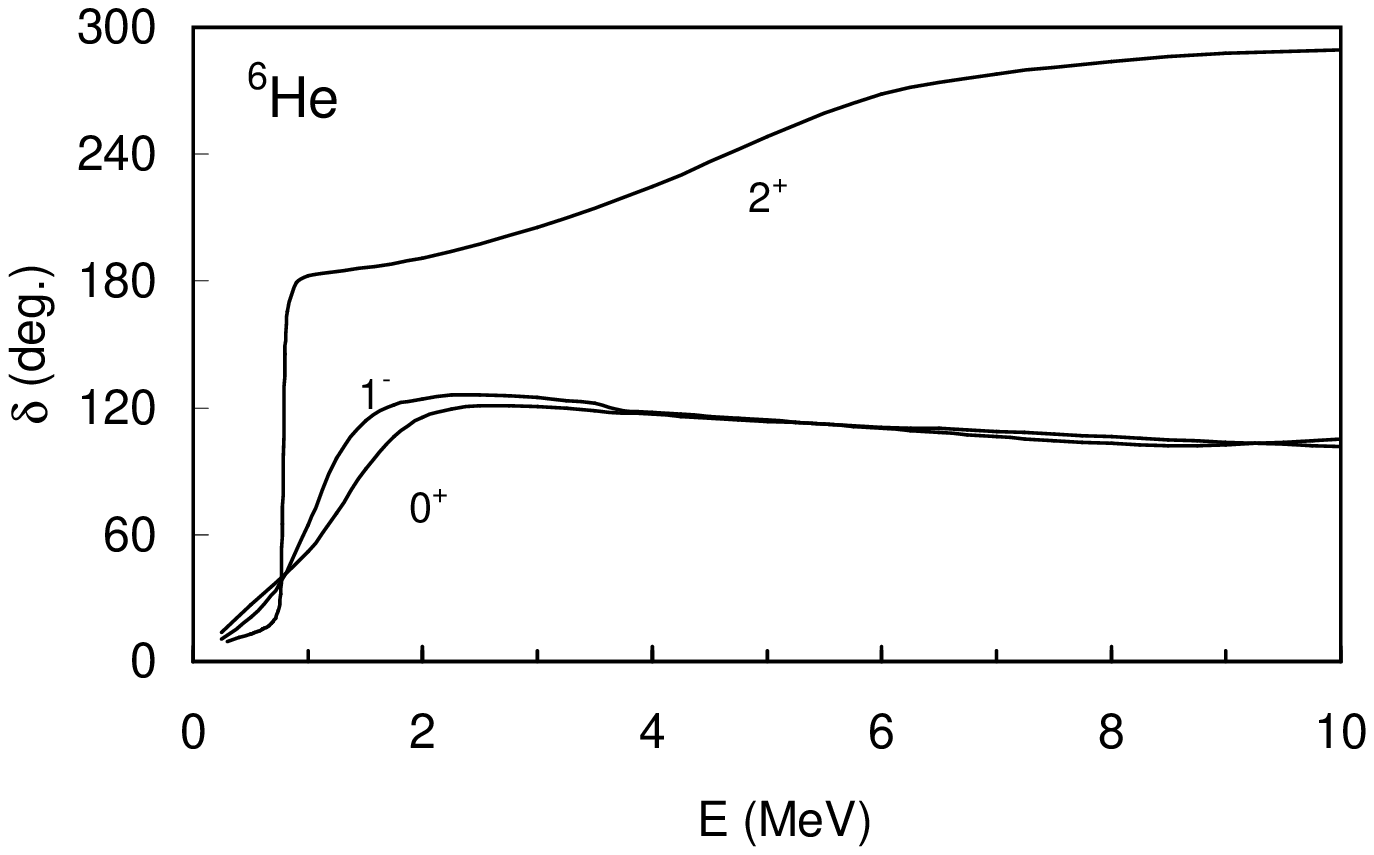} 
\caption{\label{fig_3b3} Eigenphases of $^6$He for different $J$ values (from \cite{DTB06}).}
\end{center}
\end{figure}

In the future, this formalism could be applied to the $3\alpha$ system. Current experimental \cite{IAF04} and 
theoretical \cite{DB87b,KK07,AGJ07} results are rather controversial concerning the existence (or non-existence) 
of broad $0^+$ and $2^+$ resonances above the $3\alpha$ threshold. This issue is crucial in 
nuclear astrophysics, since the Hoyle state $(0^+_2)$ in $^{12}$C is known to be the main resonance 
in Helium burning. If the properties of this resonance are well known, the $^{12}$C level scheme above this 
resonance is still not clear. In this context, the calculation of $3\alpha$ phase shifts would help 
clarifying the situation. However, existing $\alpha+\alpha$ local potentials do not provide a 
satisfactory description of the $^{12}$C spectroscopy \cite{SMO08}. Non-local potentials are more 
promising \cite{SMO08,PM08}, but although bound states can be easily investigated with Lagrange 
meshes \cite{DTB06}, their application to three-body continuum states remains a challenge for the future.

\section{Conclusion}
\label{sec:conc}

The $R$ matrix method was born sixty years ago with the important 
but rather limited goal of describing resonances in nuclear reactions. 
Today it has evolved into powerful tools covering several 
subfields of atomic, molecular and nuclear collisions. 
The literature is so enormous that it is not possible to master it 
and to cover it in a single review. 

We have emphasized a fact that is often unknown to $R$ matrix practitioners: 
two variants exist. 
The phenomenological $R$ matrix remains close to the original spirit 
and is very much used in nuclear physics to parametrize low-energy cross sections. 
Its main merits are that all parameters are real and that they have a physical meaning.
Although resonances often play a crucial role in these parametrizations, 
non-resonant cross sections are accurately described as well. 
The calculable $R$ matrix is an efficient technique to solve the stationary 
or time-dependent Schr\"odinger equation in various situations 
as well as its relativistic extensions. 
It underwent most of its developments in atomic physics but we have shown 
in section \ref{sec:recent} that it can also be useful in nuclear physics. 

Because of the variety and complexity of its applications, 
the $R$-matrix theory has been and is still sometimes misunderstood or misjudged 
for different reasons. 
A first often-made criticism concerns the role of the channel radius. 
The $R$-matrix formulas depend on this radius which has no strict physical meaning. 
This criticism can not be addressed to the calculable $R$ matrix where the independence 
of the physical results on this radius, provided that it is large enough, 
is a useful validity test. 
In the phenomenological $R$-matrix however, the channel radius is indeed a parameter 
whose value is disputable. 
This arises from the truncation of the $R$ matrix to a small number of poles. 
In spite of this truncation, this approximation often provides excellent fits to the data. 
The standard option is to optimize the channel radius together with the other parameters. 
This radius should however be larger than the sum of the radii of the colliding nuclei. 
Its value should always be mentioned because the other parameters depend on its choice. 

Another criticism deals with a reputation of poor convergence of the calculable $R$ matrix. 
As we have shown, this reputation is undeserved. 
Its origin lies in the choice of a common logarithmic derivative for the basis states 
in the founding papers. 
While this choice is acceptable (although with some discontinuity drawback) for an 
infinite basis, it leads to inaccuracies when the basis is truncated. 
The introduction of the Bloch operator has opened a way to the use of finite bases 
providing a sufficient variety of behaviours at the boundary for which this problem disappears, 
a fact not yet known enough. 
Modern $R$-matrix codes can employ different types of such bases. 
They reach an excellent convergence and do not require the use of the Buttle correction. 

The calculable $R$ matrix provides bound-state and scattering-state wave functions 
that can be used in a variety of applications 
with sometimes technical complications due to the existence of two regions
to define the wave function. 
In atomic, molecular and nuclear physics, the challenge is now to reach the same level of accuracy 
for processes with more than two unbound particles in the final states. 
Advances have been made for double ionization by a photon in atomic and molecular physics 
and for the breakup of two- and three-body halo nuclei in nuclear physics. 
Progresses of $R$-matrix theory in these directions should still be expected. \\
\\
\section*{Acknowledgments}
We are grateful to our present and former colleagues of the department "Physique Nucl\'eaire Th\'eorique
et Physique Math\'ematique" for many common works and helpful discussions.
This text presents research results of the Belgian program P6/23 on
interuniversity attraction poles initiated by the Belgian-state
Federal Services for Scientific, Technical and Cultural Affairs (FSTC). \\

\setcounter{section}{0}
\setcounter{equation}{0}
\renewcommand{\theequation}{A\arabic{equation}}
\renewcommand{\thesection}{\Alph{section}}
\section{Appendix: Collision matrix and $K$ matrix}
\label{sec:A}
System \rref{2.45} possesses $N$ linearly independent solutions that vanish at the origin. 
A matrix solution $\ve{u}$ is obtained by putting those independent solutions as columns 
of a square matrix. 
By multiplication on the right by any invertible matrix, one obtains another matrix solution 
of system \rref{2.45} which is physically equivalent. 
The most general asymptotic expression of such a matrix generalizes \rref{2.44} as 
\beq
\ve{u} \arrow{r}{\infty} \ve{v}^{-1/2} (\ve{I} - \ve{O} \ve{U}) \ve{C},
\eeqn{K.1}
where $\ve{U}$ is the collision matrix,  
$\ve{I}$ and $\ve{O}$ are complex conjugate diagonal matrices involving 
incoming and outgoing Coulomb functions \rref{2.8} on the diagonal, 
and $\ve{v}$ is a diagonal matrix of velocities.
Complex matrix $\ve{C}$ is arbitrary non singular. 
In \rref{2.44}, we have chosen $\ve{C}$ diagonal for simplicity. 

With $\ve{C} = i (1 + \ve{U})^{-1} \ve{C}'$ where matrix $\ve{C}'$ is also non singular, 
an equivalent asymptotic form which is often used is obtained with \rref{2.8} as 
\beq
\ve{u} \arrow{r}{\infty} \ve{v}^{-1/2} (\ve{F} + \ve{G} \ve{K}) \ve{C}'
\eeqn{K.2}
where $\ve{F}$ and $\ve{G}$ are real diagonal matrices involving 
regular and irregular Coulomb functions $F_l$ and $G_l$ on the diagonal. 
This asymptotic form is real if $\ve{C}'$ is real.  
It is then the most general asymptotic form of real solutions.  

Matrices $\ve{U}$ and $\ve{K}$ are related by 
\beq
\ve{U} = (1 - i\ve{K})^{-1} (1 + i\ve{K})
\eeqn{K.3}
or 
\beq
\ve{K} = i (1 - \ve{U}) (1 + \ve{U})^{-1}.
\eeqn{K.4}
Matrix $\ve{K}$ is real and symmetric if $\ve{U}$ is unitary and symmetric. 
\setcounter{equation}{0}
\renewcommand{\theequation}{B\arabic{equation}}
\section{Appendix: Proof of relation \rref{32.13}}
\label{sec:B}
Let $\ve{B}$ be an invertible $N \times N$ matrix and $u$ and $v$ be vectors with $N$ components. 
The inverse of the square matrix 
\beq
\ve{A} = \ve{B} + uv^T
\eeqn{A.1}
is given by
\beq
\ve{A}^{-1} = \ve{B}^{-1} - \frac{\ve{B}^{-1} uv^T \ve{B}^{-1}}
{1 + v^T \ve{B}^{-1} u},
\eeqn{A.2}
where the denominator is a scalar.
A corollary of (\ref{A.2}) reads 
\beq
\ve{A}^{-1} u = \frac{\ve{B}^{-1} u}{1 + v^T \ve{B}^{-1} u}.
\eeqn{A.3}
Another corollary is the relation 
\beq
(v^T \ve{A}^{-1} u)^{-1} = 1 + (v^T \ve{B}^{-1} u)^{-1}
\eeqn{A.4}
from which (\ref{32.13}) follows. 
\setcounter{equation}{0}
\renewcommand{\theequation}{C\arabic{equation}}
\section{Appendix: Matrix elements for various basis functions}
\label{sec:C}
Here we present the matrix elements used for different basis functions in Sect.~4. Unless specified otherwise the
matrix elements of the kinetic energy are given for $l=0$ and in units of $\hbar^2/2\mu$.
\begin{enumerate}
\item{{\em Sine functions}\\
The overlap matrix elements between basis functions \rref{eq3.4.14} are given by
\beq
\la \varphi_i | \varphi_j \ra = \frac{a}{2} \delta_{ij}.
\eeqn{eqmatsin}
The matrix elements for the kinetic energy are simple, 
\beq
\la \varphi_i | T_0 | \varphi_j \ra = \frac{\pi^2}{2a} \left(i-\frac{1}{2}\right)^2 \delta_{ij}.
\eeqn{eqmatsin2}
Because of property \rref{eq_sine} at the boundary, matrix elements of the Bloch operator ${\cal L}(0)$
vanish. The matrix elements of $1/r^2$ are also analytical but involve
Sine Integral functions. For the potential, a numerical treatment is necessary.
}
\item{{\em Gaussian functions} \\
Let us define
\beq
I_k(\nu)&=&\int_0^a r^k \exp(-\nu\,r^2)\, dr \nonumber \\
&=& \gamma((k+1)/2,\nu \, a^2)/2\nu^{(k+1)/2},
\eeqn{eq_g1}
where $\gamma$ is the incomplete gamma function and $a$ is implied.
The overlap matrix elements between basis functions \rref{eq3.4.5} are given by
\beq
\la \varphi_i | \varphi_j \ra =I_{2l+2}(\nu_i+\nu_j),
\eeqn{eq_g2}
with $\nu_i=1/b_i^2$.
For the kinetic energy, we have
\beq
\la \varphi_i | T_l+\cL(0) | \varphi_j \ra &=&4\nu_i\nu_j I_{2l+4}(\nu_i+\nu_j)\nonumber \\
&&-2(l+1)(\nu_i+\nu_j)I_{2l+2}(\nu_i+\nu_j)\nonumber \\
&&+(l+1)(2l+1)I_{2l}(\nu_i+\nu_j).
\eeqn{eq_g3}
For a Gaussian potential and for the Coulomb potential, the matrix elements read
\beq
\la \varphi_i | \exp(-(r/r_0)^2) | \varphi_j \ra =I_{2l+2}(\nu_i+\nu_j+1/r_0^2),
\eoln{eq_g4a}
\la \varphi_i | 1/r | \varphi_j \ra =I_{2l+1}(\nu_i+\nu_j),
\eeqn{eq_g4b}
and therefore do not require any numerical integration.
Of course, other potentials can be considered, but the matrix elements must be, in general, 
obtained from a numerical integration.}
\item{{\em Lagrange functions} \\
Let us start with matrix elements in interval $(0,a)$. 
The regularization coefficient $n$ in \rref{eq3.4.9} is taken as $n=1$. 
This ensures that the wave function vanishes at the origin and allows that 
the Coulomb potential is treated accurately at the Gauss approximation. 
Using wave functions (\ref{eq3.4.11}) 
and the corresponding Gauss approximation 
the matrix elements take a very simple form
\beq
\la \varphi_i | \varphi_j \ra&=&\delta_{ij}
\eeqn{eq_lag1a}
\beq
\la \varphi_i |V| \varphi_j \ra & =& V(a x_i)\,\delta_{ij}
\eeqn{eq_lag1b}
i.e.\ they only require the evaluation of the potential at the mesh points.
For the kinetic energy, a simple calculation \cite{BGS02} provides for $i=j$ 
\beq
\la \varphi_i |T_0 + \cL(0) | \varphi_i \ra 
= \frac{(4N^2+4N+3) x_i(1-x_i) - 6 x_i + 1}{3a^2 x_i^2 {(1-x_i)}^2}
\eeqn{eq_lag2}
and for $i \ne j$,
\beq
\la \varphi_i |T_0 + \cL(0) | \varphi_j \ra 
 = \frac{(-1)^{i+j}}{a^2[x_ix_j(1-x_i)(1-x_j)]^{1/2}}
\eol \times 
\left[ N^2+N+1 + \frac{x_i+x_j-2x_ix_j}{(x_i-x_j)^2} -\frac{1}{1-x_i}-\frac{1}{1-x_ j} \right].
\eeqn{eq_lag3}
Thanks to the Bloch operator, this matrix element is symmetric. 
For $l\neq 0$, the centrifugal term is included in the potential \rref{eq_lag1b}.

Next we consider $(a_1,a_2)$ intervals which are used in the propagation method. 
The basis functions \rref{eq3.4.9} with $n=0$ are extended to
\beq
\varphi_i (r)=(-1)^{N+i}\left( x_i (1-x_i) \Delta a \right) ^{1/2}
\frac{P_N\left( (2r-a_1-a_2)/\Delta a\right) }{r-x_i\Delta a -a_1},
\eeqn{eq_lag4}
with $\Delta a=a_2-a_1$, and the Lagrange condition becomes
\beq
\varphi_i(a_1+ x_j\Delta a)=(\lambda_i\Delta a )^{-1/2}\delta_{ij}.
\eeqn{eq_lag5}

The matrix elements of the potential read
\beq
\la \varphi_i |V| \varphi_j \ra & =& V(a_1+x_i\Delta a )\,\delta_{ij},
\eeqn{eq_lag5b}
and are still given by a simple evaluation of the potential at the mesh points.
The matrix elements of the kinetic energy are given at the Gauss approximation by 
\beq
\la \varphi_i |T_0  | \varphi_i \ra &=&\frac{1}{3\Delta a^2 x_i(1-x_i)}
\left[ N^2+N+6-\frac{2}{x_i(1-x_i)}\right ] 
\eeqn{eq_lag6a}
for $i=j$ and 
\beq
\la \varphi_i |T_0  | \varphi_j \ra &=&\frac{(-1)^{i+j}}{\Delta a^2}
\sqrt{\frac{x_j(1-x_j)}{x_i(1-x_i)}}\frac{2x_ix_j+3x_i-x_j-4x_i^2}{x_i(1-x_i)(x_j-x_i)^2}
\eeqn{eq_lag6b}
for $i\neq j$. 
The matrix elements of the Bloch operators read 
\beq
\la \varphi_i |{\cal L}_{a_2}  | \varphi_j \ra &=&\frac{(-1)^{i+j}}{\Delta a^2}
\sqrt{\frac{x_ix_j}{(1-x_i)(1-x_j)}}\left[ N(N+1)-\frac{1}{1-x_j}\right]
\eeqn{eq_lag6c}
and 
\beq
\la \varphi_i |{\cal L}_{a_1}  | \varphi_j \ra &=&\frac{(-1)^{i+j}}{\Delta a^2}
\sqrt{\frac{(1-x_i)(1-x_j)}{x_ix_j}}\left[-N(N+1)+\frac{1}{x_j}\right] .
\eeqn{eq_lag6}
Although this does not appear clearly, one can verify that the matrix elements of operator 
$T_0+{\cal L}_{a_2}-{\cal L}_{a_1}$ are symmetric in accord with the fact that this operator 
is Hermitian over the region $(a_1,a_2)$.
}

\end{enumerate}

\section*{References}
\providecommand{\newblock}{}

\end{document}